\documentclass[iop,twocolumn]{aastex63}
\hypersetup{citecolor=blue,urlcolor=red}
\shorttitle{Correlation analysis between several observables of OJ~287 radio jet}
\shortauthors{Q. Yuan et al.}
\usepackage{multirow}
\usepackage{hyperref}
\usepackage{amssymb}
\usepackage{url}
\usepackage{subfiles}
\makeatletter

\newcommand{\Rmnum}[1]{\expandafter\@slowromancap\romannumeral #1@}

\makeatother

\begin{document}

\title{Correlation analysis between OJ~287 radio jet observables}

\correspondingauthor{M. Zhang}
\email{mgnahz@gmail.com}

\author[0000-0003-4671-1740]{Q. Yuan}
\affiliation{Xinjiang Astronomical Observatory, Chinese Academy of Sciences, 150 Science 1-Street, Urumqi 830011, China}
\affiliation{University of Chinese Academy of Sciences, 19A Yuquan Road, Beijing 100049, China}

\author[0000-0002-8315-2848]{M. Zhang}
\affiliation{Xinjiang Astronomical Observatory, Chinese Academy of Sciences, 150 Science 1-Street, Urumqi 830011, China}
\affiliation{Key Laboratory for Radio Astronomy, Chinese Academy of Sciences, 2 West Beijing Road, Nanjing 210008, China}
\affiliation{Xinjiang Key Laboratory of Radio Astrophysics, 150 Science 1-Street, Urumqi 830011, China}

\author{X. Liu}
\affiliation{Xinjiang Astronomical Observatory, Chinese Academy of Sciences, 150 Science 1-Street, Urumqi 830011, China}
\affiliation{Key Laboratory for Radio Astronomy, Chinese Academy of Sciences, 2 West Beijing Road, Nanjing 210008, China}
\affiliation{Xinjiang Key Laboratory of Radio Astrophysics, 150 Science 1-Street, Urumqi 830011, China}

\author{P.F. Jiang}
\affiliation{Xinjiang Astronomical Observatory, Chinese Academy of Sciences, 150 Science 1-Street, Urumqi 830011, China}
\affiliation{University of Chinese Academy of Sciences, 19A Yuquan Road, Beijing 100049, China}

\author{G. I. Kokhirova}
\affiliation{Institute of Astrophysics, National Academy of Sciences of Tajikistan, 22 Bukhoro Street, Dushanbe 734042, Tajikistan}

\begin{abstract}

We collected the archival data of blazar OJ~287 from heterogeneous very long baseline interferometry (VLBI) monitoring programs at 2.3 GHz, 8.6 GHz, 15 GHz and 43 GHz.
The data reduction and observable extraction of those multi-band multi-epoch observations are batch-processed consistently with our automated pipeline.
We present the multivariate correlation analysis on the observables at each band. 
We employ the cross-correlation function to search the correlations and the Monte Carlo (MC) technique to verify the certainty of correlations.
Several correlations are found.
The foremost findings are the correlations between the core flux density and the jet position angles on different scales, which validated the plausible predictions of the jet with precession characteristics.
Meanwhile, there is a variation in the offset between the core EVPA and the inner-jet position angle over time at 15~GHz and 43~GHz.

\end{abstract}


\section{introduction}

Blazars are a particular class of active galactic nucleus (AGN) with their jet oriented at a very small angle with respect to the line of sight~\citep{urry.95.pasp}.
The relativistic jet propagates from the vicinity of the actively accreting supermassive black hole (SMBH) in the center. The blazar subclass consists of BL Lacertae objects (BL Lacs) and flat-spectrum radio quasars (FSRQs) identified by their optical spectra. 
Blazars show flux variability at almost all wavelengths of the electromagnetic (EM) spectrum and significant polarization variability in both optical and radio bands. 

The VLBI is a powerful technique to increase the radio telescope's resolution by forming a synthetic aperture with long baselines. 
More and more high-resolution VLBI observations have revealed the swinging innermost jet position angle and complex innermost jet structure in the plane of the sky~\citep{lister.13.aj,jorstad.17.apj}.
\citet{lister.21.apj} analyzed a large data set and concluded that most of the jets show variations between 10$^\circ$ and 50$^\circ$ in their inner-jet position angle over time.
The essential reason for this jet bending phenomenon in the innermost region may be the wobbling flow instability near the jet base, which includes Lense-Thirring precession of the black hole and the accretion disk~\citep{caproni.04.apj}, jet precession caused by the binary black hole system~\citep{britzen.18.mn,dey.21.mn} or the warping of the accretion disk~\citep{lai.03.apjl}, magnetohydrodynamic (MHD) flow instabilities~\citep{matveyenko.15.al}.

Long-term observed flux density variations can be co-modulated by multiple mechanisms induced by intrinsic and extrinsic factors, which are often difficult to distinguish~\citep{bottcher.19.galaxies}.
Extrinsic factors mainly include the geometrical effects caused by non-radial jet motion accompanied by the viewing angle variations and consequent Doppler factors variations.
Meanwhile, the non-radial jet motion essentially forms the intrinsic helical or curved structure, and the projection on the sky plane is S-shaped or some complex morphology.
If the above geometric effect is the main reason for the variability in radio emission and jet morphology, a correlation between the flux density and the jet position angle could be predicted in that case.
Precession is a special and important mechanism that causes geometric effects.
So, the source OJ~287 with long-term multi-band VLBI monitoring and well-studied precession characteristics is selected for correlation analysis.

Through the time-varying properties of the jet direction caused by the jet precession,
we can study whether the rotation of the core EVPA is consistent with that of the jet position angle at the core scale.
That is, whether both maintain relatively constant directions during long-term monitoring, as the theory studies suggest~\citep{laing.80.mn,lyutikov.05.mn}.
It should be mentioned that the results of various VLBI studies on radio core EVPA and local position angle alignment in AGN jets are observationally controversial~\citep{agudo.18.mn,hodge.18.apj}.

OJ~287 is a low synchrotron peaked (LSP) BL Lac object which is highly active at all wavelengths and famous for its repetitive optical flares with roughly a 12-year period~\citep{sillanpaa.88.apj} and variable jet morphology~\citep{gomez.22.apj}.
As seen in many other BL Lac objects, OJ~287 has a one-sided core-jet structure.
Up to now, the jet of OJ~287 has been extensively investigated at several wavelengths and with different long-term VLBI monitoring programs, like the MOJAVE\footnote{The {\em Monitoring Of Jets in Active galactic nuclei with VLBA Experiments} is a VLBA program carried out at the $K_u$ band (15.3 GHz) to monitor radio brightness and polarization variations in AGN jets. Approximately 1/3 of these were observed from 1994--2002 as part of the 2 cm Survey.}, the 2 cm survey~\citep{kellermann.98.aj} and the VLBA-BU-BLAZAR program\footnote{{\em Boston University Gamma-ray Blazar monitoring program with the VLBA at 43 GHz}}~\citep{jorstad.17.apj}.
\citet{tateyama.04.apj} reported the position angle of the more collimated unresolved jet rotated clockwise by \textasciitilde30$^\circ$ from 1994 to 2002 due to the ballistic precession of the jet. 
\citet{agudo.12.apj} found a sharp innermost jet position angle swing during 2004 and 2006. 
In recent years, \citet{cohen.17.galaxies} suggested that the jet is rotating on a 30-year period by analyzing the jet ridge lines.
And~\citet{britzen.18.mn} found that the OJ~287 jet is precessing on a time-scale of \textasciitilde22 years.
The radio emission is highly polarized, the time-domain variations of the degree of the linear polarization and the EVPA have been extensively studied in radio bands~\citep{roberts.87.apj,cohen.18.apj}.

The structure of this paper is as follows:
In Section~\ref{obs}, we describe the observations, the data reduction techniques, and the methods to derive observables.
In section~\ref{method}, we describe the method used in the correlation analysis in detail.
Section~\ref{res} shows the results of correlations between observables.
Section~\ref{dis} includes some discussions of current correlation analysis results and the significance and merits of correlation research.
The final section gives a summary of our findings. 
All position angles are expressed in degrees from north to south, between 0$^\circ$ to $-$180$^\circ$ in the clockwise and 0$^\circ$ to 180 $^\circ$in the counterclockwise direction.

\section{observations \& data reduction }\label{sec:obs}
\label{obs}

\subsection{Monitoring observations}

A collection of archived multi-band multi-epoch VLBI data from several geodetic/astrometric and astrophysical VLBI programs are analyzed in this study.  
It includes the $S$/$X$-band (2.3~GHz/8.6~GHz) data from Research and Development VLBA (RDV) program,
the $K_u$-band (15~GHz) data from the MOJAVE program and the $Q$-band (43~GHz) data from the VLBA-BU-BLAZAR program.

The RDV program utilizes simultaneous 2.3~GHz/8.6~GHz global-VLBI observations to monitor the radio source structures with geodetic antennas around the world.
We use the data observed from July 1994 to December 2003 until the source was dropped from the astrometric monitoring due to the structural complexity.
The RDV data were acquired through collaboration with the Bordeaux VLBI Image Database (BVID).
For this source, the database maintained by the MOJAVE team comprises 127 observed epochs over a span of 25 years from April 1995 to June 2020.
The MOJAVE database consists of two parts: the 2 cm Survey and the MOJAVE program initiated in 2002 as the successor to the former\footnote{Hereinafter, unless otherwise specified, the expression with 'MOJAVE' refers to the combination of two observations.}.
MOJAVE program observations were carried out in dual polarization mode using frequencies centered at 15.4~GHz, with a bandwidth of 32~MHz.
We use the self-calibrated visibility data from the streamed release of the MOJAVE database in our analyzing pipeline. 
In addition, we also use the 43~GHz data from the VLBA-BU-BLAZAR program, which includes data from June 2007 and onwards.
The VLBA-BU-BLAZAR observations were made in continuum mode, with both left and right circular polarizations at a central frequency of 43.1~GHz. 
It also provided self-calibrated visibility data. 
In this paper, we analyze their released data onwards until January 2020.

\subsection{Data reduction}

The data are reduced using the Astronomical Image Processing System ({\textsc{AIPS}) package and
our \textsc{SAND} pipeline~\citep{zhang.18.mn} was used to automate the procedures of mapping and model-fitting, as well as the observable extraction and post-processing.

\subsection{Imaging}

The CLEANed images are produced with the \textsc{AIPS} task IMAGR using the BGC CLEAN algorithm~\citep{clark.80.aa}. 
We choose to use uniform weighting and a proportional-to-beam cell size of 0.5, 0.15, 0.08 and 0.02~mas pixels$^{-1}$ from the lowest frequency (2.3~GHz) to the highest frequency (43~GHz), respectively.
We use restoring beam smaller than the nominal CLEAN beam obtained by fitting a Gaussian to the central lobe of the dirty beam to create slightly super-resolved images.
In most cases, our 15~GHz and 43~GHz observations have full Stokes parameters successfully correlated, 
thence the pipeline will automatically produce polarization maps. 
Fig.~\ref{fig:samples} displays a sample of quad-band CLEANed images contours with the Gaussian model-fitting components superimposed at each frequency.
In the same band, all images have been convolved with the same restoring beam, i.e., the FWHM dimensions of circular restoring beam size are 3, 0.7, 0.4, 0.2~mas from the lowest frequency to the highest frequency, respectively.

\subsubsection{Model fitting}

Our pipeline utilizes the task SAD to extract sources in the CLEANed images, and fit Gaussian components to those. 
The task can search for potential sources with bright peaks above a certain flux threshold.
We use the CPARM parameters to set the threshold level in descending order of powers of the root mean square (RMS) noise.
The background noise level is obtained by examining pixels well away from the sources in each Stokes I image with the task IMSTAT.
The components are fitted within rectangular islands, and it may be possible to get bad solutions, so the spurious components are rejected by setting DPARM parameters. 

\begin{figure*}
\centering
\includegraphics[width=5.6cm]{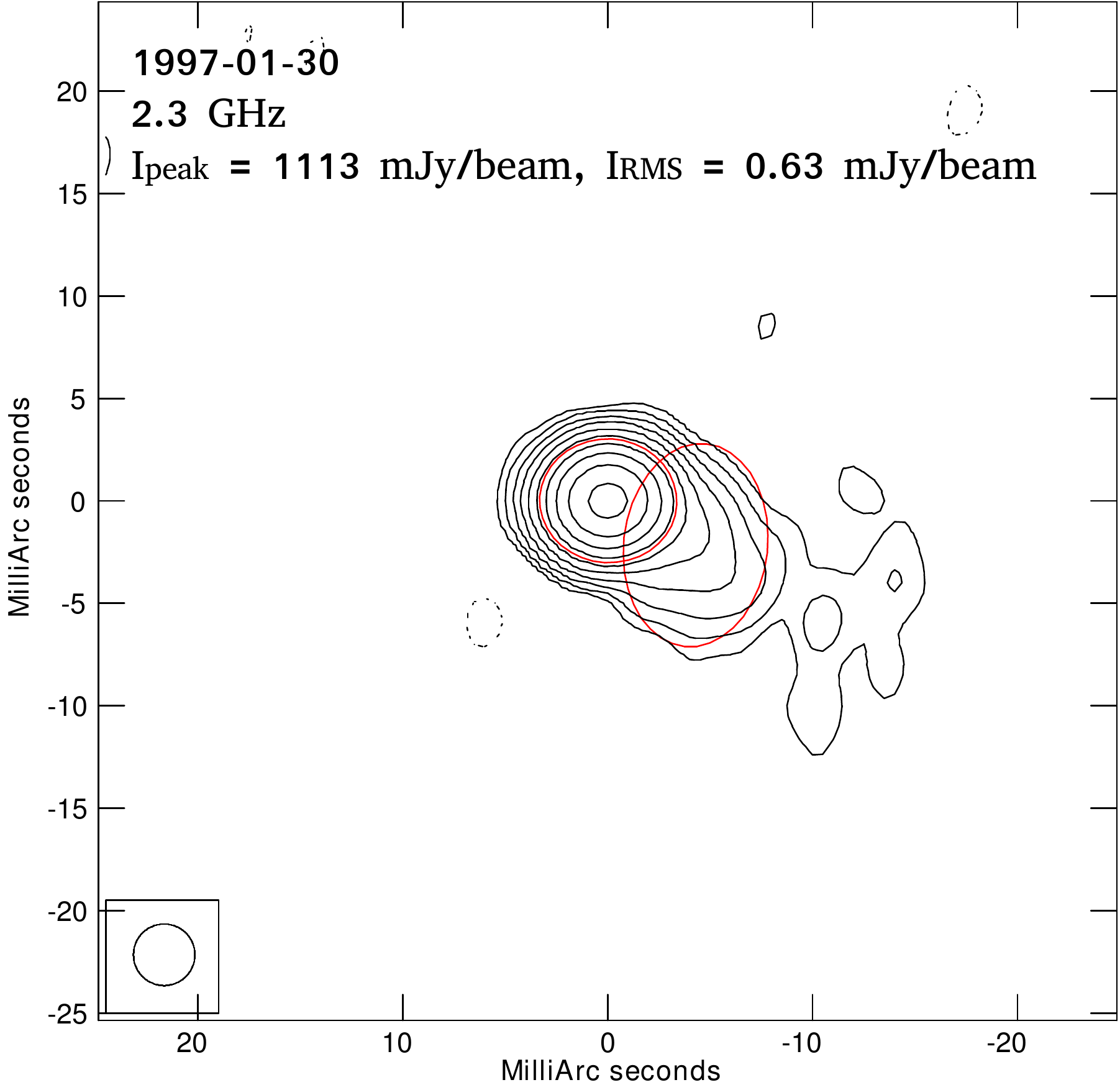}
\includegraphics[width=5.6cm]{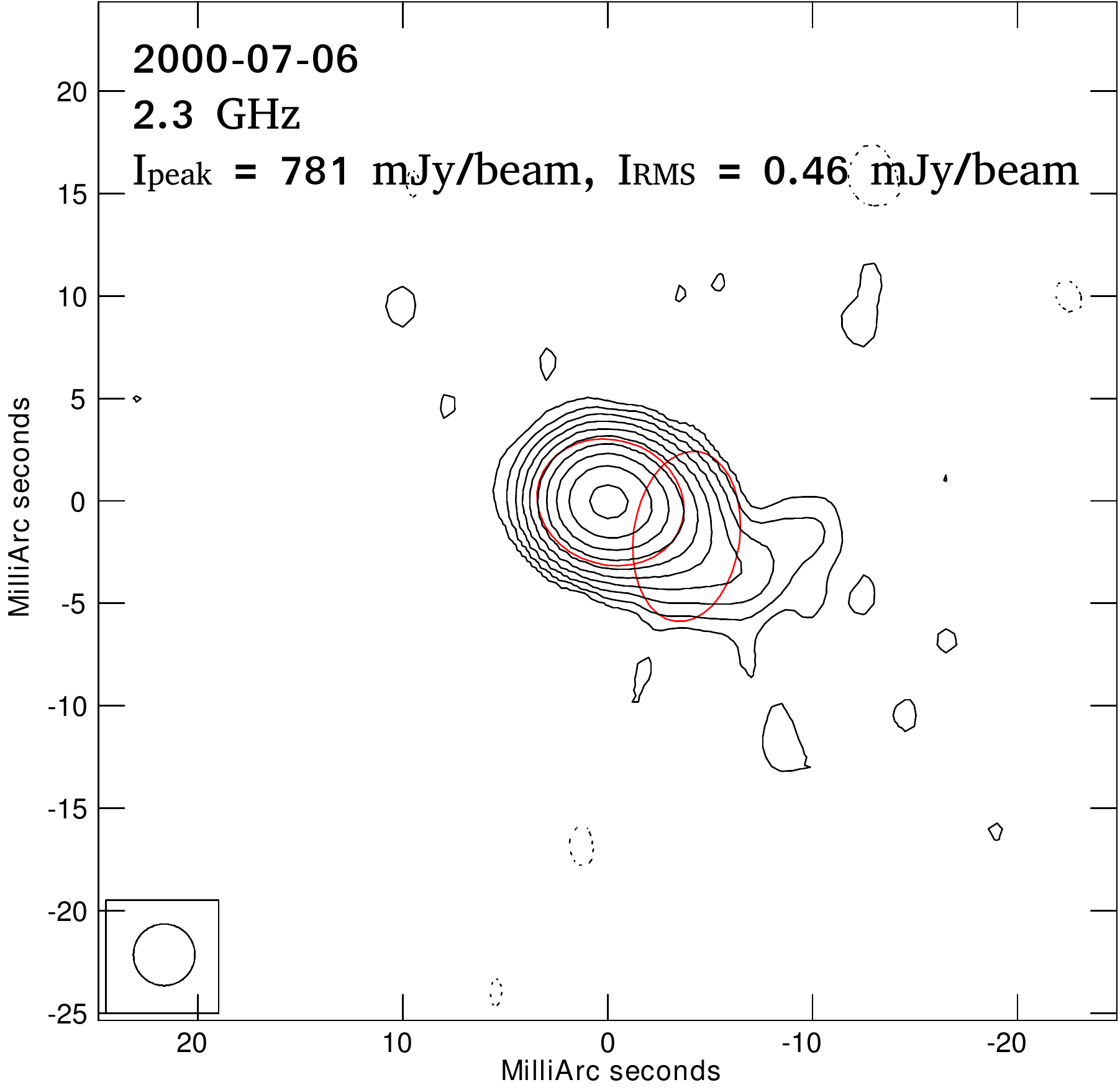}
\includegraphics[width=5.6cm]{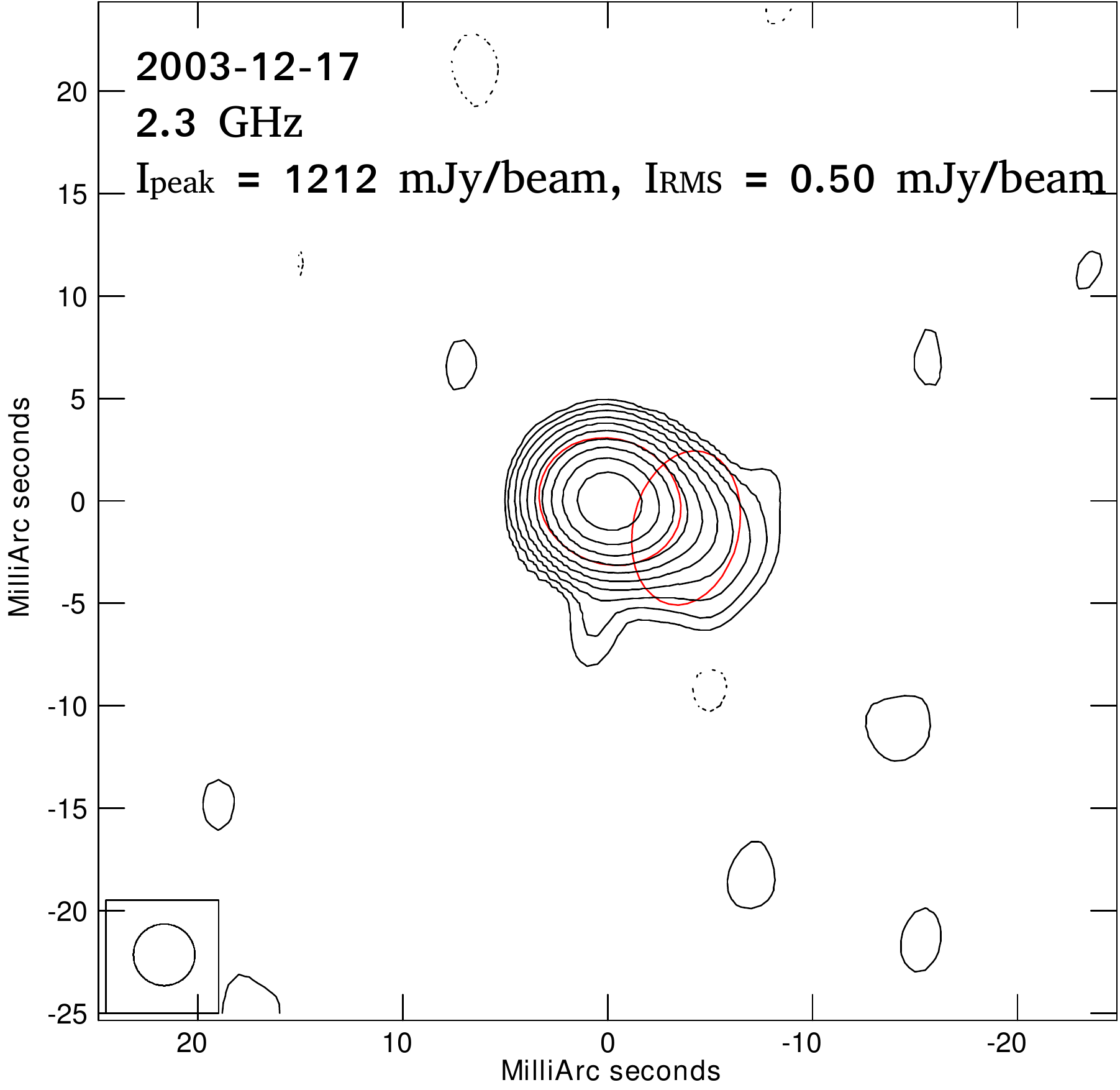}\\
\includegraphics[width=5.6cm]{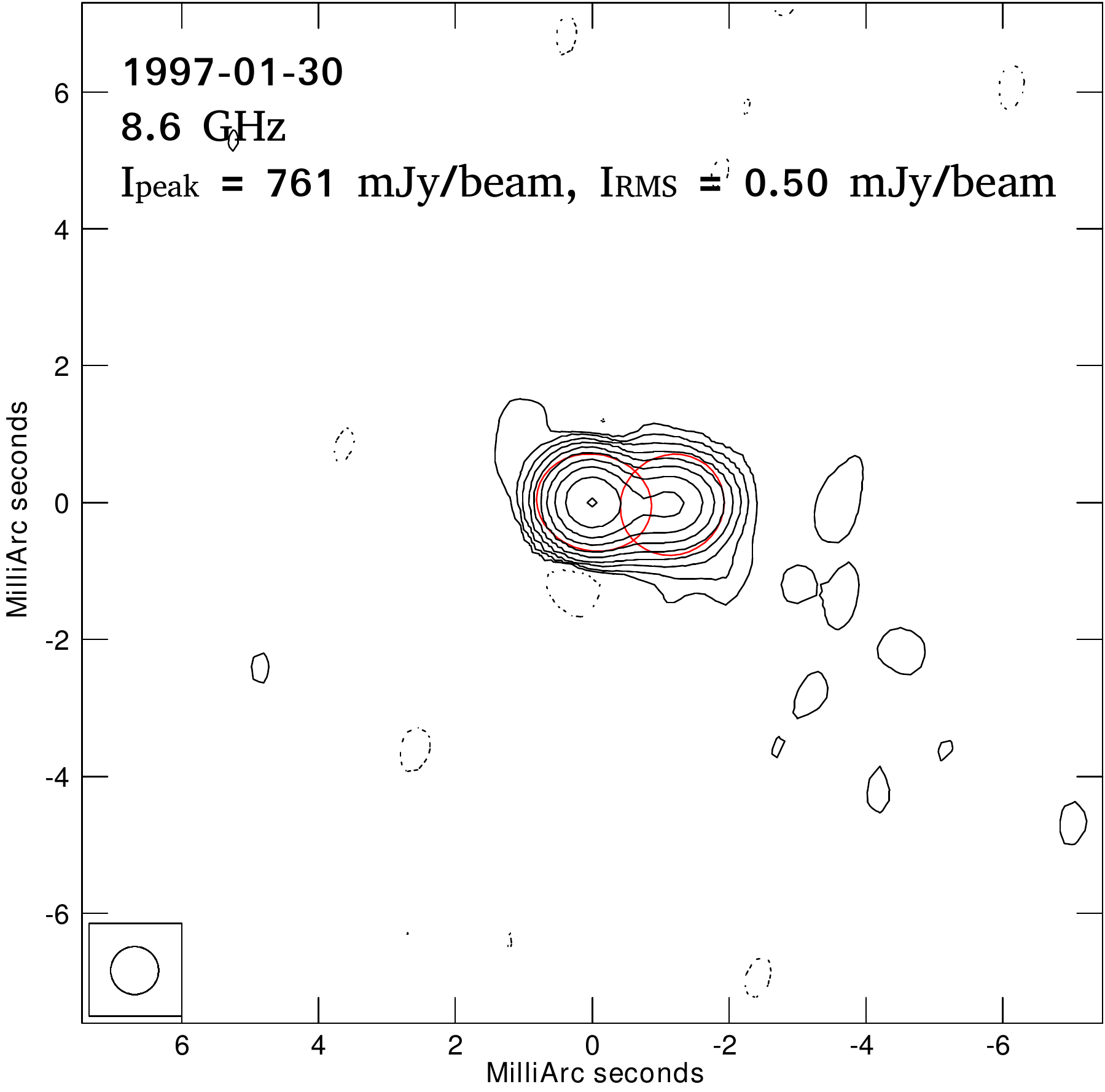}
\includegraphics[width=5.6cm]{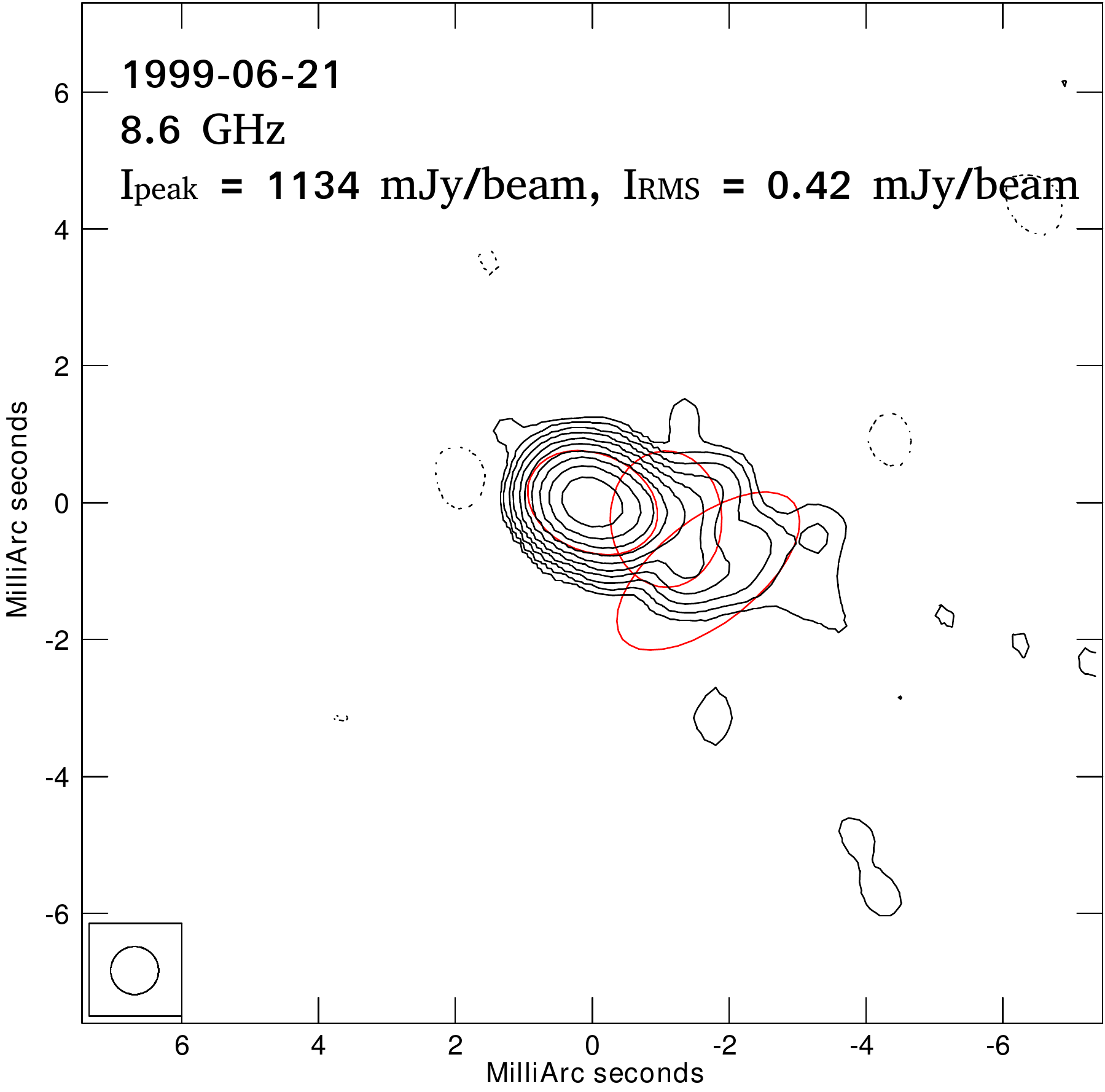}
\includegraphics[width=5.6cm]{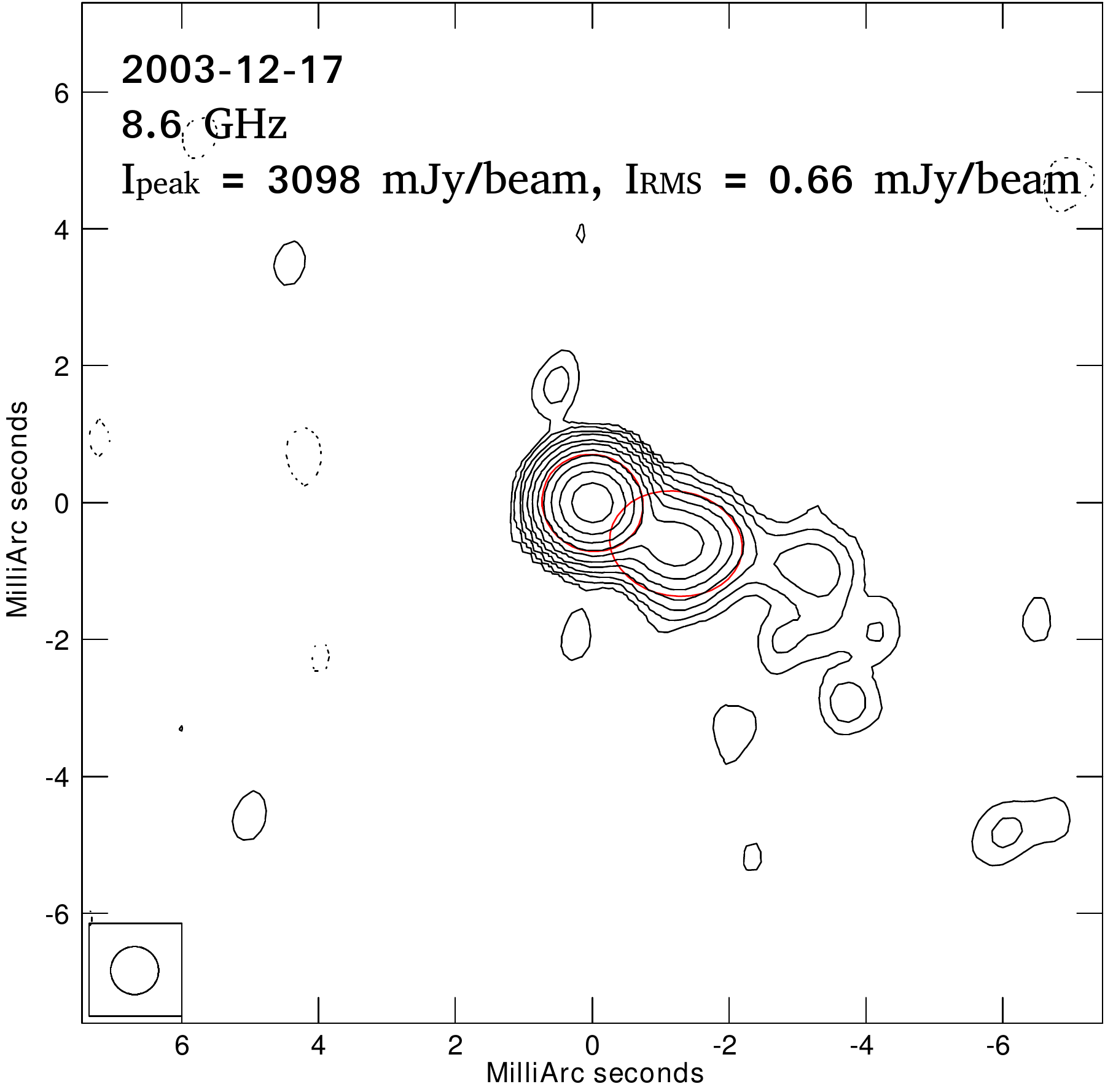}\\
\includegraphics[width=5.6cm]{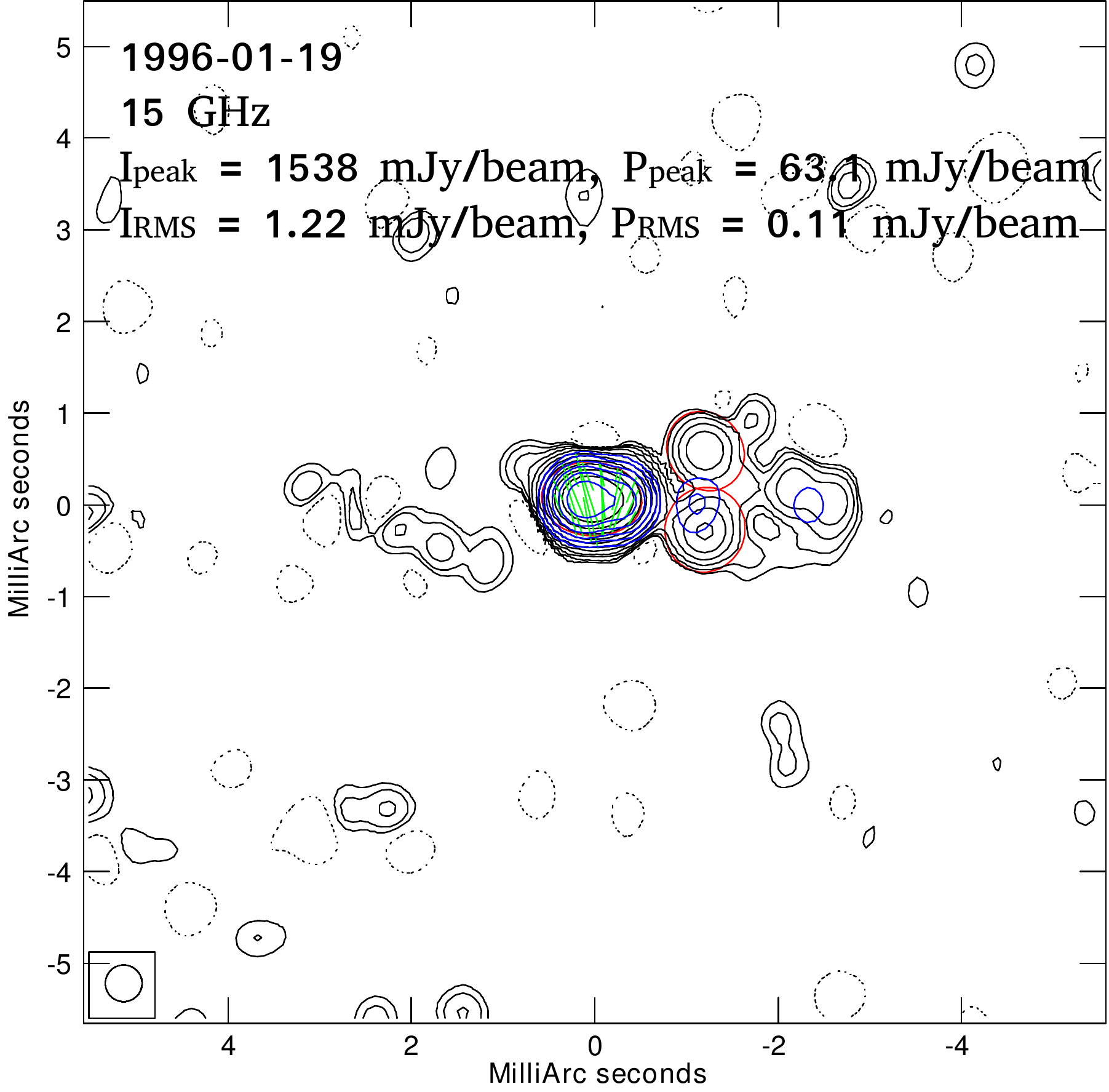}
\includegraphics[width=5.6cm]{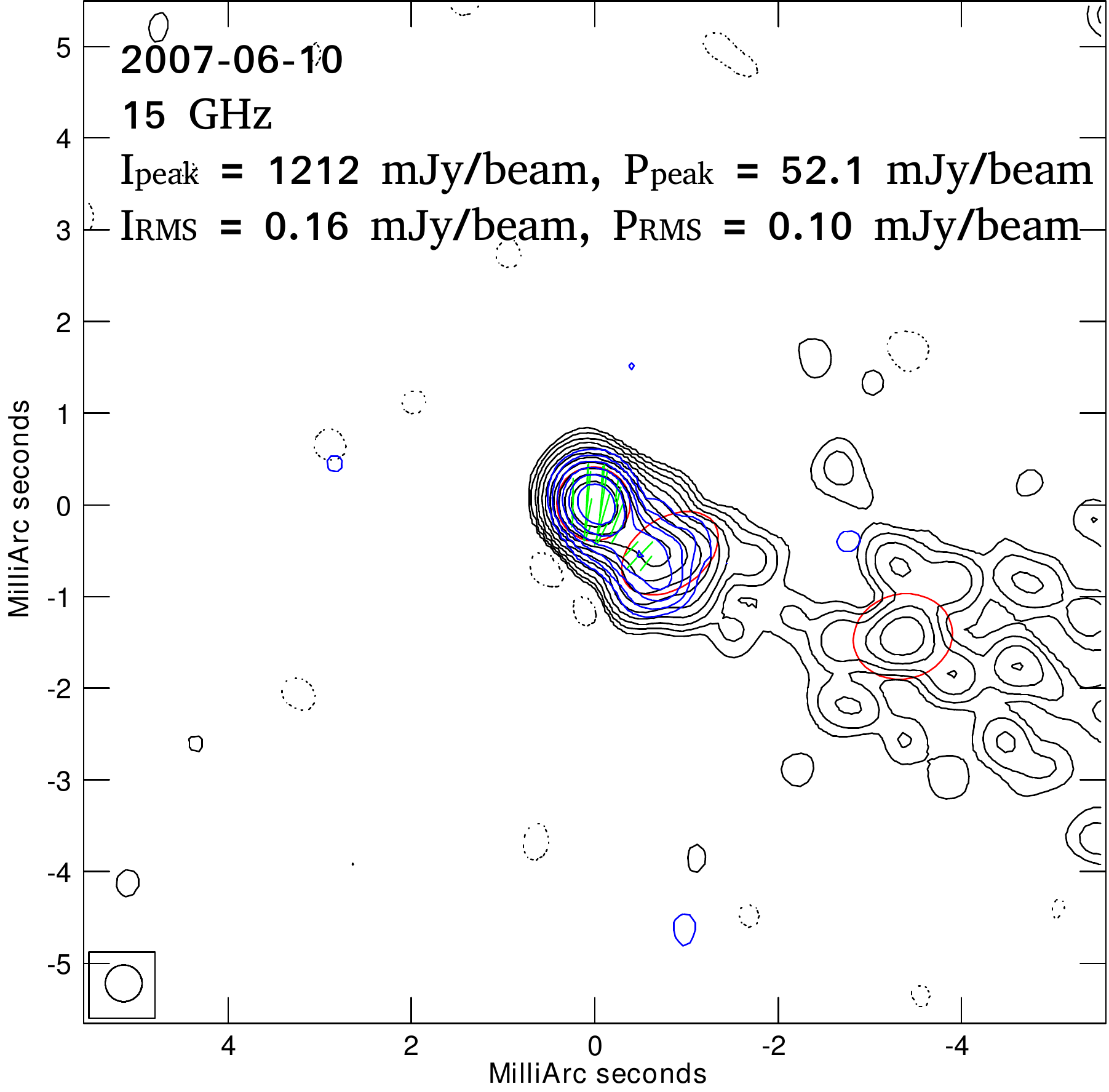}
\includegraphics[width=5.6cm]{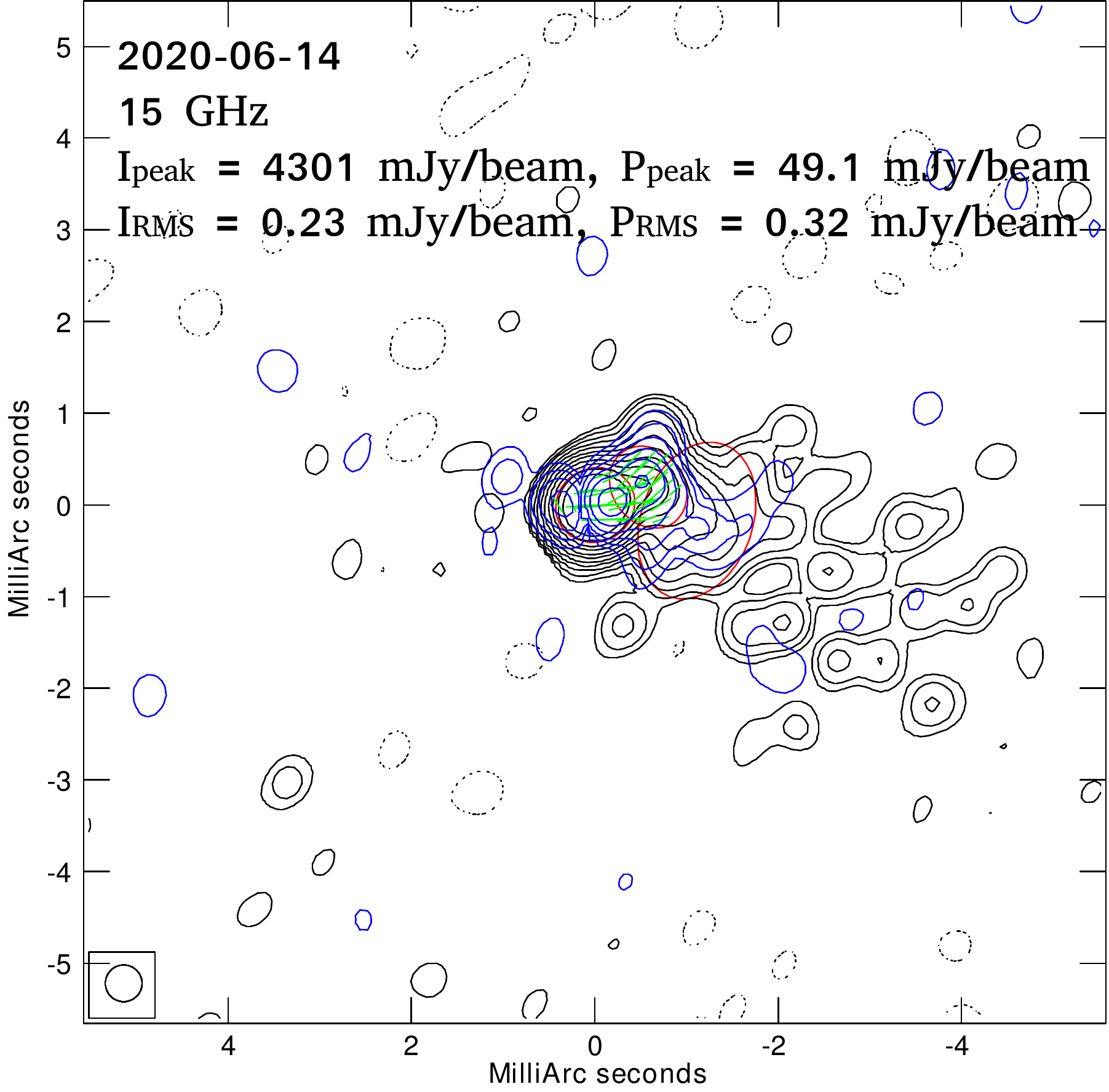}\\
\includegraphics[width=5.6cm]{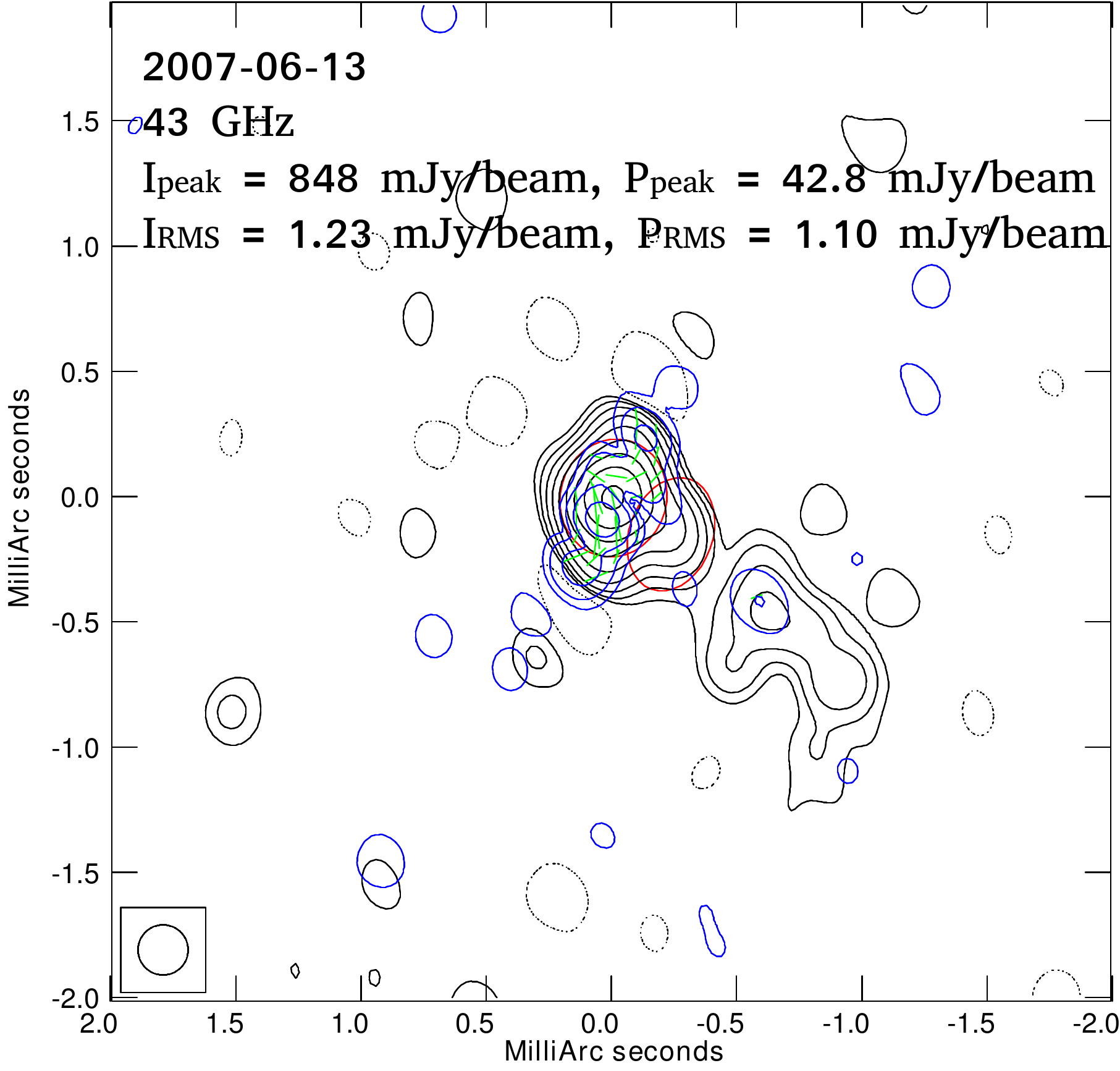}
\includegraphics[width=5.6cm]{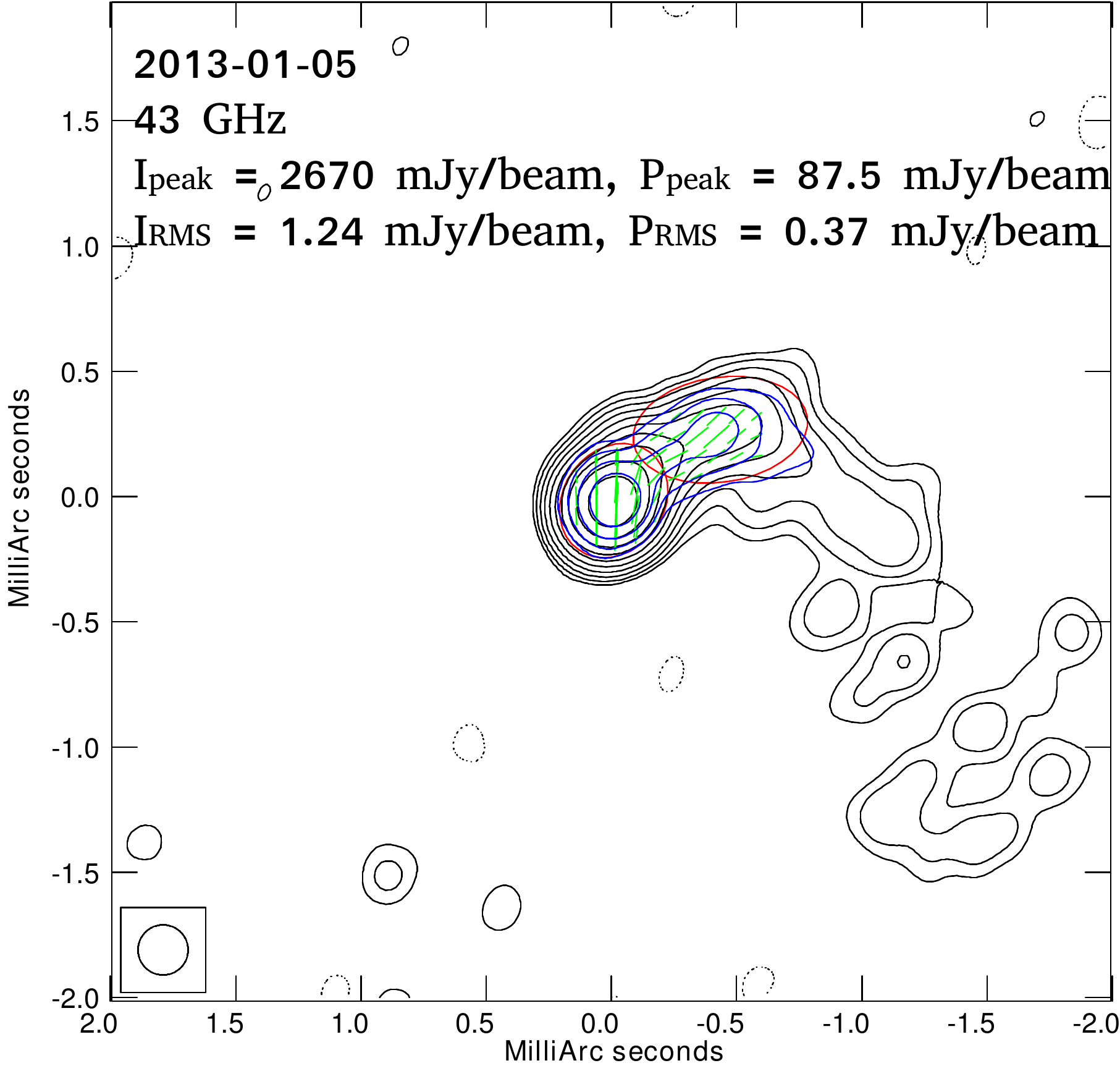}
\includegraphics[width=5.6cm]{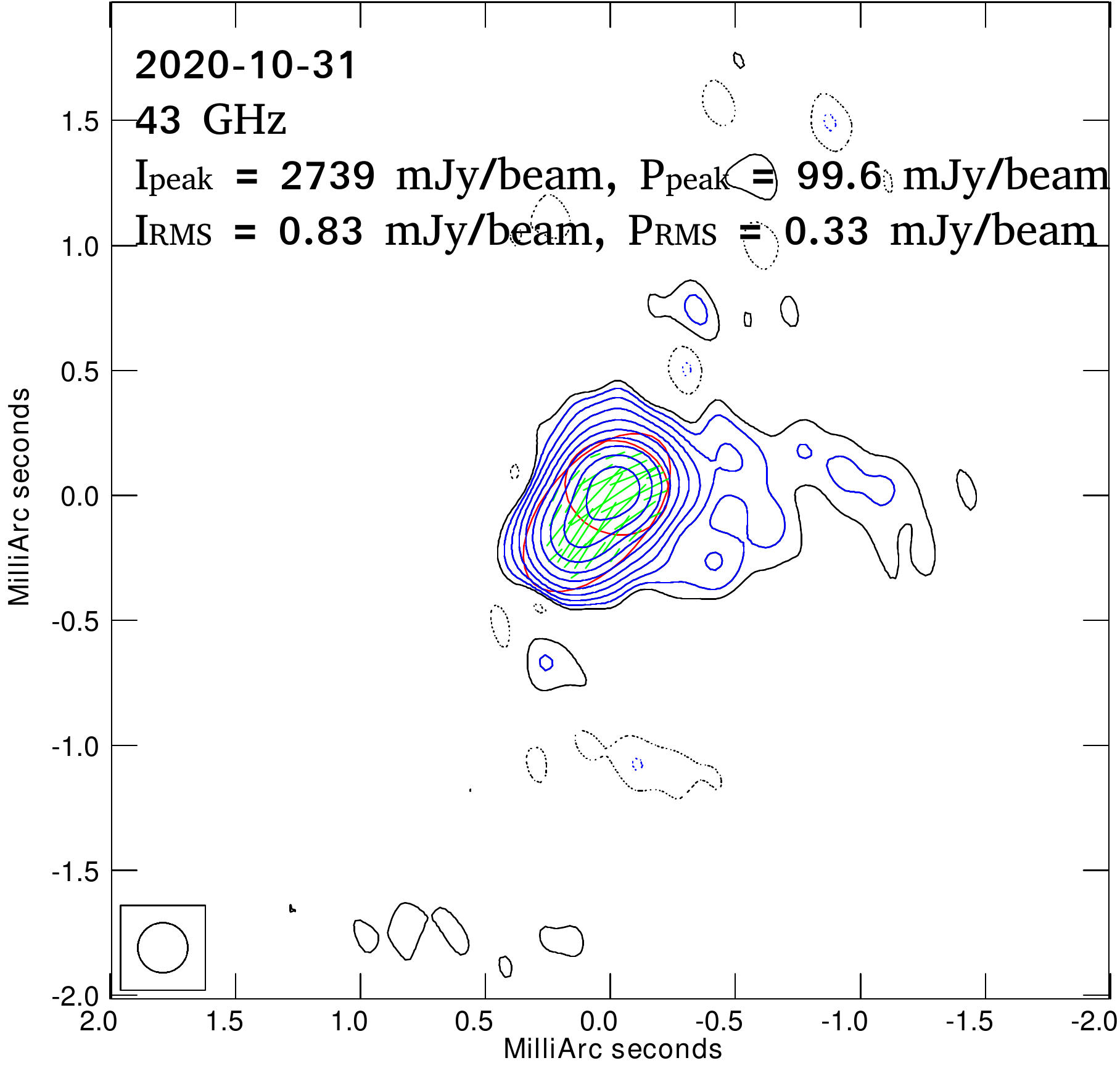}\\
\caption{CLEANed images of OJ~287 for the first, median, and last epoch of multi-epoch data at different frequencies. 
The lowest I contours are plotted at the $\pm$ 3 $\sigma$ ($\sigma$ is the off-source noise) level of all images and the positive contour levels increase by a factor of 2. 
The linearly polarized intensity contours in blue overlayed in I contours and the lowest P contours are plotted at the $\pm$ 6 $\sigma$ level of all images and the positive contour levels also increase by a factor of 2. 
The center and semi-major and semi-minor axes of the red ellipse represent the position and size (FWHM) of the Gaussian components fitted in the image plane, respectively. 
The solid green lines indicate the electric polarization vector directions, the length of this vector is proportional to the intensity of the linear polarization with 1 mas corresponding to 0.0625 Jy/beam at 15~GHz and 0.25 Jy/beam at 43~GHz.
The FWHM restoring beam is represented by the circle in the bottom left corner of each panel.}
\label{fig:samples}
\end{figure*}

\begin{figure*}
    \centering
    \includegraphics[width=5.6cm]{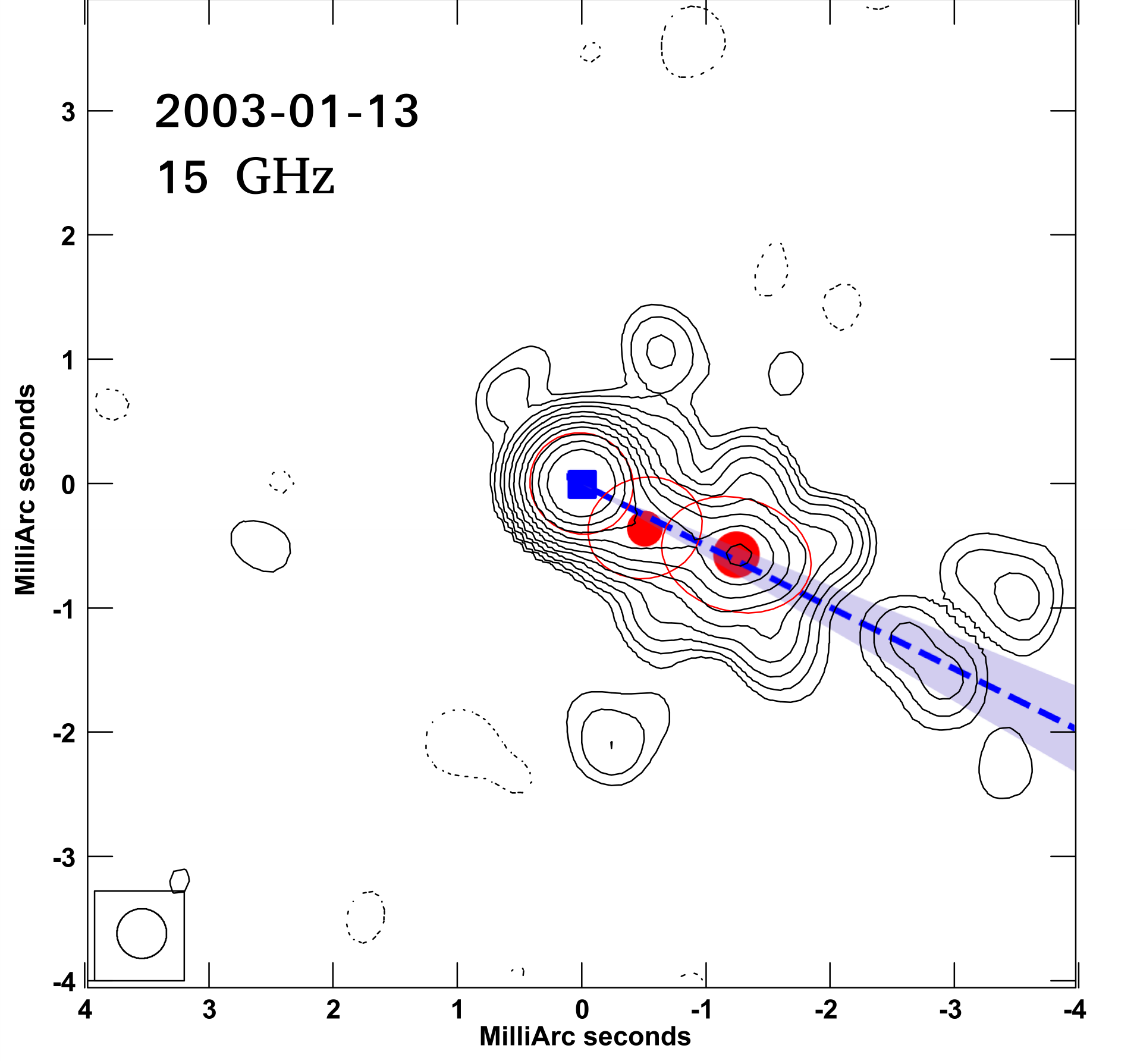}
    \includegraphics[width=5.6cm]{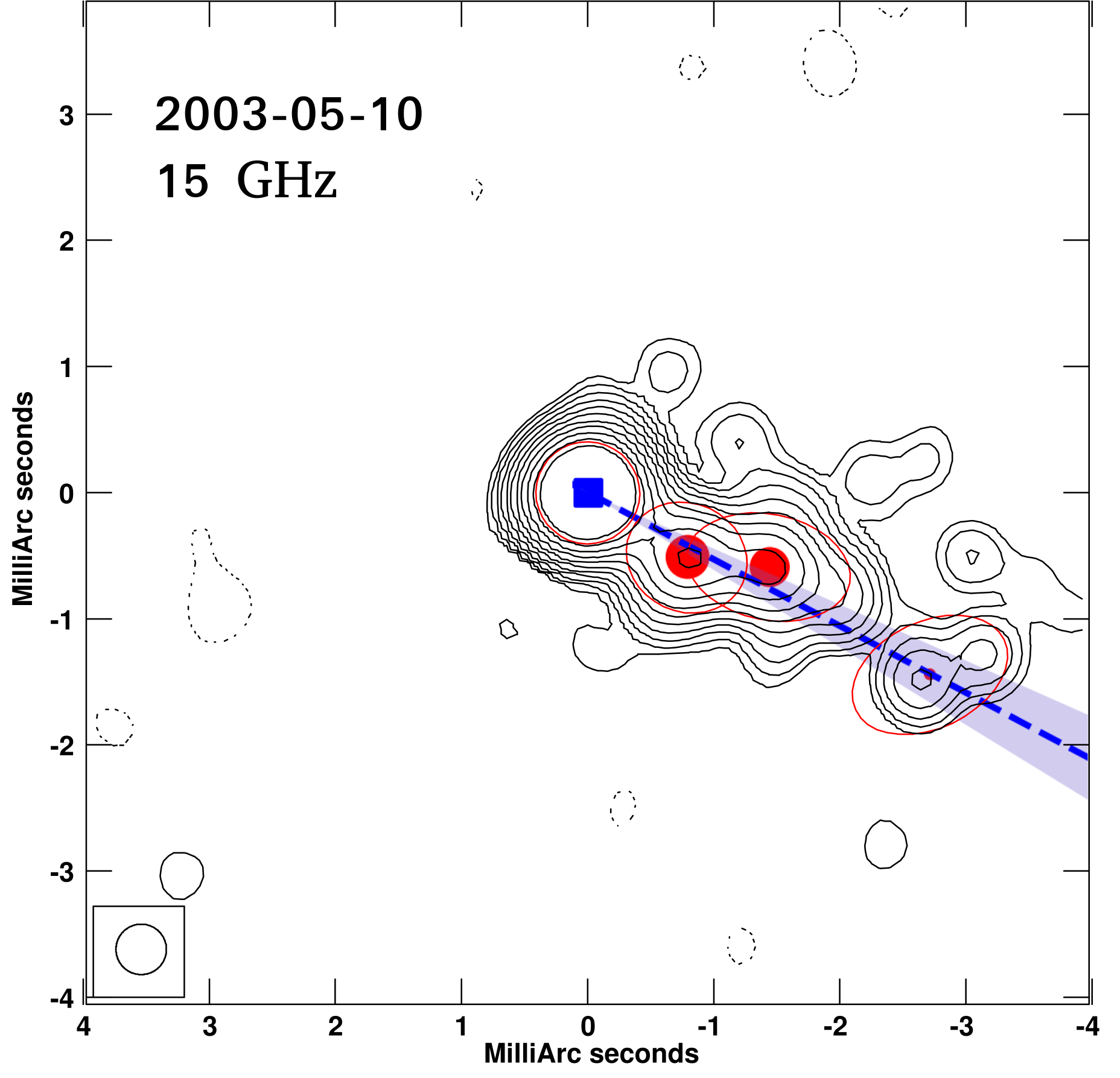}
    \includegraphics[width=5.6cm]{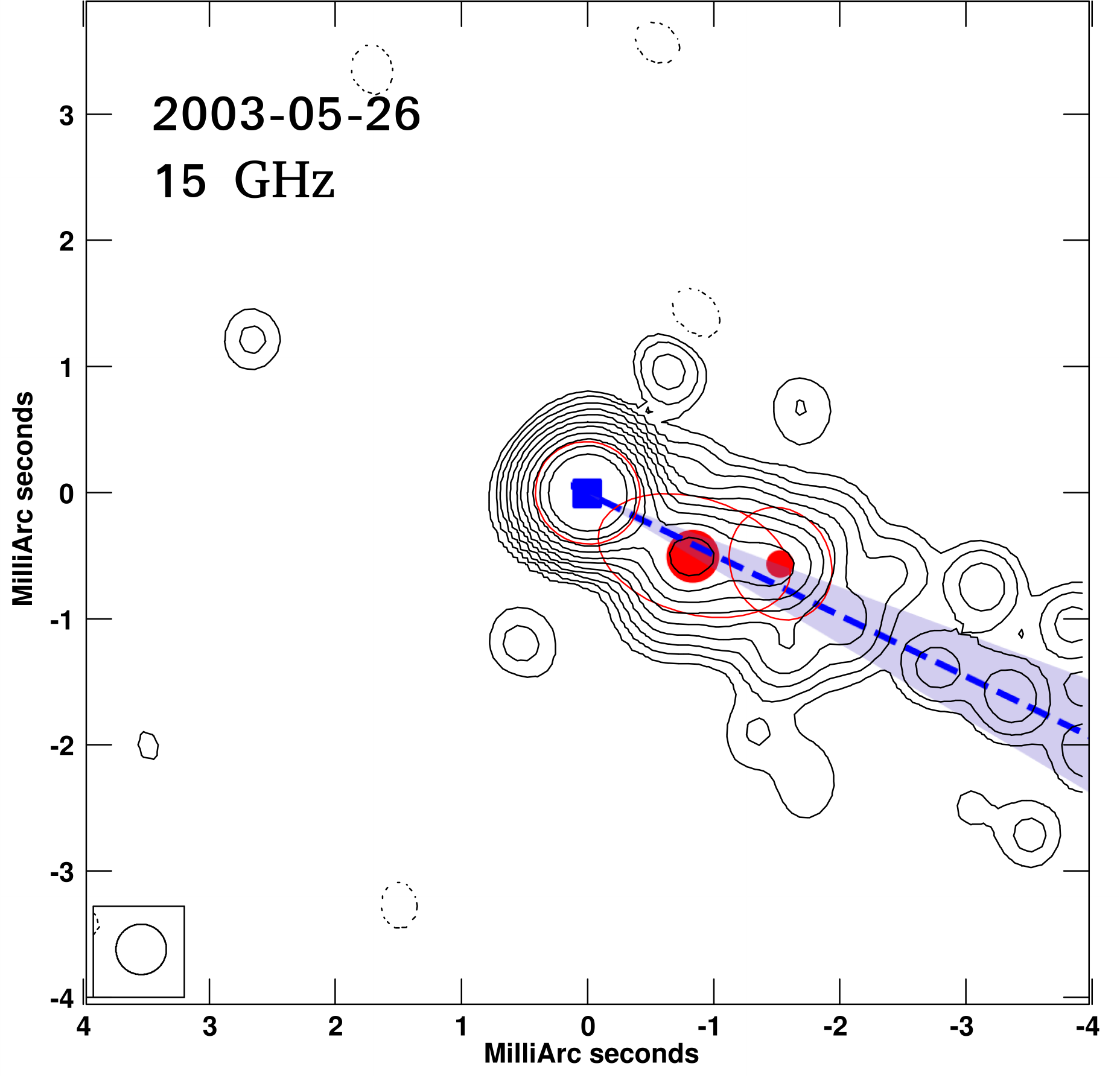}\\
    \includegraphics[width=5.6cm]{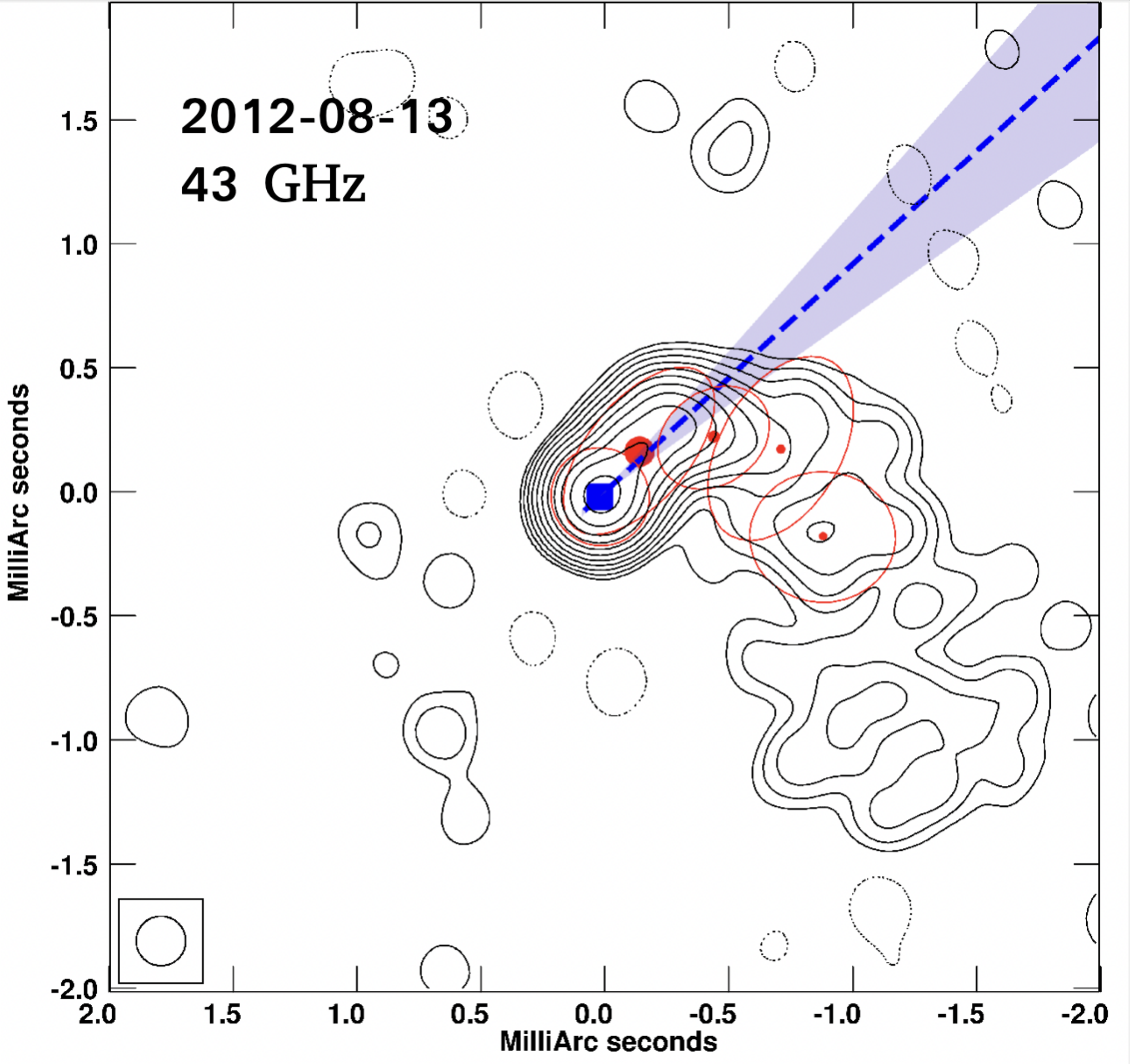}
    \includegraphics[width=5.6cm]{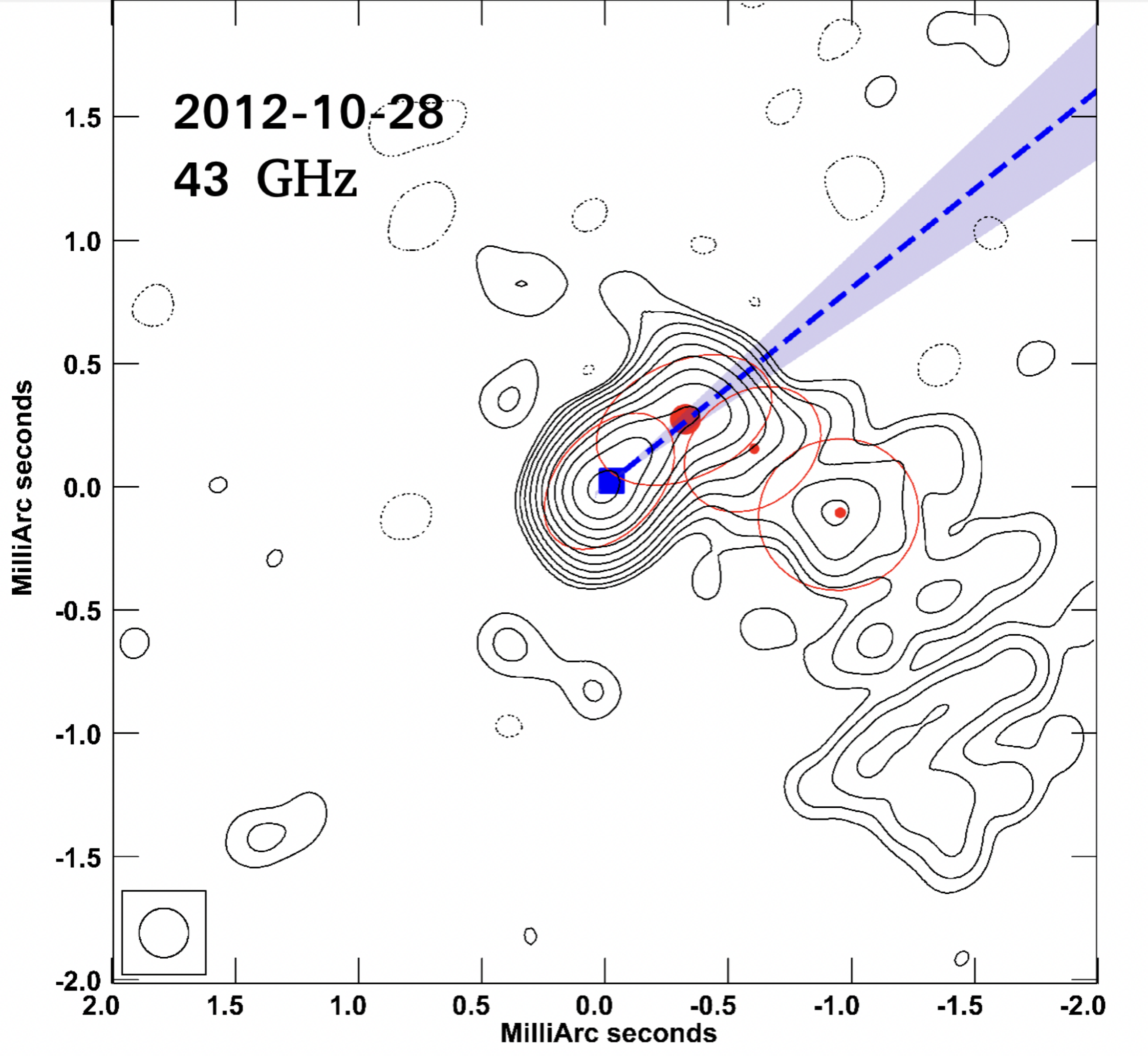}
    \includegraphics[width=5.6cm]{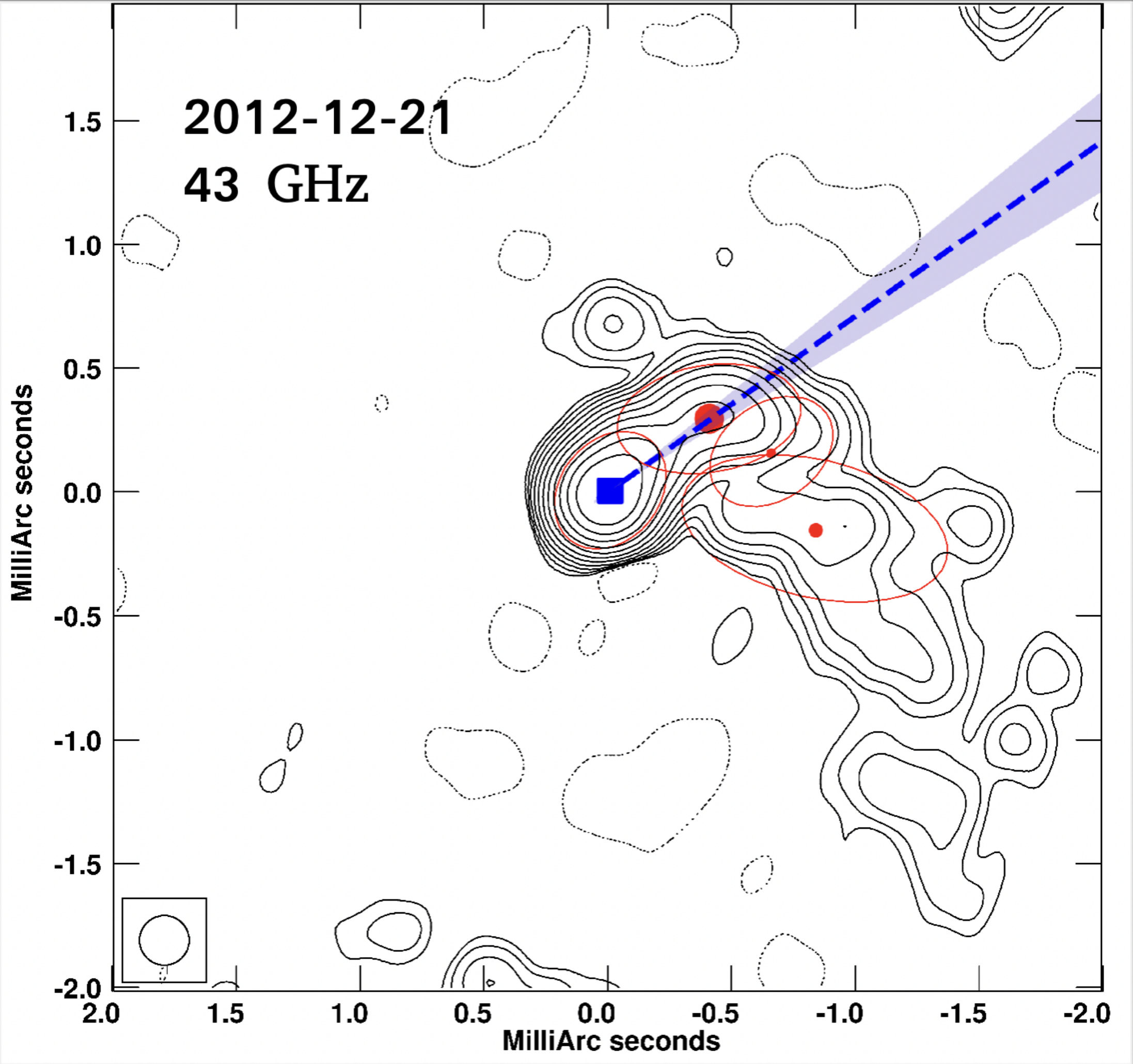}\\
    \caption{A sample of CLEANed images of OJ~287 at 15~GHz and 43~GHz. 
    Symbols are the same as those in Fig~\ref{fig:samples}. 
    The inset overlaid on the CLEANed images are examples of the linear regression method we used to determine the jet position angle.
    The blue square denotes the core component, jet components are denoted by red circles whose diameters are proportional to the flux density.
    The dashed lines show the results of our line fitting, while the lighter blue shaded area corresponds to the k $\pm$ $\sigma_k$ region.
    }
    \label{fig:odr}
\end{figure*}

\subsection{Properties}

Usually, the most upstream visible component is brighter than downstream components, which can be assigned as the 'core'. 
However, there can be confused situations when the image configuration is complex as in the high-frequency bands.
In this paper, we usually defined the brightest and compact emission feature as the VLBI core, but not always.
For special cases of complex situations in high frequency, we carefully identify the core by taking flux density, degree of compact, and position of components in adjacent epochs into consideration.

\subsubsection{Core measurements}

In Table ~\ref{tab:coreobs}, we list the general properties of the fitted core component. 
A previous study by \citet{liu.12.apss} used the position angle of the major axis of the fitted core component to determine and define the inner-jet position angle.
In our research, we also follow this definition. 
Obviously, if the ratio between the major axis and the minor axis of the fitted core component is close to 1, 
the major axis position angle can be an arbitrary value. 
In this case, the outlier will be substituted with the mean value of two adjacent epochs' inner-jet position angles.
As we know, the non-directional major axis position angle is centrosymmetric, so any position angle is equivalent to itself with an $n\pi$ flip.
To ensure that the correct inner-jet position angle is chosen, we select the multi-epoch position angle based on the \emph{jet direction matching criteria}. 

\begin{table*}
    \centering
    \caption{Core component properties. Columns are as follows: 
    (1) date of observation;
    (2) observing frequency in GHz;
    (3) flux density of the fitted Gaussian core component; 
    (4) major axis of the fitted Gaussian core component; 
    (5) minor axis of the fitted Gaussian core component;
    (6) major axis position angle of the fitted Gaussian core component; 
    (7) EVPA of the fitted Gaussian core component.\\
    (This table is available in its entirety in machine-readable form.)}
    \label{tab:coreobs}
    \begin{tabular}{cccccccc}
       \hline
        Epoch     & Frequency  & $I_{core}$              &  Maj.     &  Min.    & P.A.                 &  EVPA       \\
                  & (GHz)      &  (Jy)                   &  (mas)    &  (mas)   & ($^\circ$)           &  ($^\circ$) \\
        \hline
        1997.079  & 2.3        &  1.265  $\pm$  0.063    & 3.35      & 3.02     & 86.7  $\pm$  0.3  & -- \\
        2000.511  & 2.3        &  0.947  $\pm$  0.047    & 3.63      & 3.05     & 73.9  $\pm$  0.2  & -- \\ 
        2003.959  & 2.3        &  1.434  $\pm$  0.072    & 3.52      & 3.05     & 69.5  $\pm$  0.2  & -- \\         
        1994.516  & 8.6        &  1.593  $\pm$  0.080    & 0.98      & 0.70     & 88.5  $\pm$  0.1  & -- \\
        1999.468  & 8.6        &  1.608  $\pm$  0.080    & 0.98      & 0.72     & 68.2  $\pm$  0.1  & -- \\
        2003.959  & 8.6        &  3.295  $\pm$  0.165    & 0.75      & 0.70     & 65.1  $\pm$  0.2  & -- \\        
        1996.049  & 15         &  2.131  $\pm$  0.107    & 0.55      & 0.40     & 85.5  $\pm$  0.1  & 15.0 $\pm$ 3.002\\
        2007.438  & 15         &  1.109  $\pm$  0.055    & 0.40      & 0.40     &    --             & $-$7.6 $\pm$ 3.004\\      
        2020.451  & 15         &  4.650  $\pm$  0.233    & 0.44      & 0.40     & $-$67   $\pm$  0.1  & $-$47.8$\pm$ 3.469\\ 
        2007.447  & 43         &  1.037  $\pm$  0.052    & 0.24      & 0.21     & $-$15.4 $\pm$  0.6  & 30.9 $\pm$ 7.171\\
        2013.038  & 43         &  3.154  $\pm$  0.158    & 0.24      & 0.20     & $-$36.6 $\pm$  0.1  & $-$1.4 $\pm$ 7.006\\
        2020.831  & 43         &  2.843  $\pm$  0.142    & 0.38      & 0.21     & $-$41.6 $\pm$  0.1  & $-$58.3 $\pm$ 7.003\\
        \hline   
    \end{tabular}
\end{table*}

\subsubsection{Core polarization properties}

If full Stokes parameters are correlated for a data record, SAND will generate the Q and U images. 
And then, the Stokes Q and U images are combined using the AIPS task COMB to create the polarized intensity and position angle maps. 
The EVPA is calculated as EVPA = (1/2)$\tan^{-1}$($U/Q$).
We averaged the EVPAs of all pixels within an area of 3 $\times$ 3 pixels centered on the core of the Stokes I image in the position angle map as the nominal core EVPA.
We used the following equations to estimate the errors in EVPA:
\begin{equation}
\sigma_{p}=\frac{\sigma_{Q}+\sigma_{U}}{2},
\sigma_{\mathrm{EVPA}}=\frac{\sqrt{Q^{2} \sigma_{U}^{2}+U^{2} \sigma_{Q}^{2}}}{2\left(Q^{2}+U^{2}\right)}=\frac{\sigma_{p}}{2 p}
\label{evpa1}
\end{equation}
where $p=\sqrt{Q^{2}+U^{2}}$, $Q$ and $U$ are the Stokes parameters in the given pixels, and $\sigma_{Q}$ and $\sigma_{U}$ uncertainties in the Stokes Q, and U data. 
$\sigma_{Q}$ and $\sigma_{U}$ should include contributions from the rms error, D-term error and CLEAN error, defined as follows:
\begin{equation}
\sigma=\left(\sigma_{\text {rms }}^2+\sigma_{\text {Dterm }}^2+\sigma_{\text {CLEAN }}^2\right)^{1 / 2},
\sigma_\mathrm{CLEAN}=1.5\sigma_{\mathrm{rms}}
\label{evpa2}
\end{equation}
where $\sigma_{\mathrm{rms}}$, $\sigma_{\mathrm{Dterm}}$, and $\sigma_{\mathrm{CLEAN}}$ denote rms noise, D-term error, and CLEAN errors, respectively.
According to \citet{hovatta.12.aj}'s simulations that considering the instrument polarization, the additional error $\sigma_{\mathrm{Dterm}}$ is defined as
\begin{equation}
\sigma_{\text {Dterm}}=\frac{0.002}{\left(N_{\text {ant}} \times N_{\mathrm{IF}} \times N_{\text {scan}}\right)^{1 / 2}}\left(I^2+\left(0.3 \times I_{\text {peak}}\right)^2\right)^{1 / 2}
\label{evpa3}
\end{equation}
where $N_{\text {ant}}$ is the number of antennas, $N_{\text {IF}}$ is the number of IFs, $N_{\text {scan}}$ is the number of scans with independent parallactic angles, and $N_{\text {peak}}$ is the peak total intensity map.
We used AIPS task PRTAN and task LISTR with optype='scan' to read $N_{\text {ant}}$, $N_{\text {IF}}$, $N_{\text {peak}}$ value from the actual recorded data epoch by epoch for $\sigma_{\mathrm{Dterm}}$ calculation.
We adopted the error from imperfect EVPA calibration of 3$^\circ$ at 15 GHz~\citep{hovatta.12.aj} and of 7 $^\circ$ at 43 GHz~\citep{kravchenko.20.aa, jorstad.05.apj}, and added this error to Equation (1)($\sigma_{\mathrm{EVPA}}$) in quadrature.
We also averaged the error over a small region of 9 pixels as the core EVPA error.
The core EVPAs measured and its error} parameters are listed in Table ~\ref{tab:coreobs}.

Polarized waves are affected by Faraday rotation when propagating through non-relativistic plasma within or external to the source. In the case of external Faraday rotation, the observed EVPA($\chi_{obs}$) and wavelength squared appear a linear dependence and follow Equation \ref{rm}

\begin{equation}
\chi_{obs} = \chi_{0} + RM \lambda^2
\label{rm}
\end{equation}
where $\chi_{0}$ is the intrinsic EVPA and RM is the rotation measure.

The rotating medium located outside the jet mainly includes a sheath surrounding the jet or the broad/narrow-line regions (BLRs/NLRs), even intergalactic and galactic plasma.
It has been reported that galactic RM of OJ~287 is 30 rad m$^{-2}$ \citep{rudnick.83.aj}, which rotates the 15~GHz EVPA and 43~GHz EVPA far less than 1$^\circ$. 
For this source, \citet{hovatta.12.aj} used VLBA observations that carried out four frequencies between 8 GHz and 15 GHz in 2006 April and gave the median RM over the core is $-$307.9 rad m$^{-2}$.
The rotation value calculated by $RM$ = $-307.9$ rad m$^{-2}$ is about 7$^\circ$ at 15~GHz.
Assuming that the RM is also available at 43 GHz, the rotation value will be smaller there.
Previous studies have shown that the opaque core of OJ~287 not show signatures of internal Faraday rotation.
For these reasons, we do not need to correct the Faraday rotation of the external to the source.
A significant focus question in analyzing polarization data is the n$\pi$ ambiguity~\citep{marscher.08.nature,larionov.08.aa,abdo.10.nature} of the observed EVPA. 
Several different methods \citep{kiehlmann.16.aa} were proposed to work out this issue.
These methods are based on the assumption of minimal variation between adjacent data points or between the current and a couple of previous data points. 
For our sparsely-sampled data, we choose the final core EVPA time series, which fall within the narrowest EVPA range, in the set of data points and their $\pm n\pi$ flips.
Admittedly, it is difficult to capture the rapid rotation of the EVPA ($\ge$180$^\circ$) in the interval of two adjacent epochs due to the sampling rate limitation.
At radio wavelengths, the rapid rotation phenomenon is rarer than that at optical wavelengths ~\citep{allel.03.apj,blinov.16.mn}. 
There are few research reports on the EVPA day-level rapid rotation event of OJ~287 in the radio band.
To date, merely \citet{kikuchi.88.aa}'s study reported the EVPA varied by 80$^\circ$ over-5 days in the radio region for OJ~287. 
To sum up the above, for the observation cadence of the datasets in this study, the method we used can better extract the correct core EVPA values.
 
\begin{figure*}
    \centering
    \includegraphics[width=17.5cm]{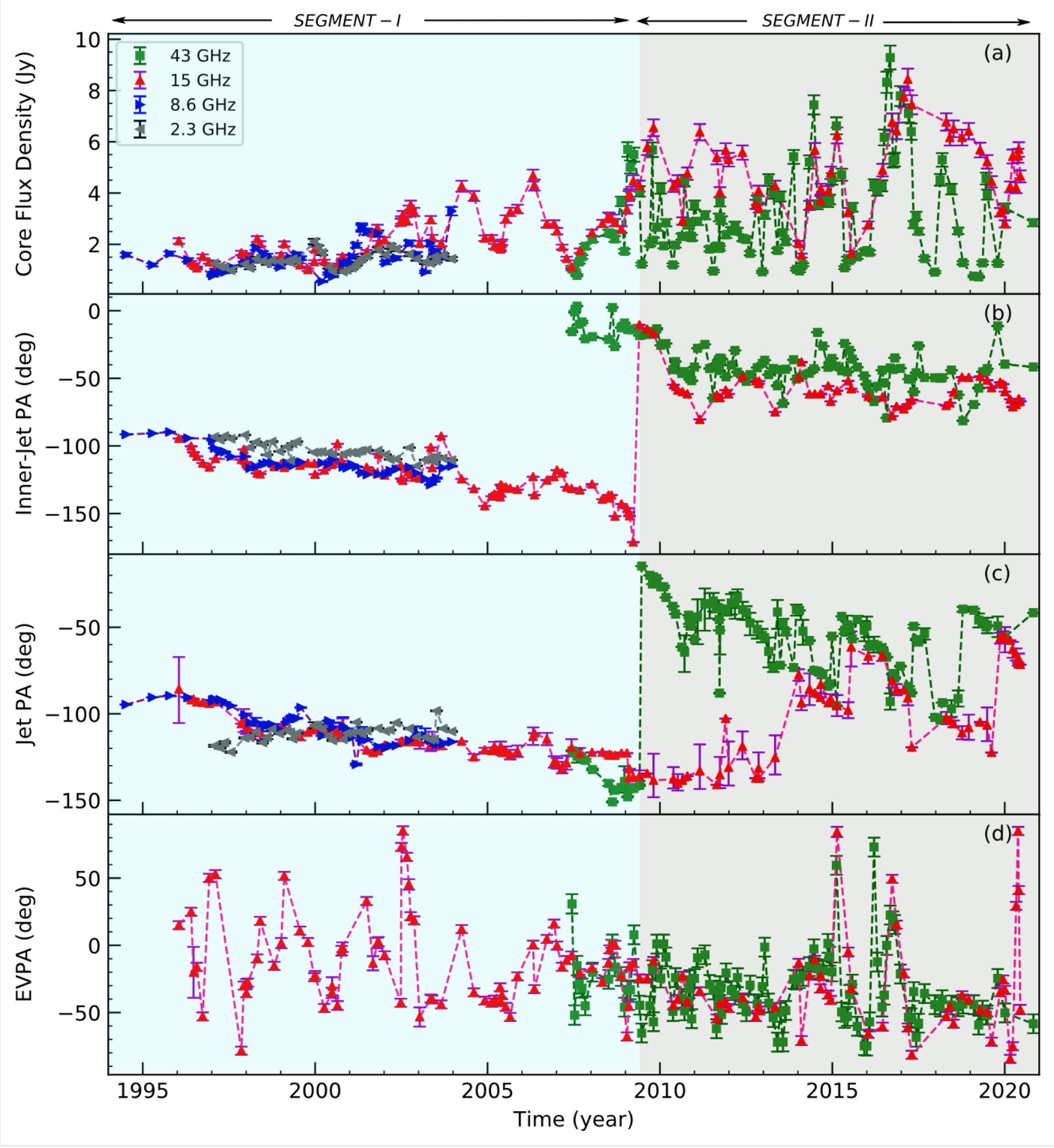}
    \caption{Evolution of the four observables of four bands with time. 
    Panel (a): Multi-band VLBI core flux density plotted as a function of time. 
    Panel (b): Multi-band inner-jet position angle plotted as a function of time.
    Panel (c): Multi-band jet position angle plotted as a function of time.
    Panel (d): Multi-band core EVPA plotted as a function of time. Two double-headed arrows with the text 'SEGMENT \Rmnum{1}' or 'SEGMENT \Rmnum{2}', and a background of different colors identify the segments considered for analysis in our work.}
    \label{fig:basic}
\end{figure*}

\subsubsection{Jet measurement}
\begin{table}
\centering
\caption{Jet components properties. 
Columns are as follows:
    (1) date of observation;
    (2) observing frequency in GHz;
    (3) fitted jet component number; 
    (4) the jet component's relative right ascension shift to the core component in mas;
    (5) the jet component's relative declination shift to the core component in mas;   
    (6) jet position angle derived by fitting regression line.
    \\(This table is available in its entirety in machine-readable form.)}
    \label{tab:jetobs}
    \begin{tabular}{ccccccc}
       \hline
        Epoch     & Frequency  &ID         & $\alpha$          & $\delta$           & JPA          \\
                  & (GHz)      &           &  (mas)            &  (mas)             & ($^\circ$)   \\
        \hline
        1997.079  & 2.3        &1          & $-4.26$              & $-2.14$               & $-117$  \\
        2000.511  & 2.3        &1          & $-3.72$              & $-1.61$               & $-114$  \\
        2003.959  & 2.3        &1          & $-3.70$              & $-1.20$               & $-109$  \\
        1994.515  & 8.6        &1          & $-0.98$              & $-0.12$               & $-94 $  \\
        1999.468  & 8.6        &1          & $-1.07$              & $-0.23$               & $-102$  \\      
        2007.447  & 8.6        &1          & $-1.22$              & $-0.60$               & $-116$  \\
        1996.049  & 15         &1          & $-1.23$              & $-0.30$               & $-85 $  \\       
        2007.438  & 15         &1          & $-0.83$              & $-0.54$               & $-119$  \\
        2007.438  & 15         &2          & $-3.37$              & $-1.45$               & $-119$  \\ 
        2020.451  & 15         &1          & $-0.57$              & $0.20$                & $-71 $  \\
        2007.447  & 43         &1          & $-0.23$              & $-0.14$               & $-121$  \\
        2013.038  & 43         &1          & $-0.43$              & $0.28$                & $-56 $  \\
        2020.831  & 43         &1          & $-0.08$              & $-0.02$               & $-41 $  \\
        \hline   
    \end{tabular}
\end{table}

In Table ~\ref{tab:jetobs}, we list the modeled properties of the recognized jet component. 
Various methods have been utilized to identify the jet position angle, 
such as using the position angle of the brightest jet component with respect to the core or taking a flux-density-weighted position angle average of all fitted jet components from the core~\citep{lister.13.aj}.
We follow the prescription of \citet{valtonen.12.mn} and \citet{moor.11.aj} to fit a regression line that is forced to go through the location of the core component.
Slightly different from previous methods, we use the orthogonal distance regression (ODR) algorithm to get the best linear fitting rather than the ordinary least-squares (OLS) fit ~\citep{moor.11.aj} or the ordinary least-squares bisector (OLS-bisector) regression ~\citep{valtonen.12.mn}.
This method can be used to take into account the errors in both relative coordinates. 
The data points are weighted by the inverse square of their positional uncertainty. 
The uncertainty of the position of the fitted Gaussian components in coordinates along their major and minor axes are estimated by the following formula:

\begin{equation}
\mu \approx \frac{\Theta_{beam}}{2} \times \frac{1}{\mathrm{SNR}} \times \sqrt{1+\left[\frac{\Theta_{image\ size}}{\Theta_{ beam }}\right]^{2}}
\end{equation}

$\mu$ contains information in two directions: $\mu (x_0)$ and $\mu (y_0)$.
$\mu (x_0)$ and $\mu (y_0)$ are the rms lengths of the major and minor axes of the error ellipse composed of position uncertainty.
SNR is the signal-to-noise of the component, which is determined by dividing the peak flux density of the fitted Gaussian component by the off-source noise.
$\Theta_{image\ size}$ is the deconvolved size (the full-width at half-maximum, FWHM) of major or minor axis which deconvolves the Clean beam $\Theta_{beam}$ from the fitted component size.
The error ellipse major-axis position angle $\phi$ is measured from north to east, then the right-ascension and declination errors, i.e.,~$\mu (\alpha)$ and $\mu (\delta)$, are calculated by~\citep{condon.97.pasp}:\
$\mu^{2}(\alpha)=\mu^{2}\left(x_{0}\right) \sin ^{2} \phi+\mu^{2}\left(y_{0}\right) \cos ^{2} \phi,$ \
$\mu^{2}(\delta)=\mu^{2}\left(x_{0}\right) \cos ^{2} \phi+\mu^{2}\left(y_{0}\right) \sin ^{2} \phi$.
The ODR result outputs the slope of the best linear regression line ($k$) and the standard error ($\sigma_k$); we calculate the jet position angle and its error with $k$ and $\sigma_k$.
Fig.~\ref{fig:odr} illustrates our fitting method for three contiguous epochs at 15~GHz and 43~GHz.
If only one jet component is fitted within a reasonable distance from the core, the position angle of this component is assumed as the jet position angle. 
All calculated jet position angles are listed in column (6) in Table~\ref{tab:jetobs}.

\subsection{Time-domain observables}
We extracted some observables, namely the core flux density, the inner-jet position angle, the jet position angle, and the core EVPA, their evolutions over time are shown in Fig.~\ref{fig:basic}.
We follow~\citet{homan.02.apj}'s work and use the dominant calibration error, estimated to be \textasciitilde5\% of the core flux density, as the core flux density error.
The inner-jet position angle error is estimated by AIPS using the formulae given from~\citep{fomalont.99.aspcs}.
The error of the jet position angle and the core EVPA is as described above.

\renewcommand{\dblfloatpagefraction}{.999}
\begin{figure*}
    \centering
    \includegraphics[width=8cm]{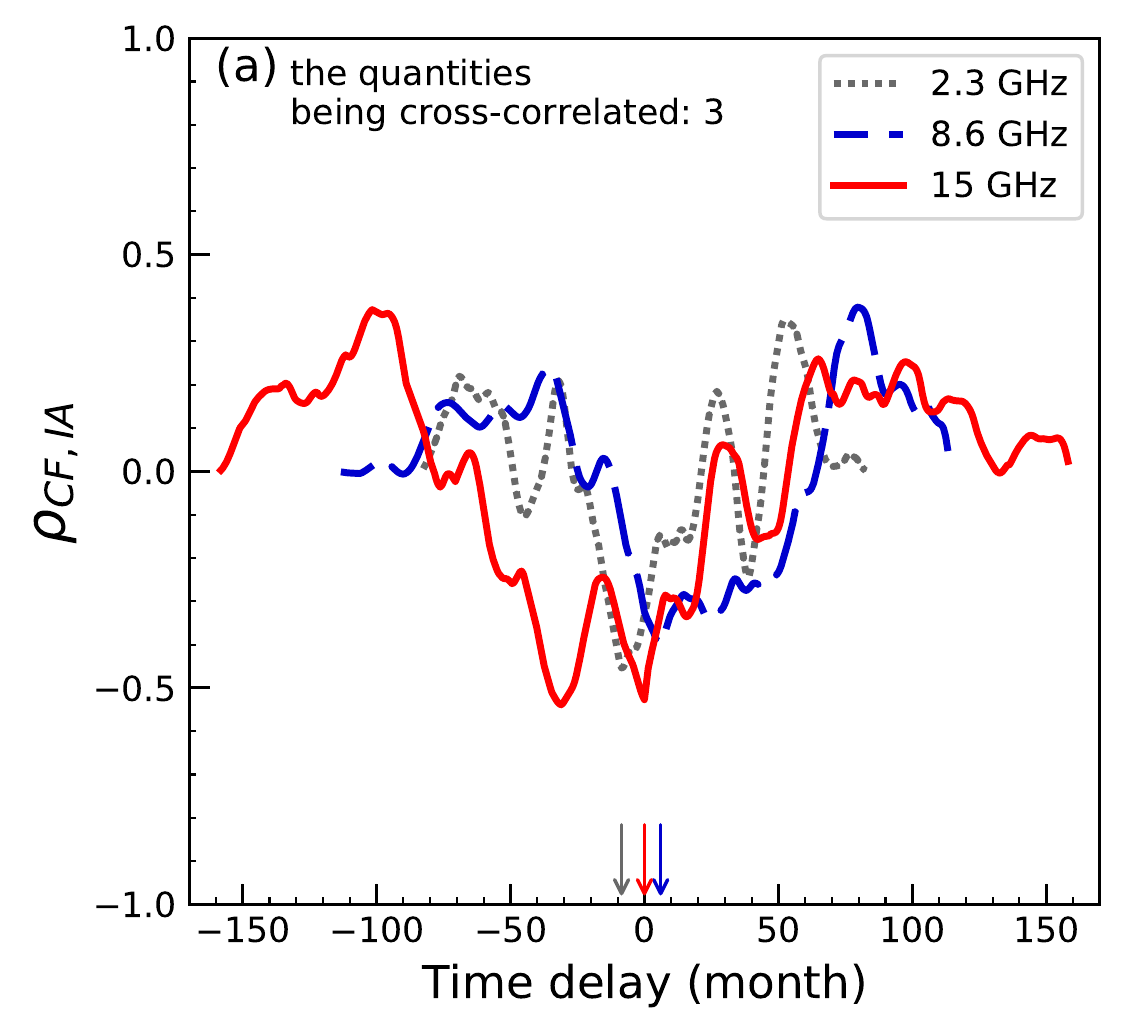}
    \includegraphics[width=8cm]{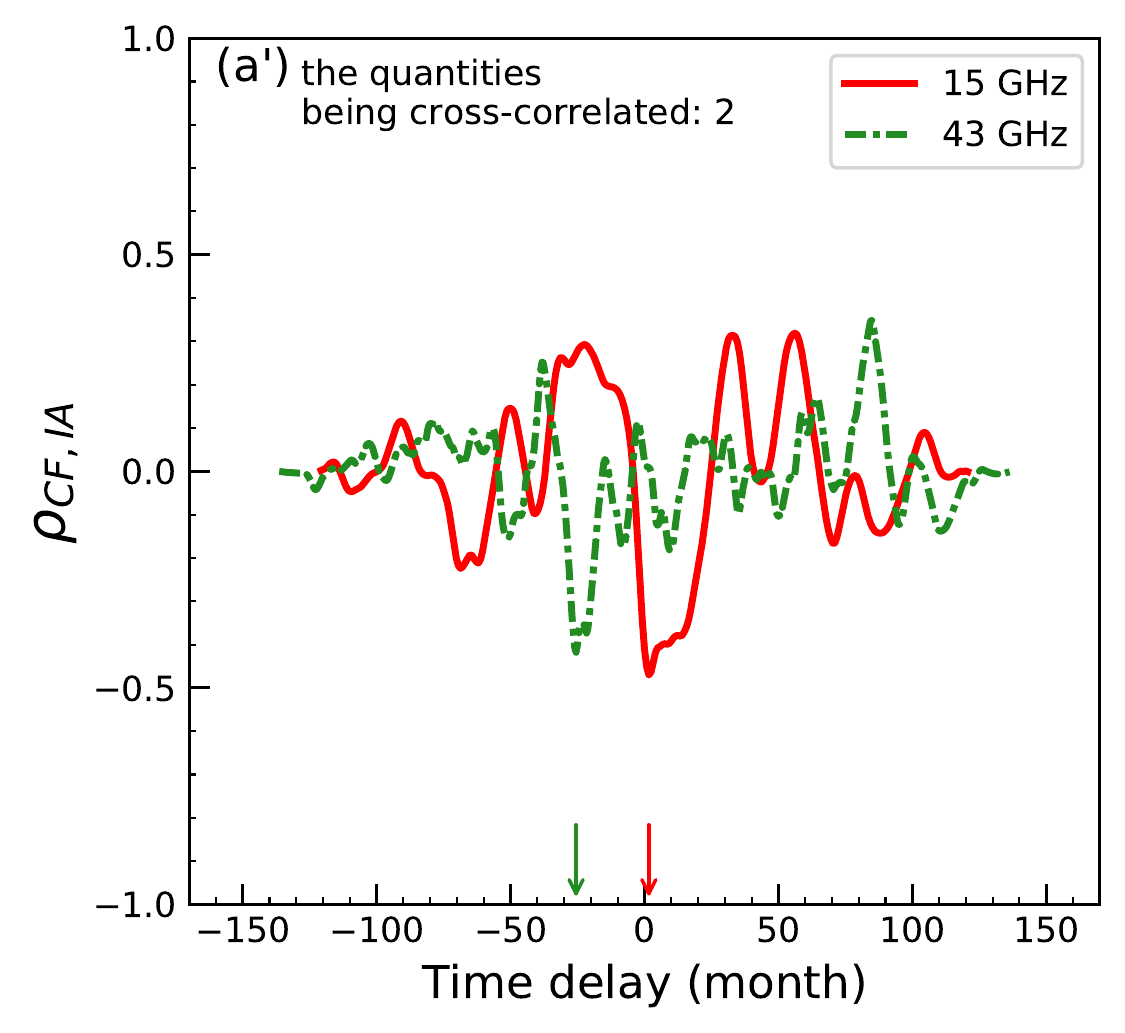}\\
    \includegraphics[width=8cm]{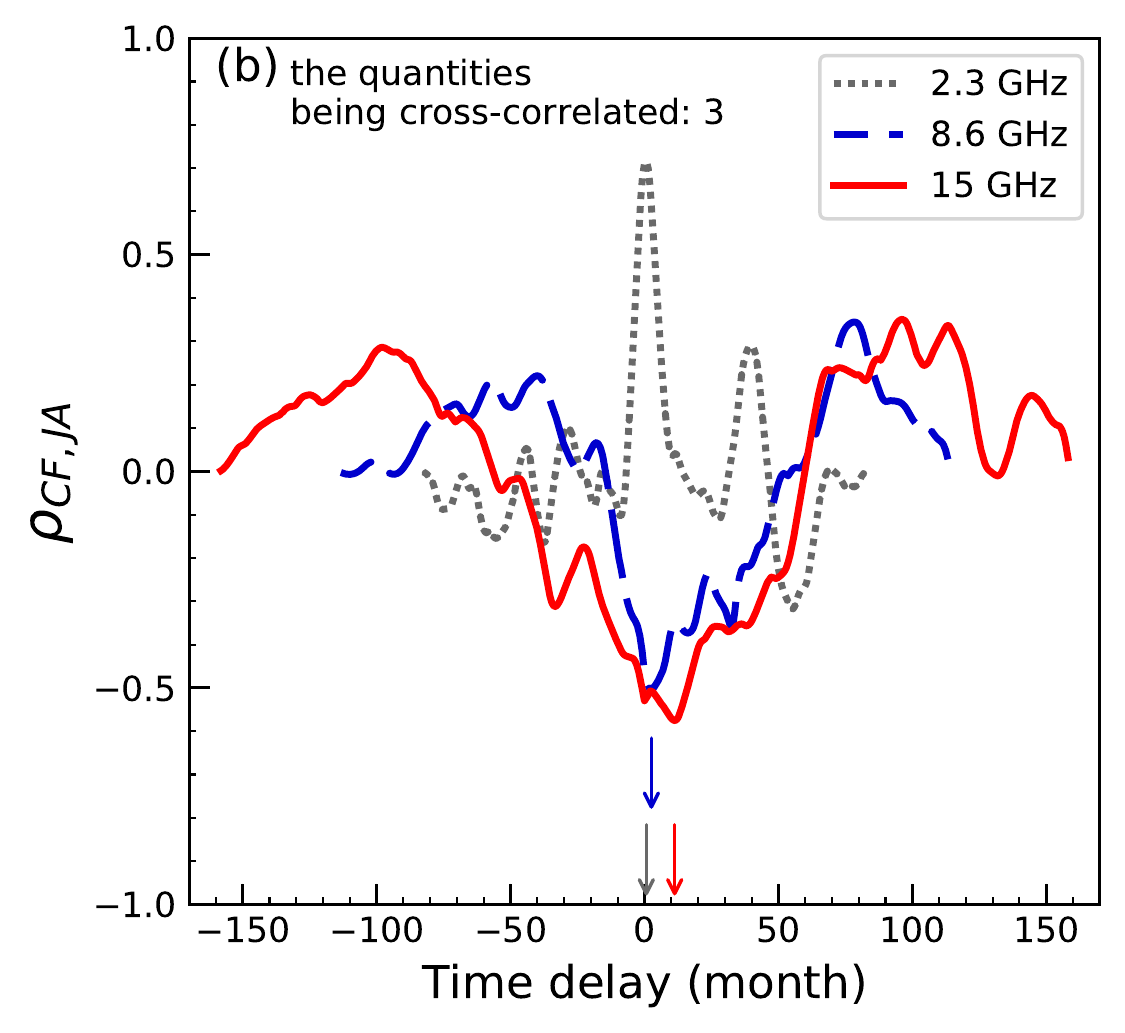}
    \includegraphics[width=8cm]{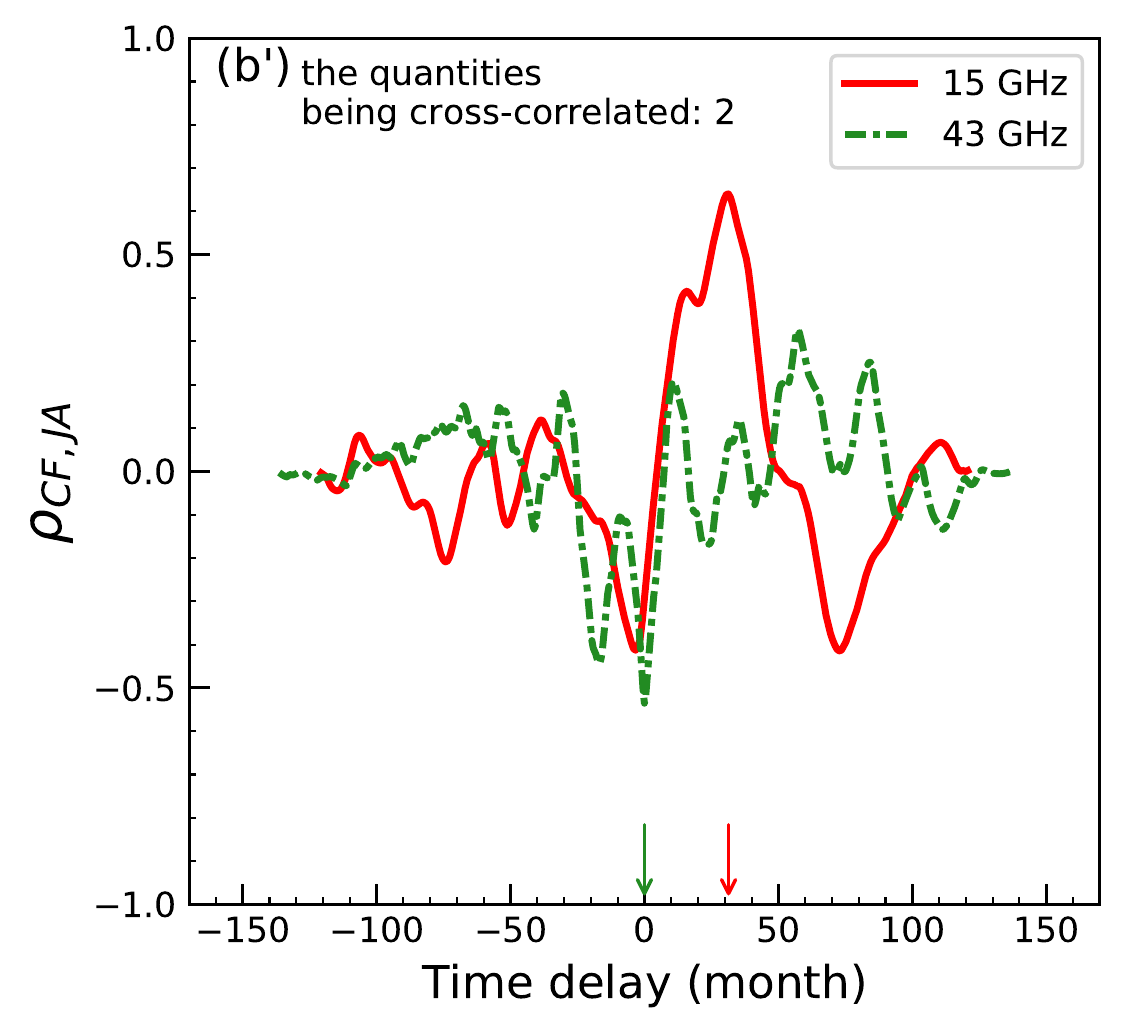}\\
    \includegraphics[width=8cm]{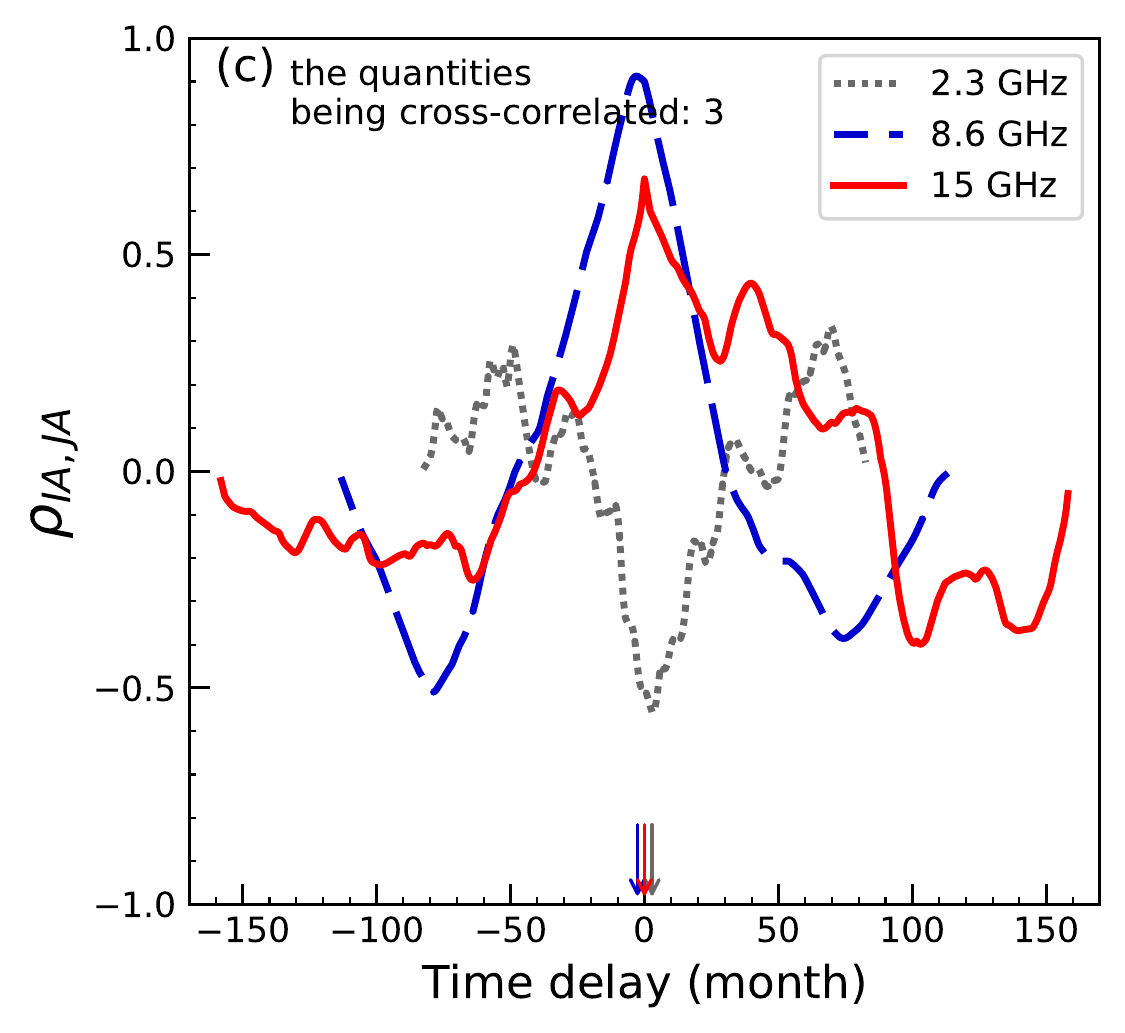}  
    \includegraphics[width=8cm]{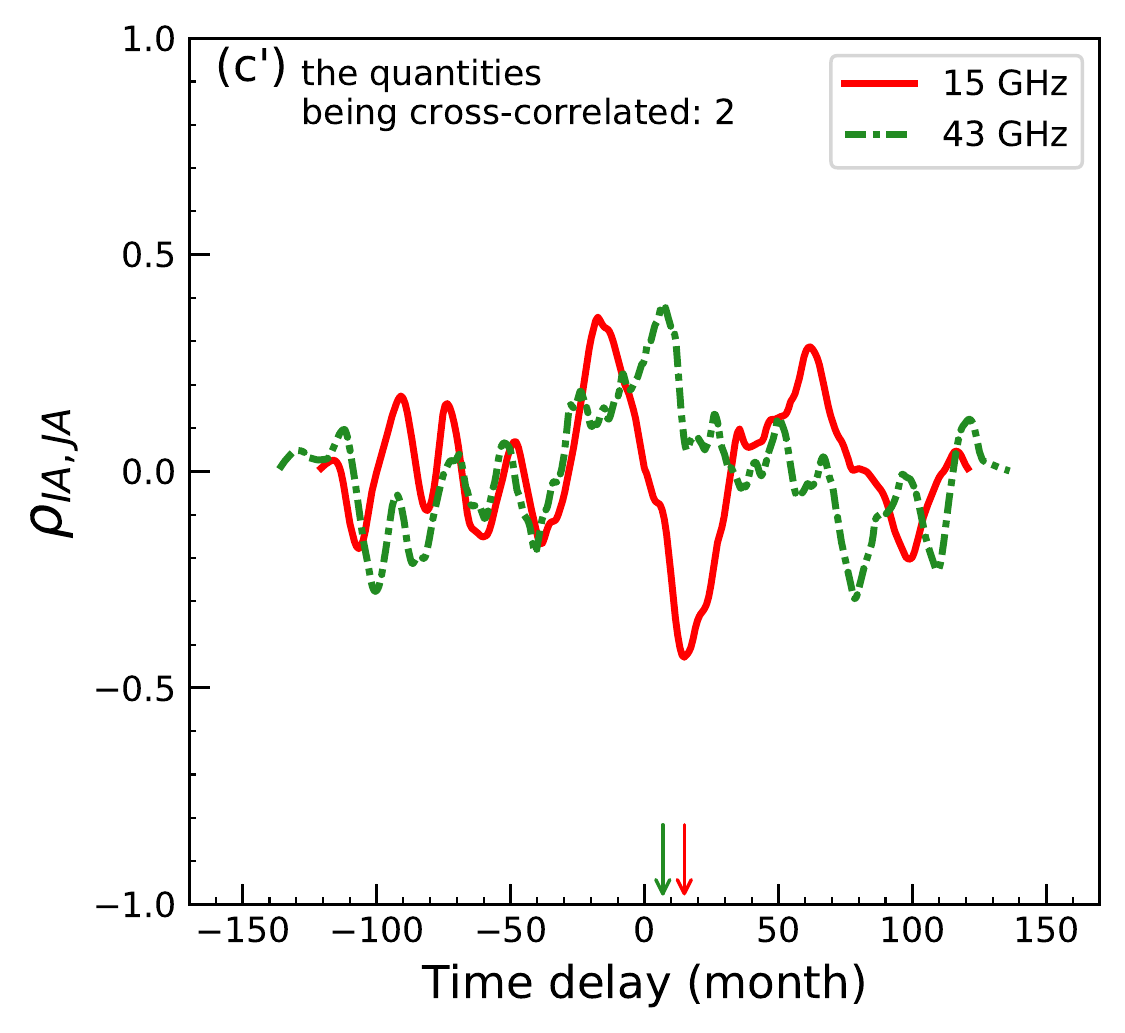}\\
    \end{figure*}
    \addtocounter{figure}{0}
  
\begin{figure*}
	\centering
    \includegraphics[width=8cm]{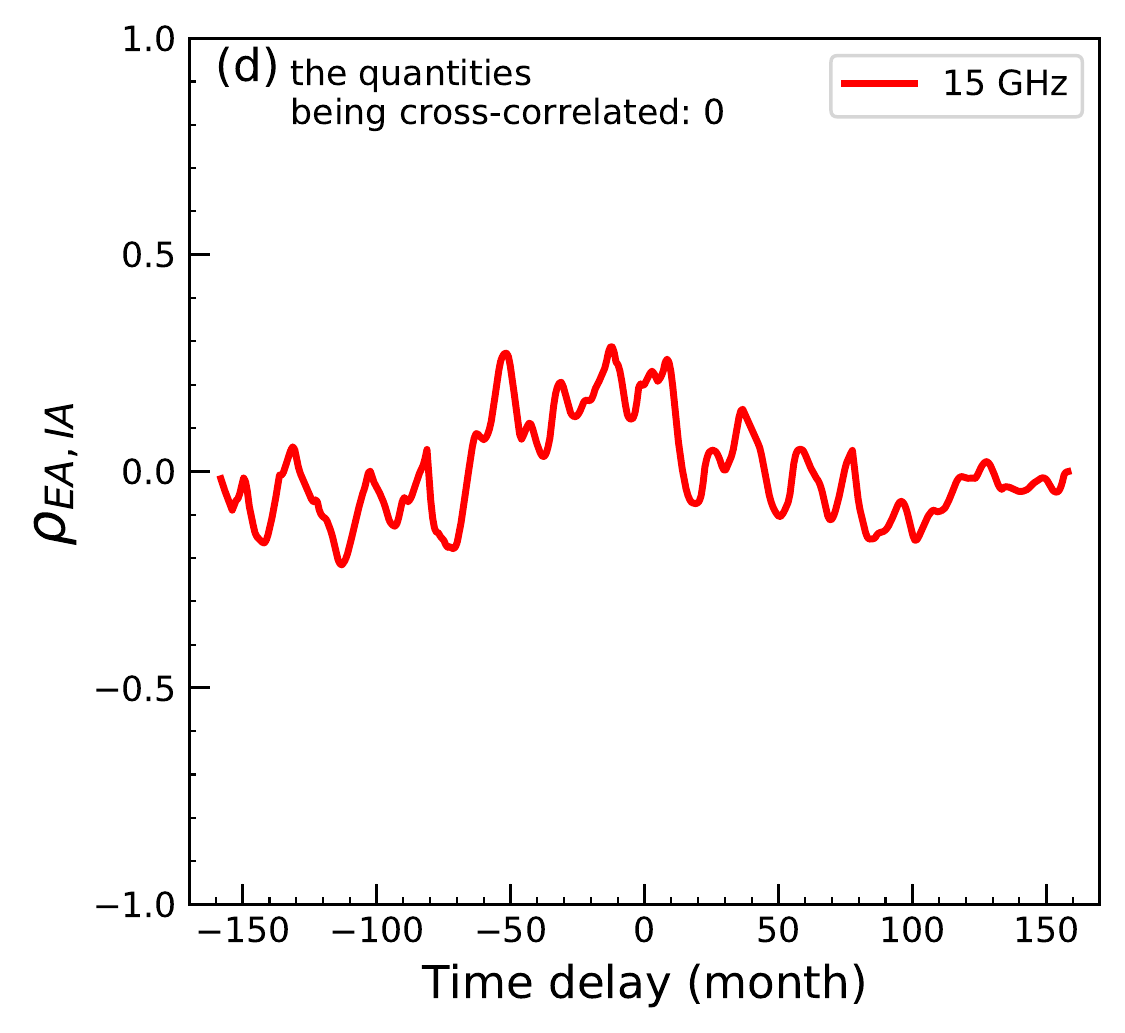}  
    \includegraphics[width=8cm]{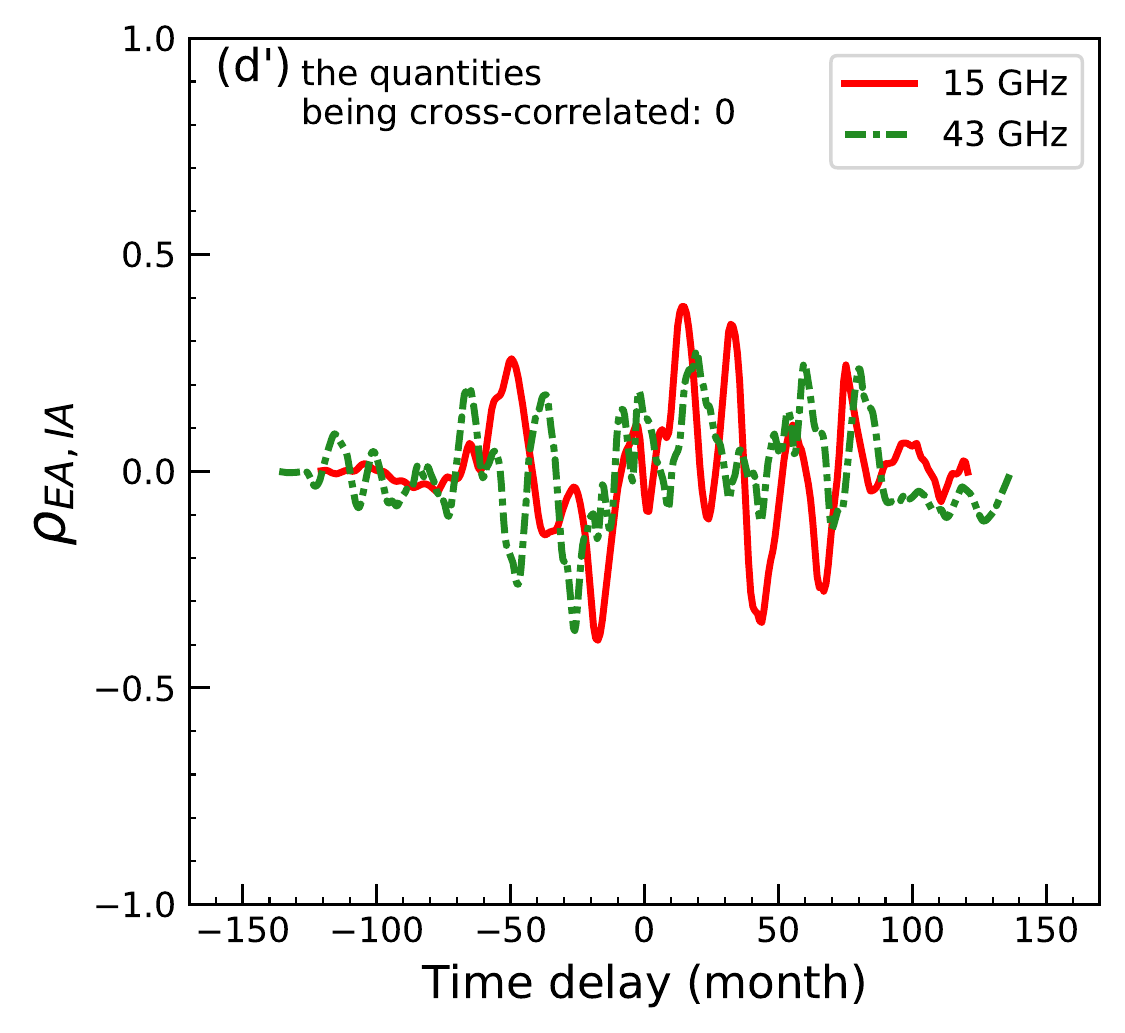}\\ 
    \caption{Cross-correlation analysis of each pair observables at 2.3~GHz, 8.6~GHz, 15~GHz and 43~GHz.
    The left panes are the results of segment \Rmnum{1}, and the right panels are the results of segment \Rmnum{2}.
        Panel (a/a'): core flux density vs. inner-jet position angle. 
        Panel (b/b'): core flux density vs. jet position angle.
        Panel (c/c'): inner-jet position angle vs. jet position angle. 
        Panel (d/d'): core EVPA vs. inner-jet position angle.
        The positive time delay corresponds to the latter observable's time series leading the former observable when doing the cross-correlation, and vice versa.
        Arrows located at time delays corresponding to cross-correlation peaks.
        }  
    \label{fig:in-band-cor}
\end{figure*}

\section{Correlation analysis method}\label{sec: conclusions}
\label{method}
The cross-correlation function (CCF) is commonly employed in the study of AGN, including probing the structure of the broad emission-line region by reverberation mapping~\citep{peterson.04.apj}, studying the continuum emission mechanism by correlating multi-band light curves~\citep{max.14.mn}, and finding correlations between the radiation variability and other AGN properties~\citep{rani.14.aa}.
We also employed the CCF to seek correlations among variations of the flux density, jet position angle, and EVPA.

When calculating the cross-correlation, each point in one time series of observables must be paired with a point in the other time series, so the data points should be regularly spaced, otherwise the pairing is only possible when the two-time series are aligned.
The datasets we collected were unevenly sampled due to observation schedules, project time allocations, and other constraints such as equipment failures and severe weather.
First, we expanded the amount of data three times by linear interpolation to make the data set even at each band.
The time interval for our interpolation seems to be a reasonable choice since it is shorter than the majority of the real data intervals, but not so short that a large number of artificial data points interfere with the analysis. 
We also compared another interpolation scheme (e.g., cubic spline interpolation) but found that this interpolation sometimes overfits, not always reliable.
We divided the entire time series into two segments for correlation analysis at the time of the large changes of the inner-jet position angle of 15 GHz or the jet position angle of 43 GHz.
Moreover, three data points after the 15 GHz inner-jet position angle change were removed, as well as several data points before the 43 GHz jet position angle change.
Our purpose in doing this is to make the time series of each segment have quasi-stationary properties as much as possible.
Then we calculated the normalized CCF between each pair of observables, with the maximum and secondary peaks obtained from the correlation coefficient sequence.
We analyze the significance of cross-correlation results by Monte Carlo (MC) trials. 
At the same time, we should not ignore the fact that since the overlap between the observables is smaller for large values of time delay, we deem these peaks relatively less credible.

We constructed 10,000 entirely simulated time series for each observable that span the approximate range of variation measured in the real data but have their own variations independent of each other in time domain. 
We repeated the step of the previous standard cross-correlation analysis for 10,000 simulated observable's time series pairs and accumulated the acquired cross-correlation coefficients for each time delay.
And for each time delay, the significance level of the real data cross-correlation coefficient is estimated based on the distribution of the simulated cross-correlation coefficients.

We also employed MC sampling to estimate the certainty of these correlations affected by all errors on the observables.
The correlation coefficient errors were determined from 10,000 MC trials by adding sampled normally distributed errors to the expected value of the observables and then performing correlation calculations for each trial.
We subtracted the correlation coefficient calculated from the expected value of the observables ($\rho_{expt.,i}$) from the MC trial's correlation coefficient ($\rho_{\sc{MC},i,n}$) to get a series of correlation coefficient error values ($\Delta\ \rho_{i,n}$), i.e., $\Delta\ \rho_{i,n}$ = $\rho_{MC,i,n}$ $-$ $\rho_{expt.,i}$, 
$i$ refers to the index of the time delay slice and $n$ denotes the $n$th MC trial.
We verified $\Delta\ \rho_i$ of each time delay slice following the normal distribution at a significance level of 5\% by the Shapiro Wilk tests \citep{shapiro.65.biom}.
When each $\rho_{MC,i,n}$ and the $\rho_{expt.,i}$ are very close, i.e., the mean and the standard deviation of the $\Delta \rho_i$ distribution close to 0,
it can be considered that the error of the observables does not affect the final correlation result. 

\section{Results of correlations between observables}
\label{res}

The correlation results which have physical meaning between each pair of observables are shown in Fig.~\ref{fig:in-band-cor} and Table ~\ref{tab:result-cc}.

It is very likely to make wrong physical interpretations based solely on the maximum peak of CCF without considering its significance.
If and only if there is no secondary peak near zero time delay falling within plus or minus 0.8 times the maximum peak's value of the CCF and this peak with a 95\% significance level, we claim that the maximum peak is significant and the two observables are correlated.
If there is a secondary peak which also has a 95\% significance level, then it indicates that the two observables are more likely to correlate at that peak.

\begin{figure*}
    \centering
    \includegraphics[width=4.4cm]{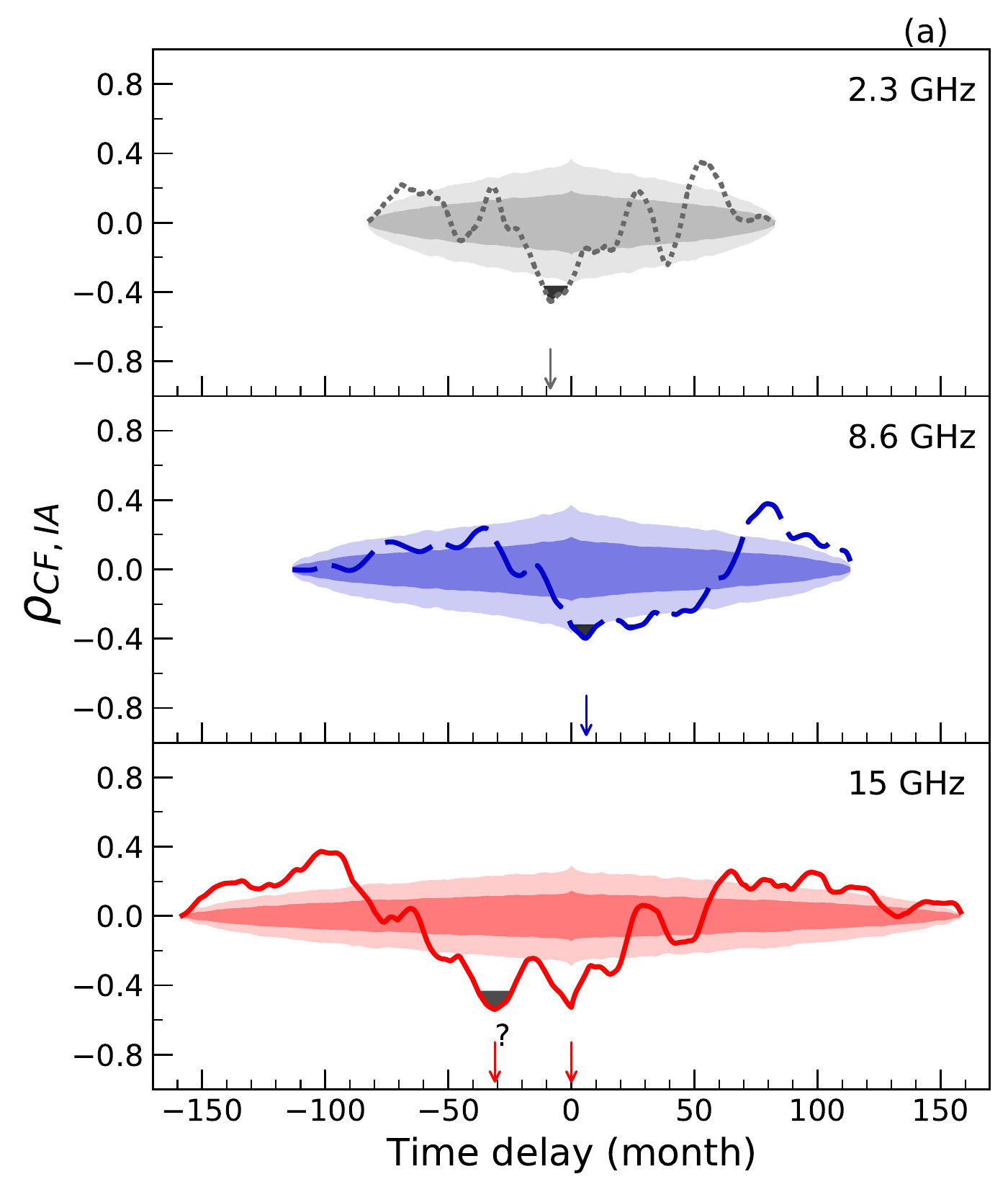}
    \includegraphics[width=4.4cm]{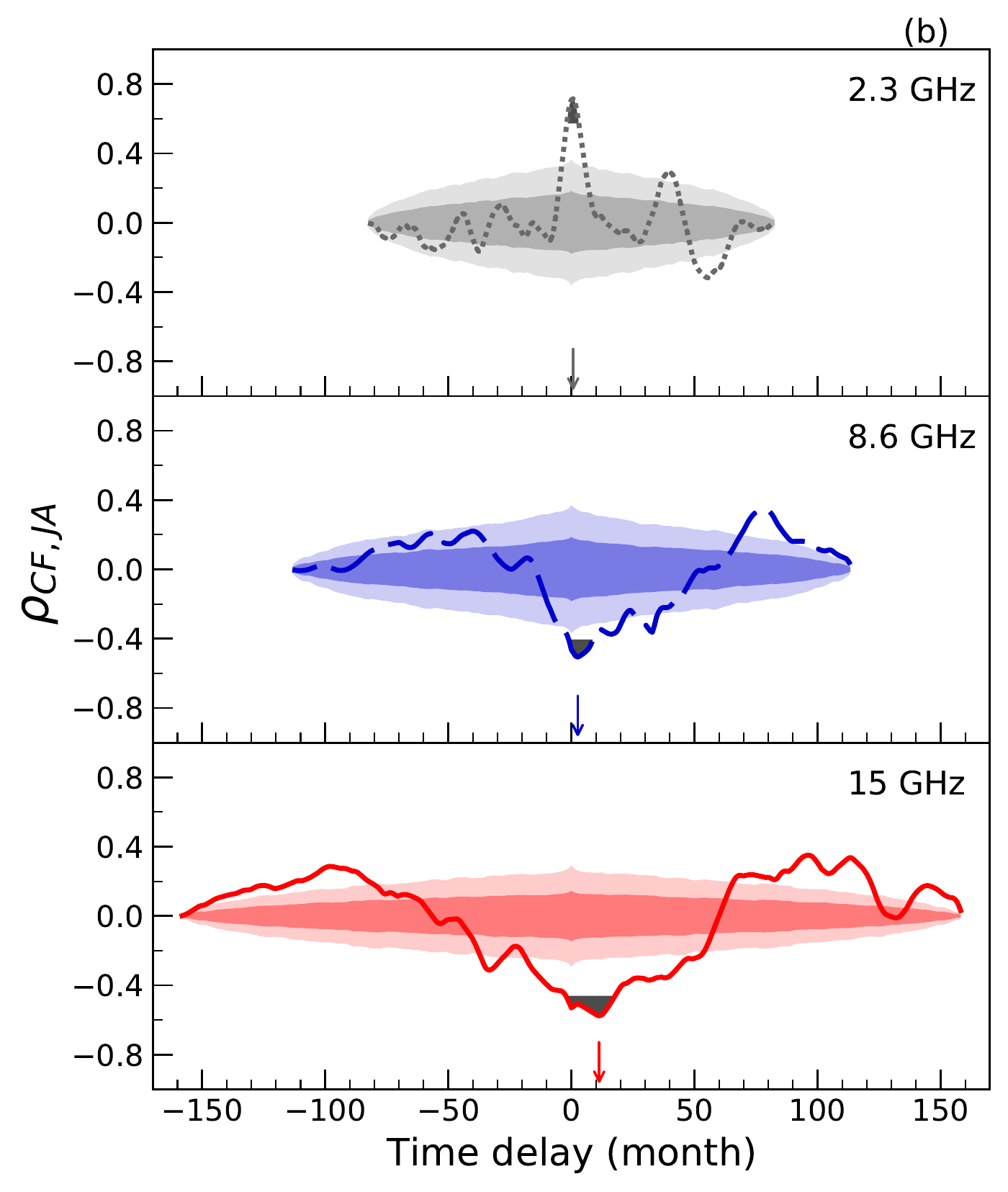}
    \includegraphics[width=4.4cm]{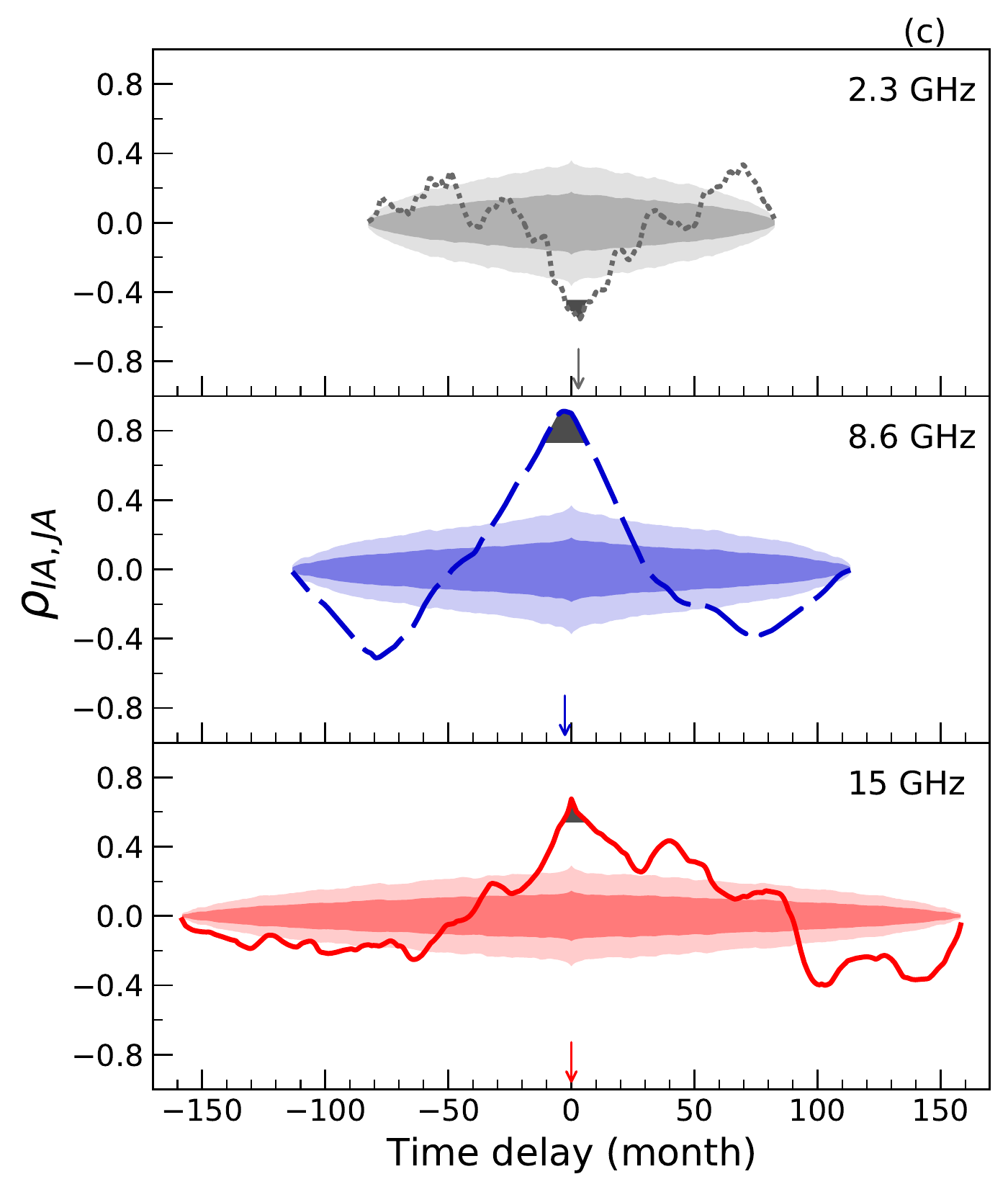}
    \includegraphics[width=4.4cm]{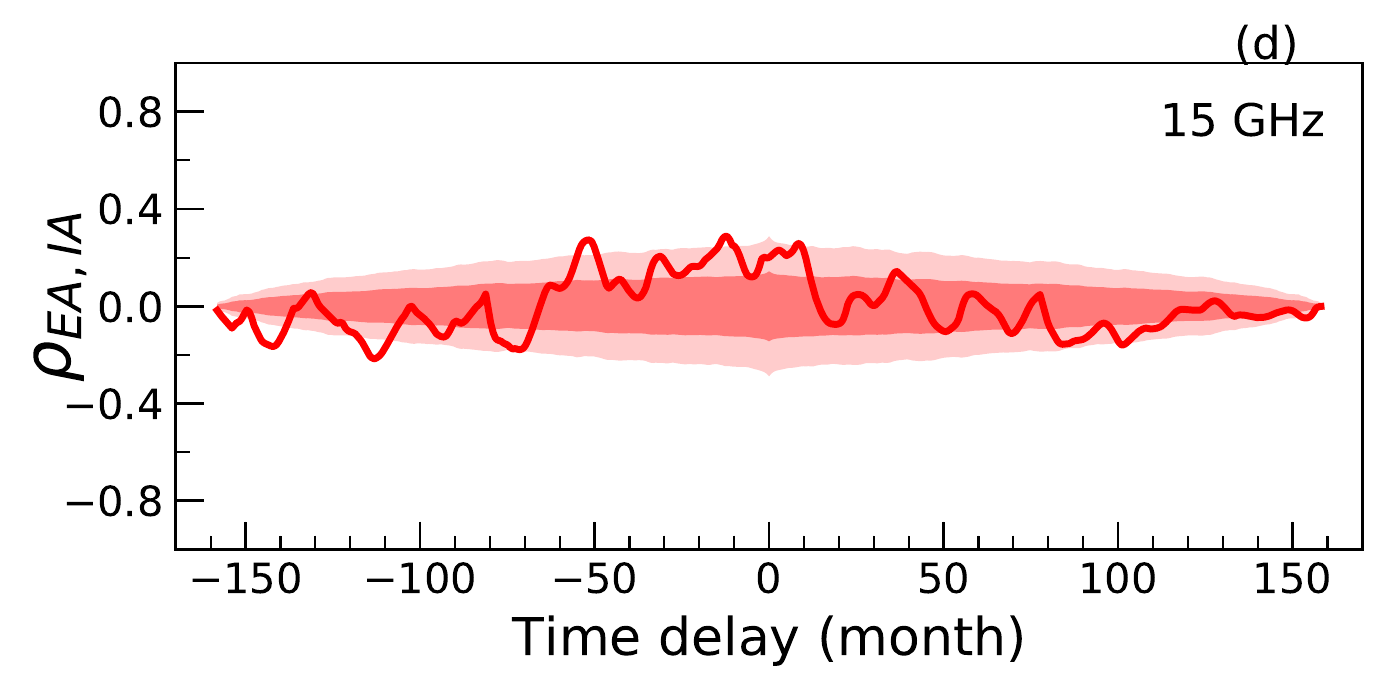}\\
    \includegraphics[width=4.4cm]{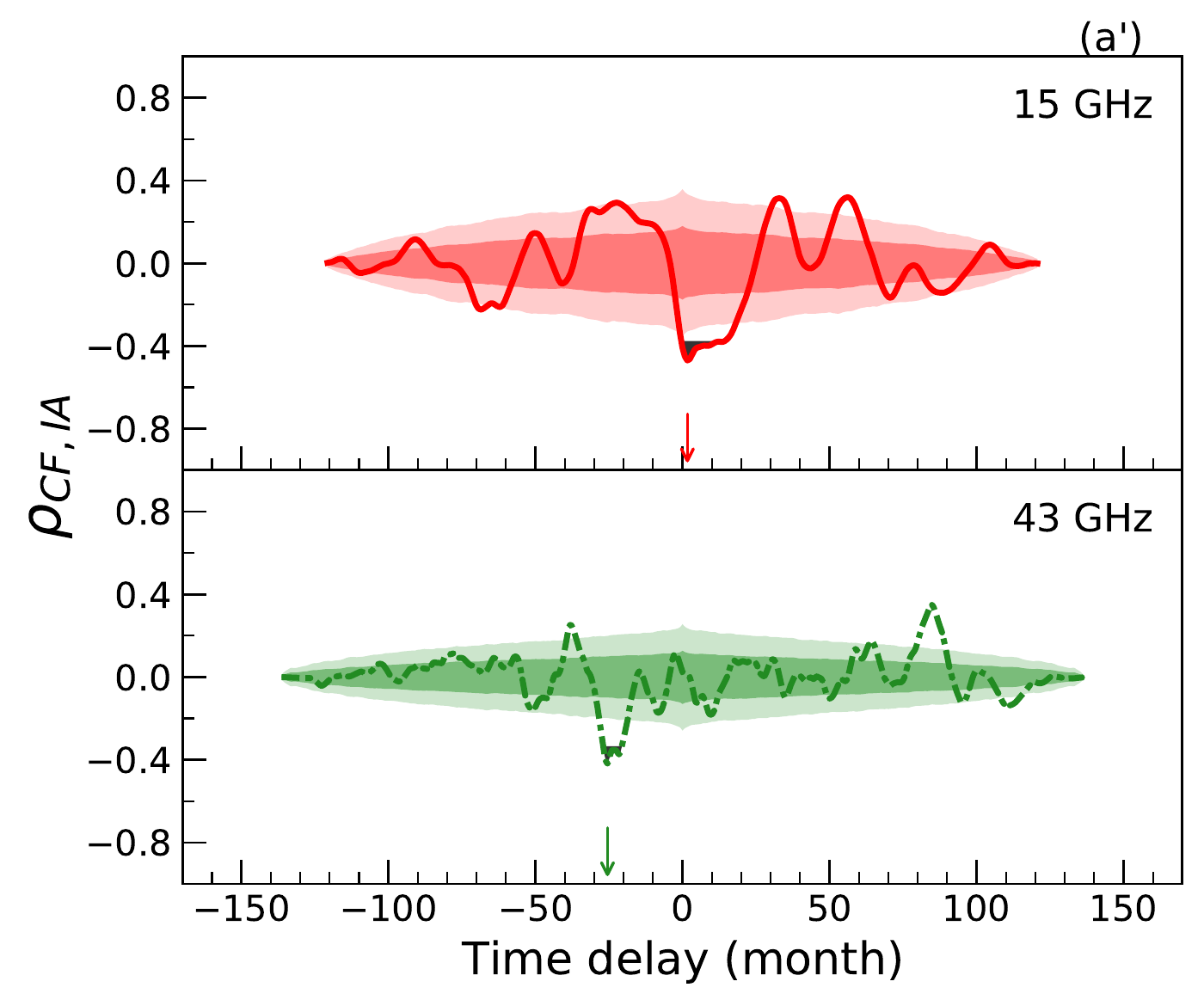}
    \includegraphics[width=4.4cm]{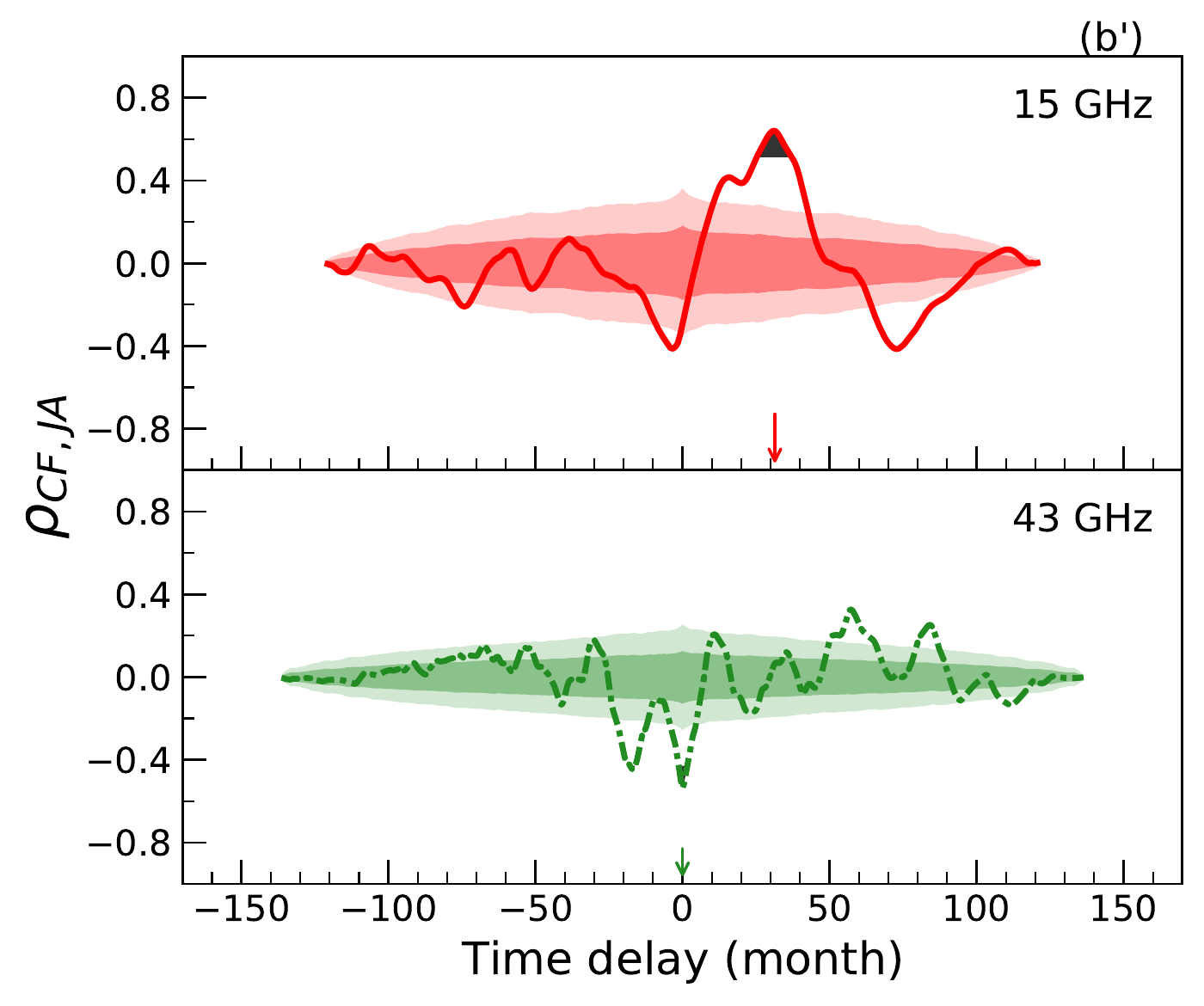}
    \includegraphics[width=4.4cm]{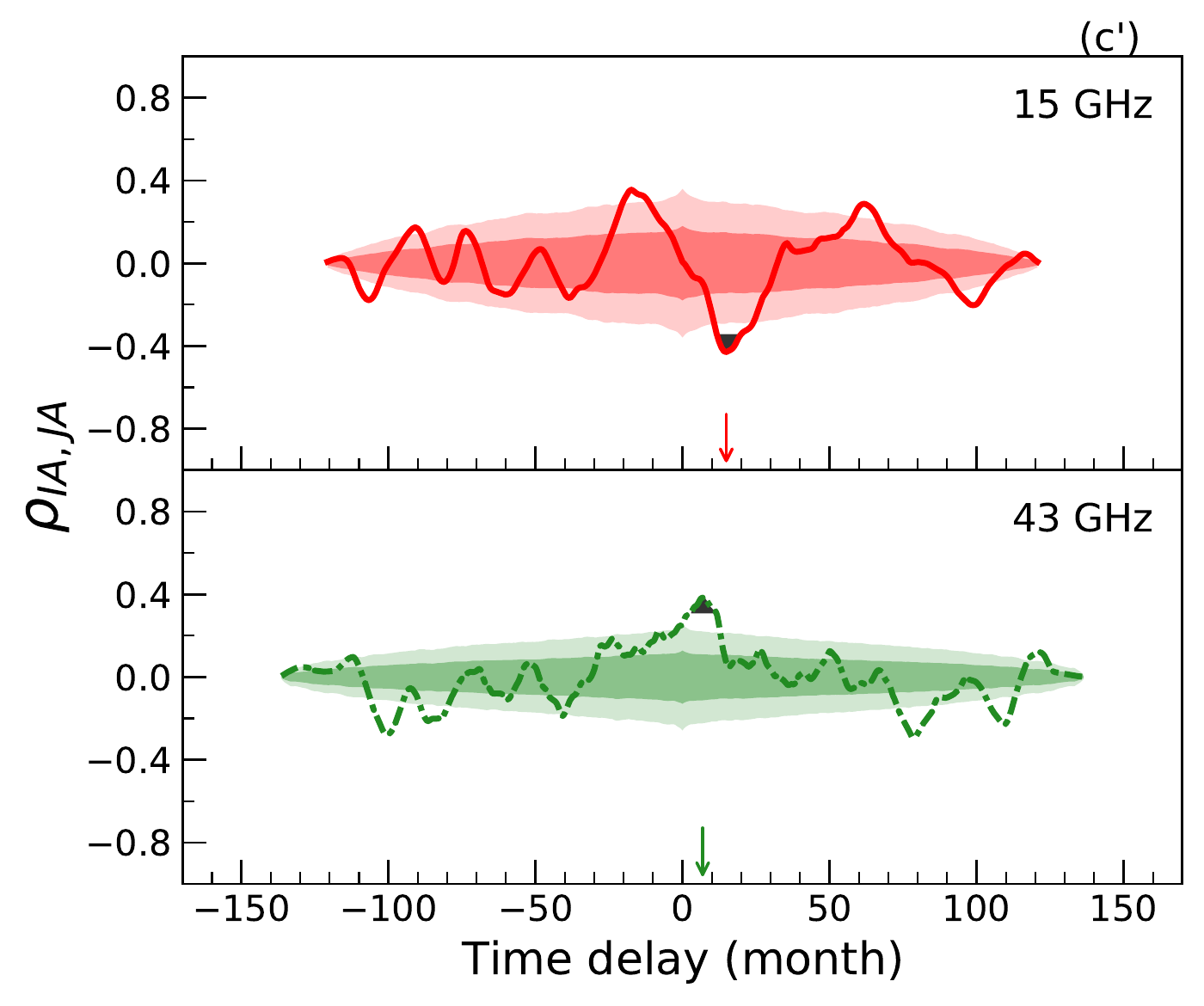}
    \includegraphics[width=4.4cm]{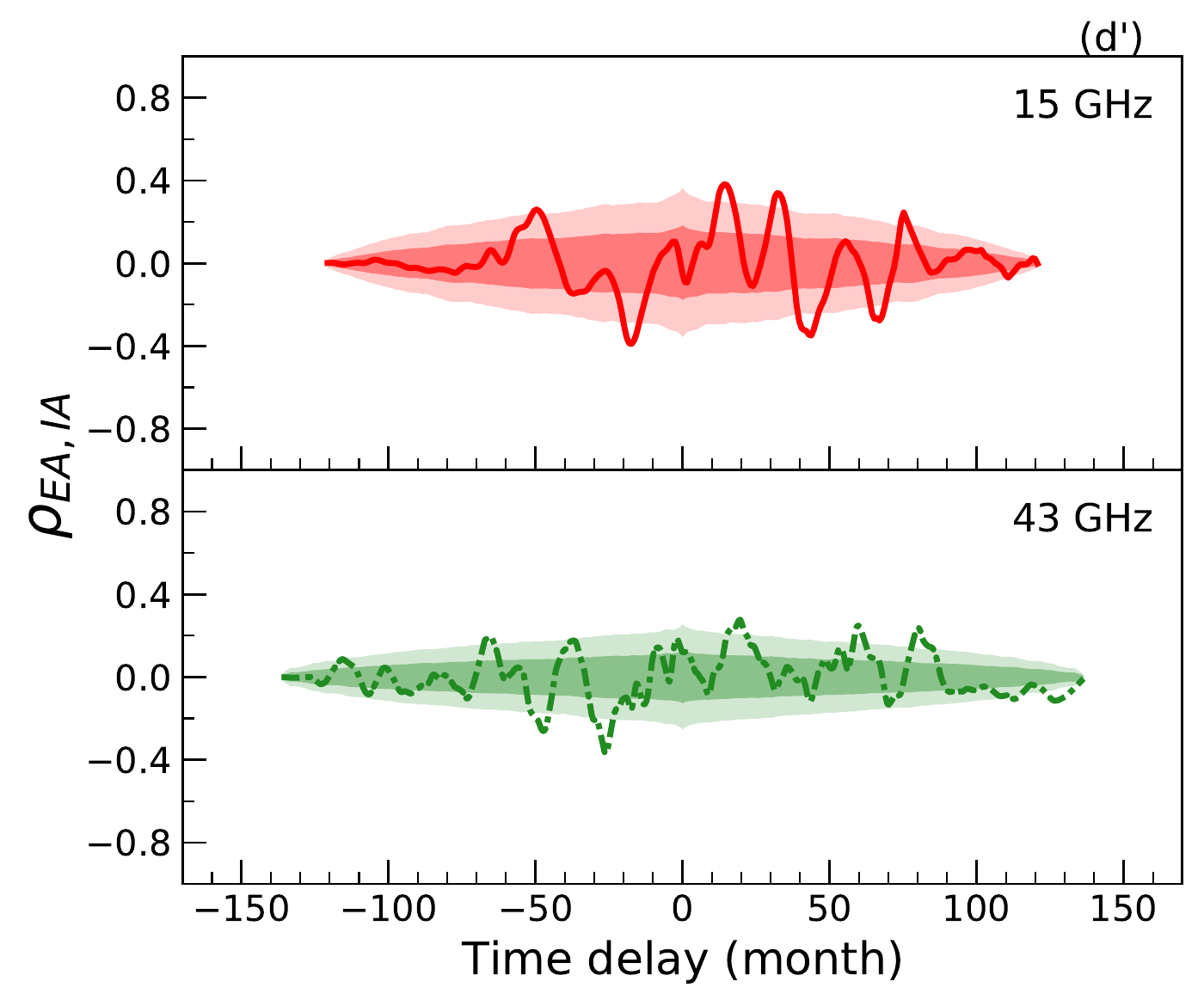}\\
    \caption{Cross-correlation significance results. The color lines represent the cross-correlation for each pair observables, while the color contours show the distribution of random cross-correlations obtained by the Monte Carlo simulation with 1$\sigma$ (dark color/inner layer near horizontal axis), 2$\sigma$ (light color).
        Panel (a/a'): core flux density vs. inner-jet position angle. 
        Panel (b/b'): core flux density vs. jet position angle.
        Panel (c/c'): inner-jet position angle vs. jet position angle. 
        Panel (d/d'): core EVPA vs. inner-jet position angle.
        Panel (a/b/c/d) is the result from segment-\Rmnum{1} data, and Panel (a'/b'/c'/d') is the result from segment-\Rmnum{2} data.
        Arrows located at time delays corresponding to cross-correlation peaks.
	The arrow with a question mark on the left side of the third panel in Panel (a) identifies the position of the maximum peak. However, based on our principle of determining the significant peak, in this case, the significant peak is considered to be at the arrow close to 0.
	The shaded area indicates parts with a correlation coefficient greater than (or less than) the 0.8 times maximum peak of the cross-correlation function.}
    \label{fig:mc1}
\end{figure*}

\begin{figure*}
    \centering
    \includegraphics[width=4.4cm]{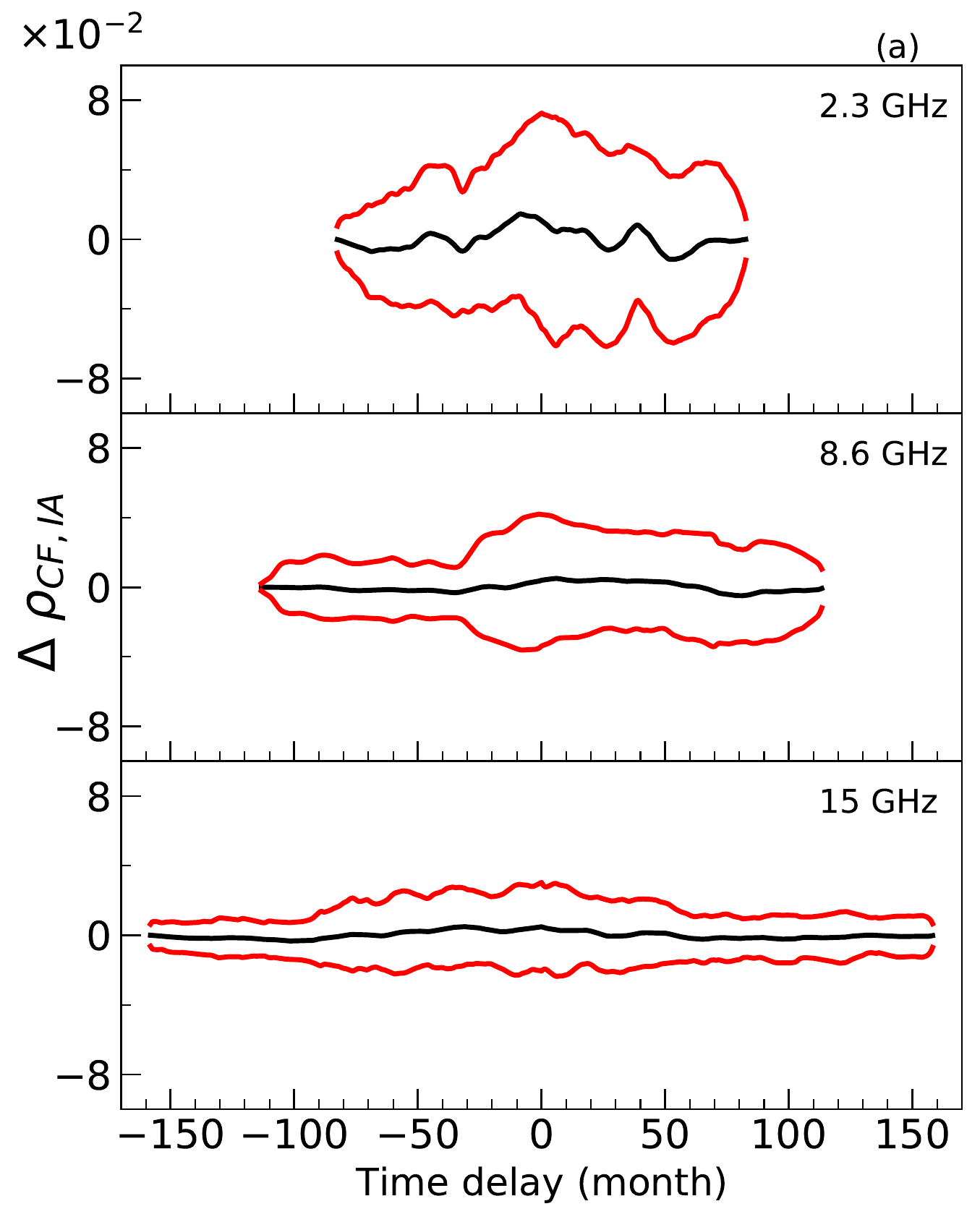}
    \includegraphics[width=4.4cm]{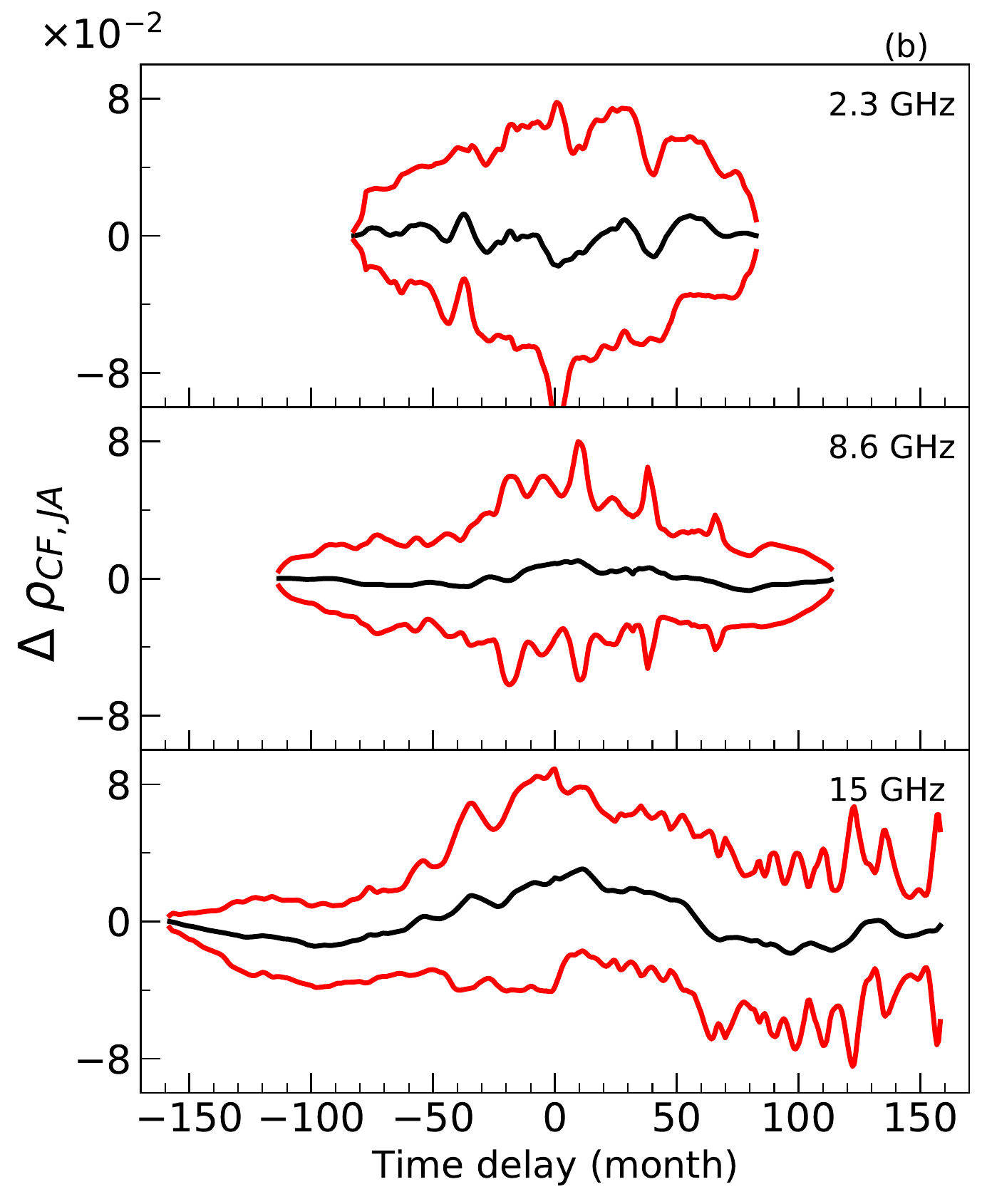}
    \includegraphics[width=4.4cm]{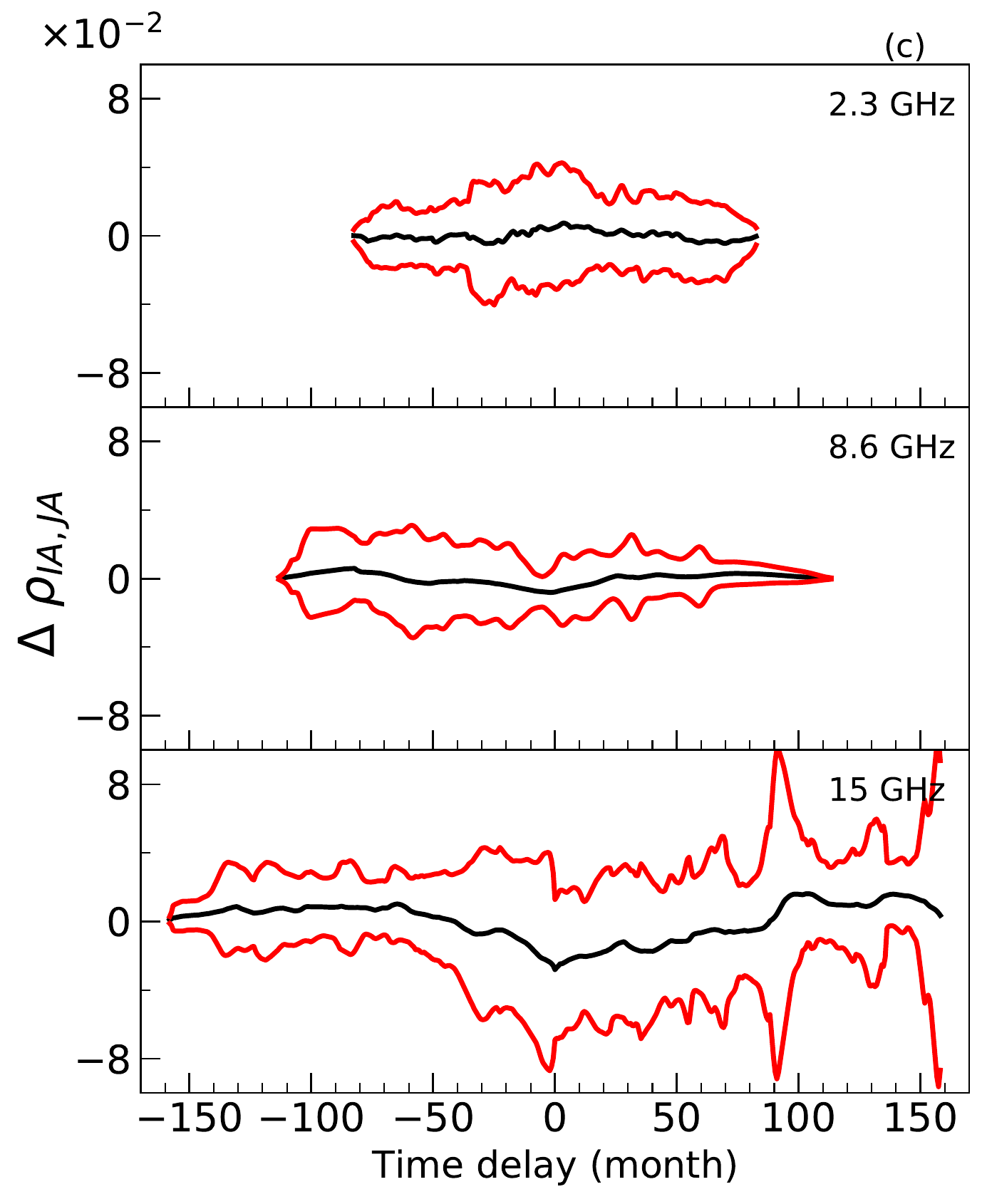}
    \includegraphics[width=4.4cm]{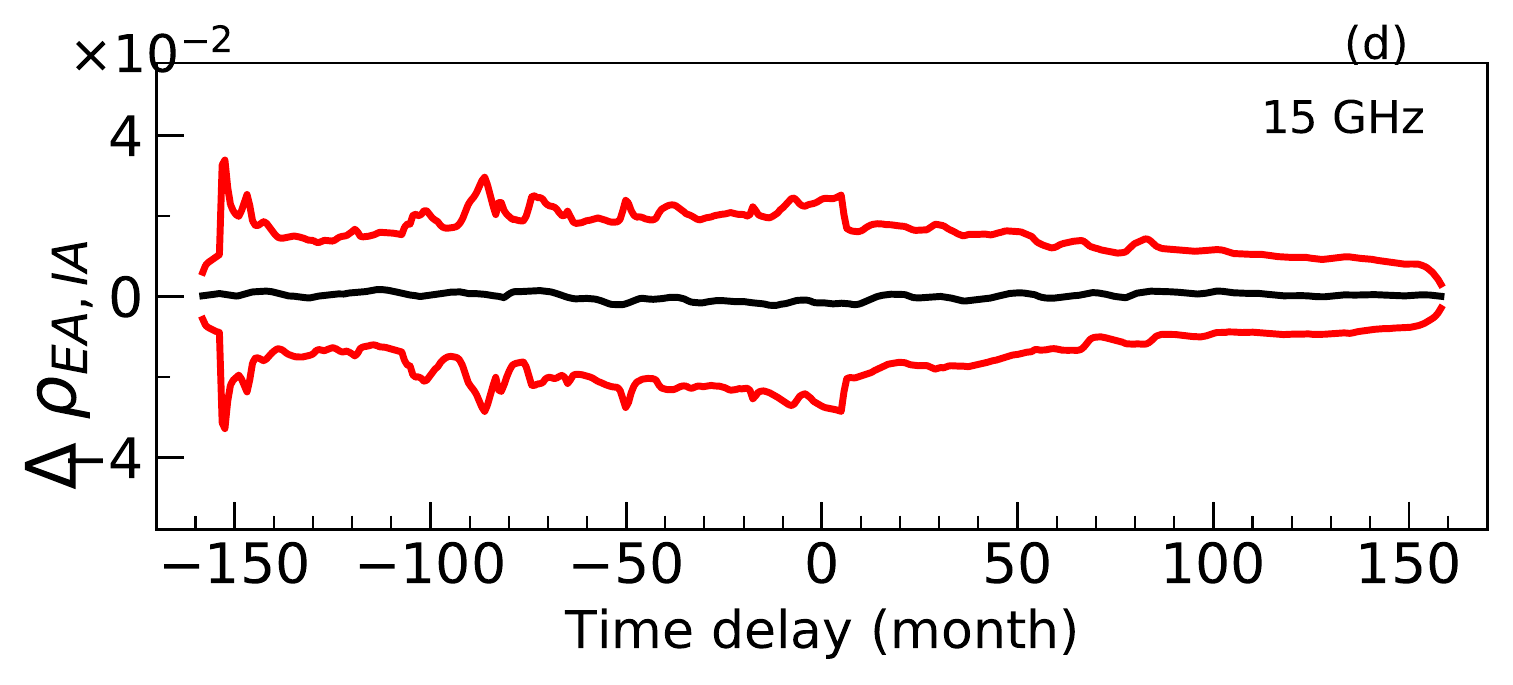}\\
    \includegraphics[width=4.4cm]{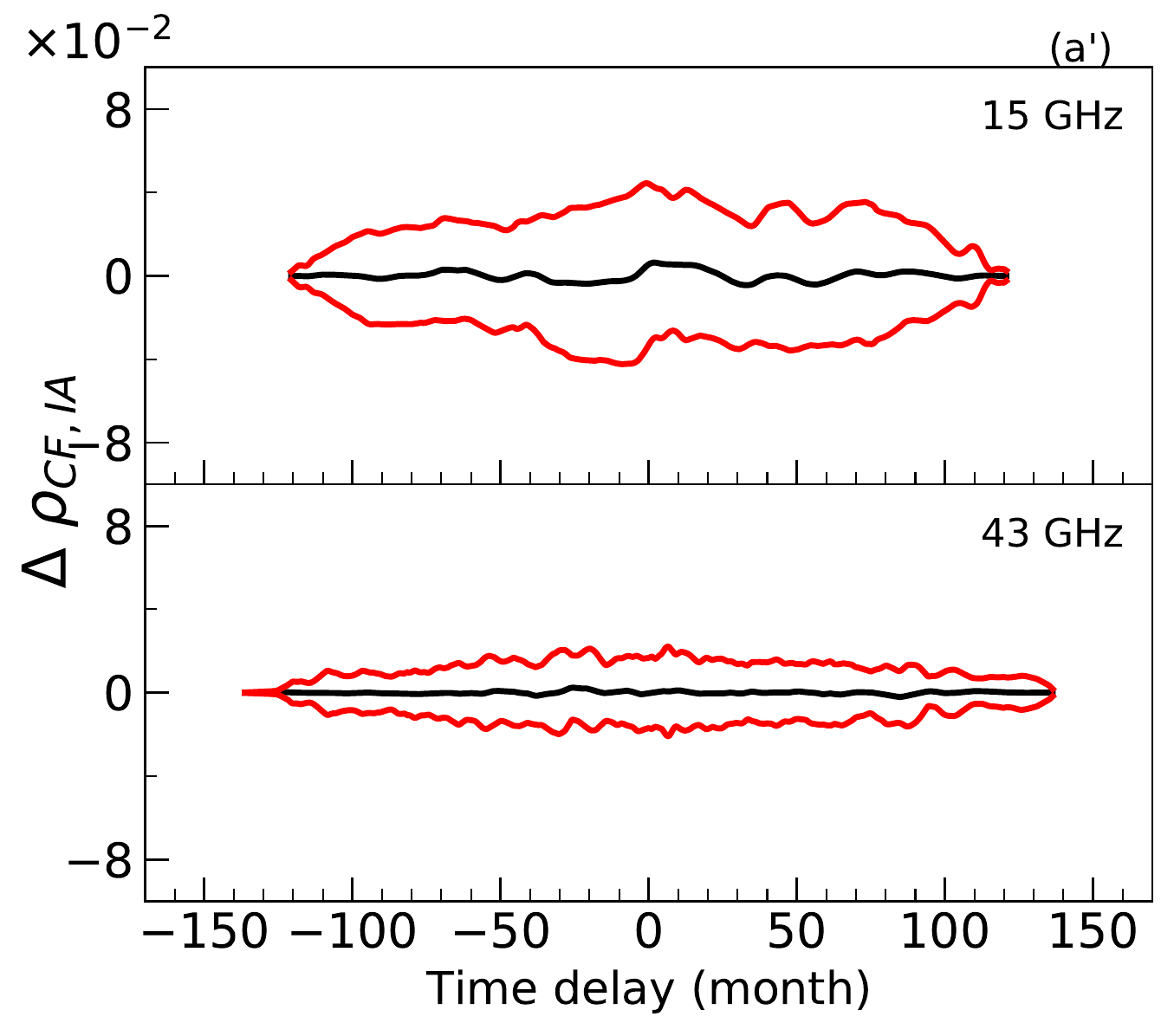}
    \includegraphics[width=4.4cm]{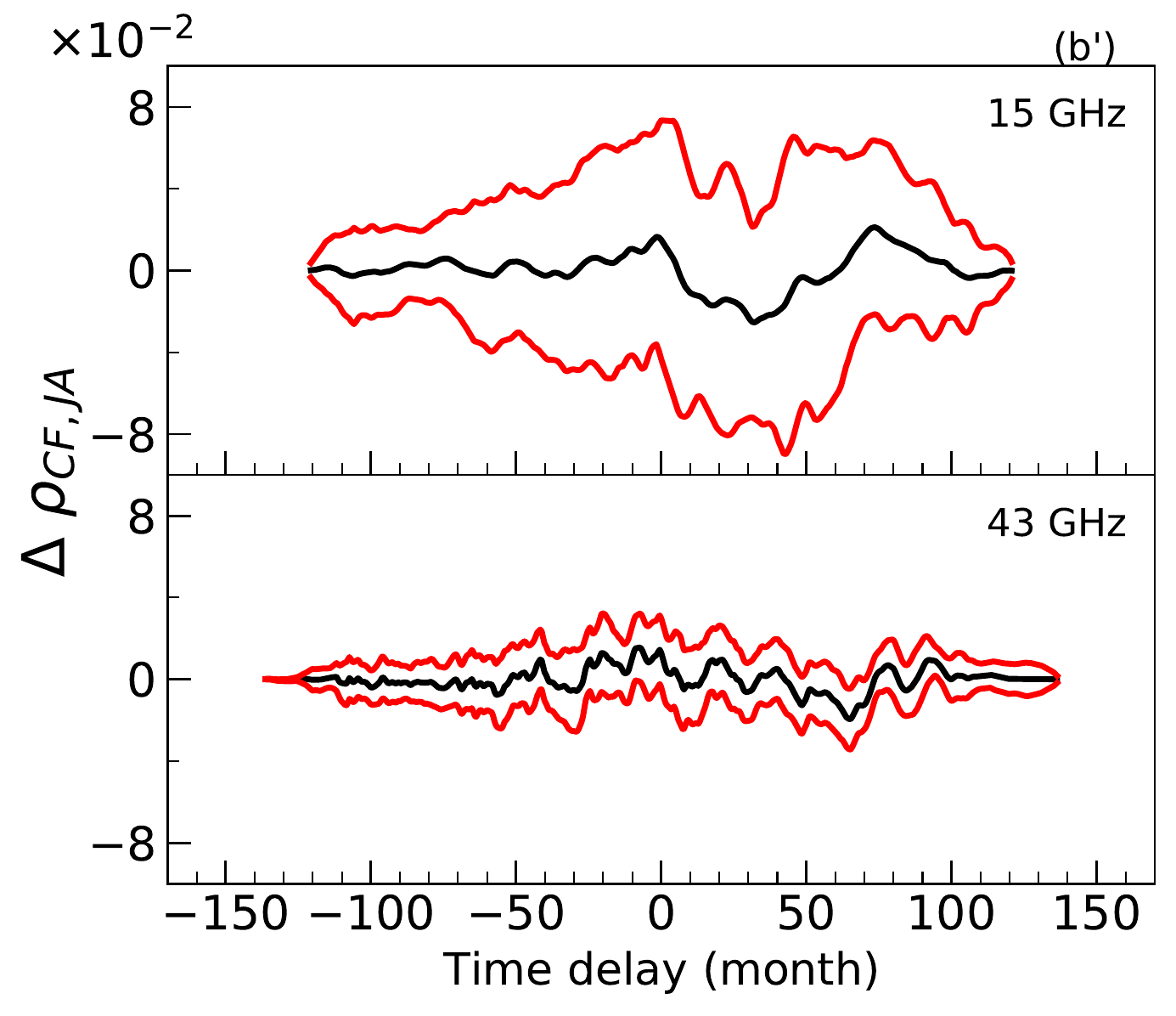}
    \includegraphics[width=4.4cm]{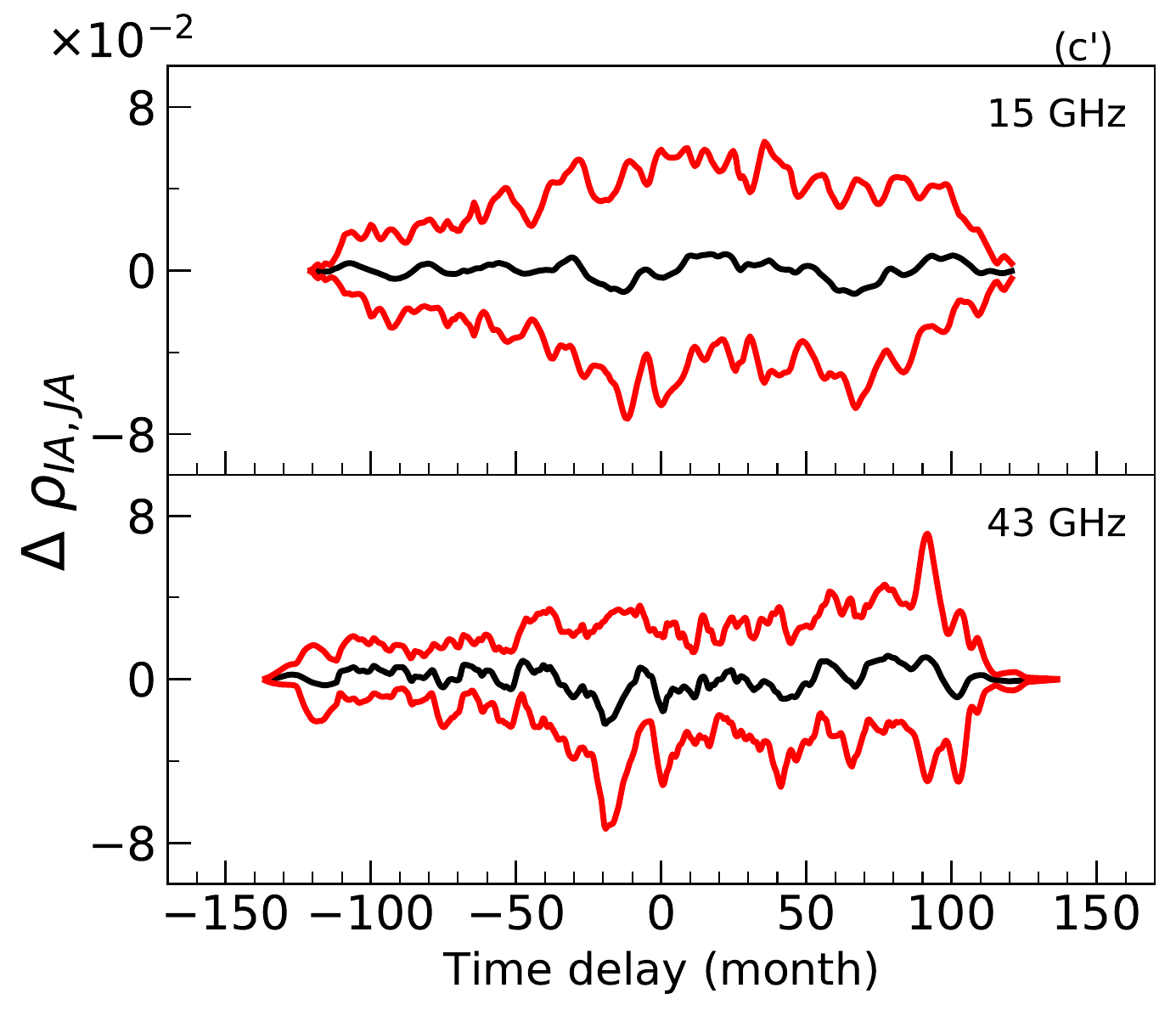}
    \includegraphics[width=4.4cm]{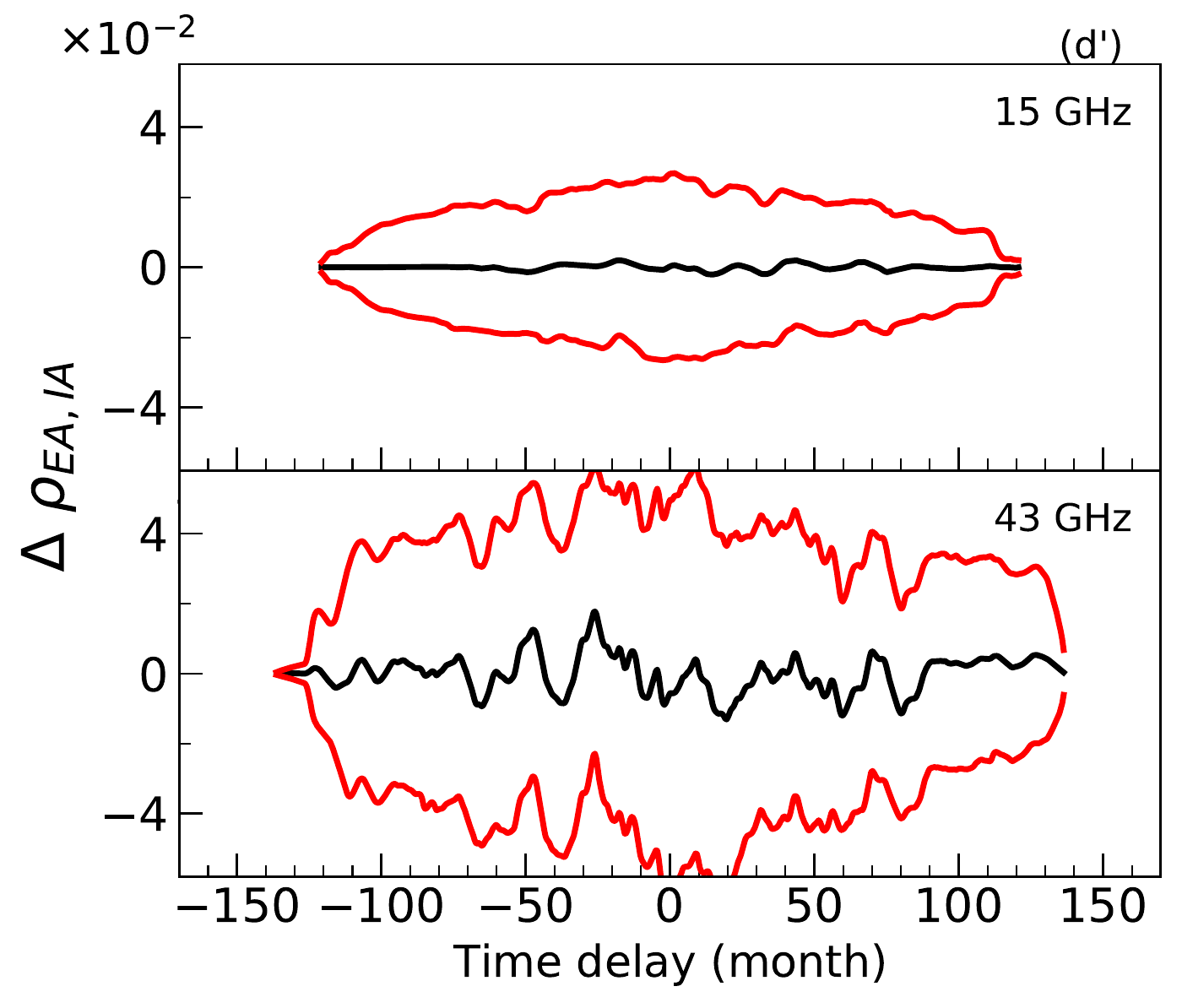}\\
    \caption{
    MC error analysis results of correlations between each pair observables.
    Panel (a/a'): The $\Delta\ \rho_{CF,IA}$ represents the correlation coefficient error values between each band's core flux density and inner-jet position angle.
    Panel (b/b'): The $\Delta\ \rho_{CF,JA}$ represents the correlation coefficient error values between each band's core flux density and jet position angle.
    Panel (c/c'): The $\Delta\ \rho_{IA,JA}$ represents the correlation coefficient error values between each band's inner-jet position angle and jet position angle.
    Panel (d/d'): The $\Delta\ \rho_{EA,IA}$ represents the correlation coefficient error values between each band's core EVPA and inner-jet position angle.
    The black line is connected by the mean values of the correlation coefficient error values of each time delay slice.
    The red lines is connected by the above mean values plus and minus 3 times standard deviation ($\sigma$), where $\sigma$ is determined by a Gaussian fit of the distribution of correlation coefficient error values over each time delay slice.
    Panel (a/b/c/d) is the result from segment-\Rmnum{1} data, and Panel (a'/b'/c'/d') is the result from segment-\Rmnum{2} data.}
    \label{fig:mc2}
\end{figure*}

The estimation of the significance of the cross-correlation achieved with the MC method is shown in Fig.~\ref{fig:mc1}.
Meanwhile, the MC trials involving the effect of the observational error on correlation indicate that the maximum peak of the $n$-th trial's correlation coefficient sequence ($\rho_{MC,n}$) and the corresponding time delay have slight changes, but the error of the observables does not affect the correlation results for each pair of observables~(Fig.~\ref{fig:mc2}).

The core flux density and the inner-jet position angle have a significant negative correlation with each other at each band. 
In segment \Rmnum{1}, at 8.6~GHz and 15 GHz, the core flux density appears to be anti-correlated with the jet position angle; and at 2.3~GHz, it shows a positive correlation between them (Fig.~\ref{fig:in-band-cor} (b)).
In segment \Rmnum{2}, the core flux density appears to be correlated with the jet position angle at 15~GHz and 43~GHz (Fig.~\ref{fig:in-band-cor} (b')).
What can be clearly seen in Fig.~\ref{fig:in-band-cor} (c), the inner-jet position angle and the jet position angle have a significant negative correlation at 2.3~GHz; a significant positive correlation at 8.6~GHz and 15~GHz.
In segment \Rmnum{2}, There is still a correlation between the two at a significance level greater than 95\%.
Fig.~\ref{fig:in-band-cor}~(d/d') show the {extremely} weak or even indistinguishable correlations between the core EVPA and the inner-jet position angle at 15~GHz and 43 GHz;
and the MC results show that we have no evidence of a correlation between the two as their peak significance is so low.

Inevitably, due to the insufficient sampling rate, interpolation of uneven sample data can affect the above statistical results.

\begin{table}
\centering
\caption{Correlation results between observables for each observing frequency.
Columns are as follow: 
(1) two observables for calculating the correlation;
(2) observing frequency;
(3--4) the correlation coefficient value of significant peaks; their corresponding time delays are in parentheses, in years.
In the first column, 'CF' represents core flux density, 'IA' represents inner-jet position angle, 'JA' represents jet position angle, 'EA' represents core EVPA.}
\label{tab:result-cc}
\begin{tabular}{c|ccc}
\cline{1-4}
                          & Frequency  & $\rho$ [segment \Rmnum{1}]  &   $\rho$ [segment \Rmnum{2}]  \\ \hline
\multirow{4}{*}{CF vs IA} & 2.3 GHz    & $-0.454$($-0.712$) &    --                                                  \\ 
                          & 8.6 GHz    & $-0.397$(0.505) &       --                                                 \\ 
                          & 15 GHz     & $-0.527$(0.000) &       $-0.470$(0.138)                                   \\ 
                          & 43 GHz     & --              &       $-0.418$($-2.125$)                                   \\ \hline
\multirow{4}{*}{CF vs JA} & 2.3 GHz    & 0.716(0.059)    &        --                                                  \\ 
                          & 8.6 GHz    & $-0.506$(0.216) &        --                                                  \\ 
                          & 15 GHz     & $-0.576$(0.941) &       0.640(2.618)                                        \\ 
                          & 43 GHz     & --              &       $-0.536$(0.000)                                     \\ \hline                                                                    
\multirow{4}{*}{IA vs JA} & 2.3 GHz    & $-0.556$(0.237) &        --                                                  \\  
                          & 8.6 GHz    & 0.912($-0.216$) &        --                                                  \\  
                          & 15 GHz     & 0.675(0.000)    &       $-0.429$(1.24)                                       \\  
                          & 43 GHz     & --              &       0.384(0.572)                                         \\ \hline
\end{tabular}
\end{table}

\section{Discussion}
\label{dis}

In this section, we present interpretations of the results of correlations between observables through two themes.
The first theme discusses the correlations between the core flux density, the inner-jet position angle, and the jet position angle. 
Another theme discusses the correlation between the core EVPA and the inner-jet position angle.

\subsection{Correlation analysis between the core flux density and the jet position angle at different spatial scales}

\subsubsection{The core flux density and the inner-jet position angle}

The higher the frequency of observation, the closer the core position is to the jet base, i.e., the 'core shift' effect~\citep{blandford.79.apj,konigl.81.apj,sokolovsky.11.aa,pushkarev.12.aa}.  
It makes sense that the radio core identified at different resolutions is seen as the different part of the body of the 'true' jet and the jet position angle at the core scale (i.e., inner-jet position angle) changes with the jet swings rather than standing still.

The results that the core flux density and the inner-jet position angle have a clear correlation conform to the prediction of the jet precession model or jet precession-nutation model proposed by~\citet{tateyama.04.apj} and~\citet{britzen.18.mn}, respectively.
This is because both models consider that the radiation variability in the radio jet could be attributed to the geometrical effects, that is, the precessing jet motion is accompanied by the viewing angle variations and consequent Doppler beaming changes.
Meanwhile, OJ 287 has the intrinsic helical shape in the source frame due to the jet precession mechanism;
and this source exhibits the bending jet morphology in the observer's frame, especially at higher observing frequencies.
In our visual inspection, we can see apparent bent jets at 43 GHz, while at lower observing frequencies, the visual bending effect is not obvious~(Fig.~\ref{fig:samples}).

Moreover, geometric effects affect the lower frequency radio emission more strongly than higher frequency radiation.
This is because the high-frequency radiations arise from higher-energy electrons with shorter cooling time scales, which can naturally lead to stronger intrinsic variability than at lower frequencies.
When all data have the same excellent data quality, our correlation analysis can confirm the influence of geometric effects in different bands.
Currently, under the premise that 2.3~GHz and 8.6~GHz data have basically the same time coverage and sampling rate, the correlation results show that the correlation between the core flux density and the inner-jet position angle at 8.6~GHz is significantly weaker than their correlation at 2.3~GHz; so we can conclude for the time being that the weakened correlation at high frequencies is most likely due to the differences in the geometric effect of different bands.

\subsubsection{The core flux density and the jet position angle}

\begin{figure*}
\centering
\includegraphics[width=8.5cm]{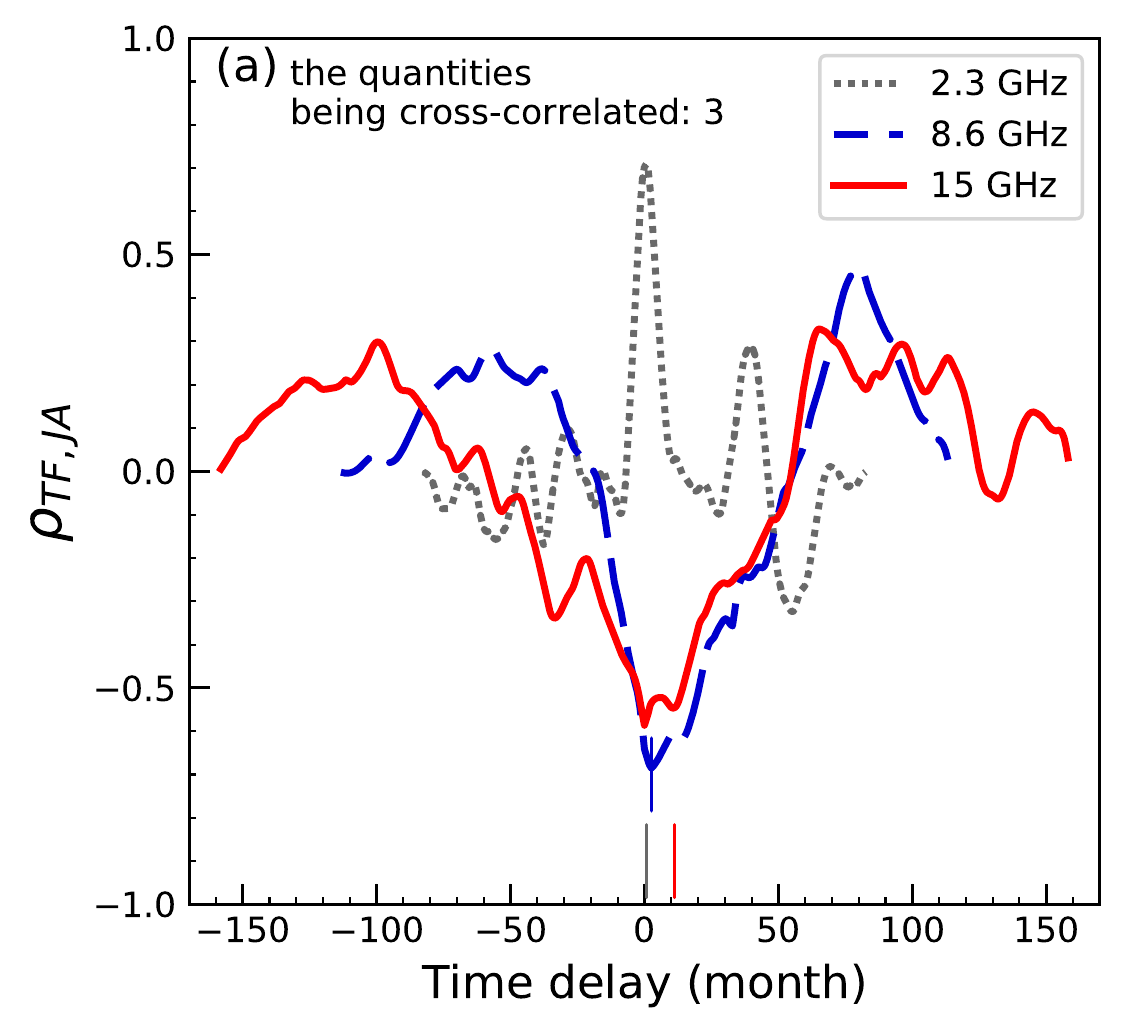}
\includegraphics[width=8.5cm]{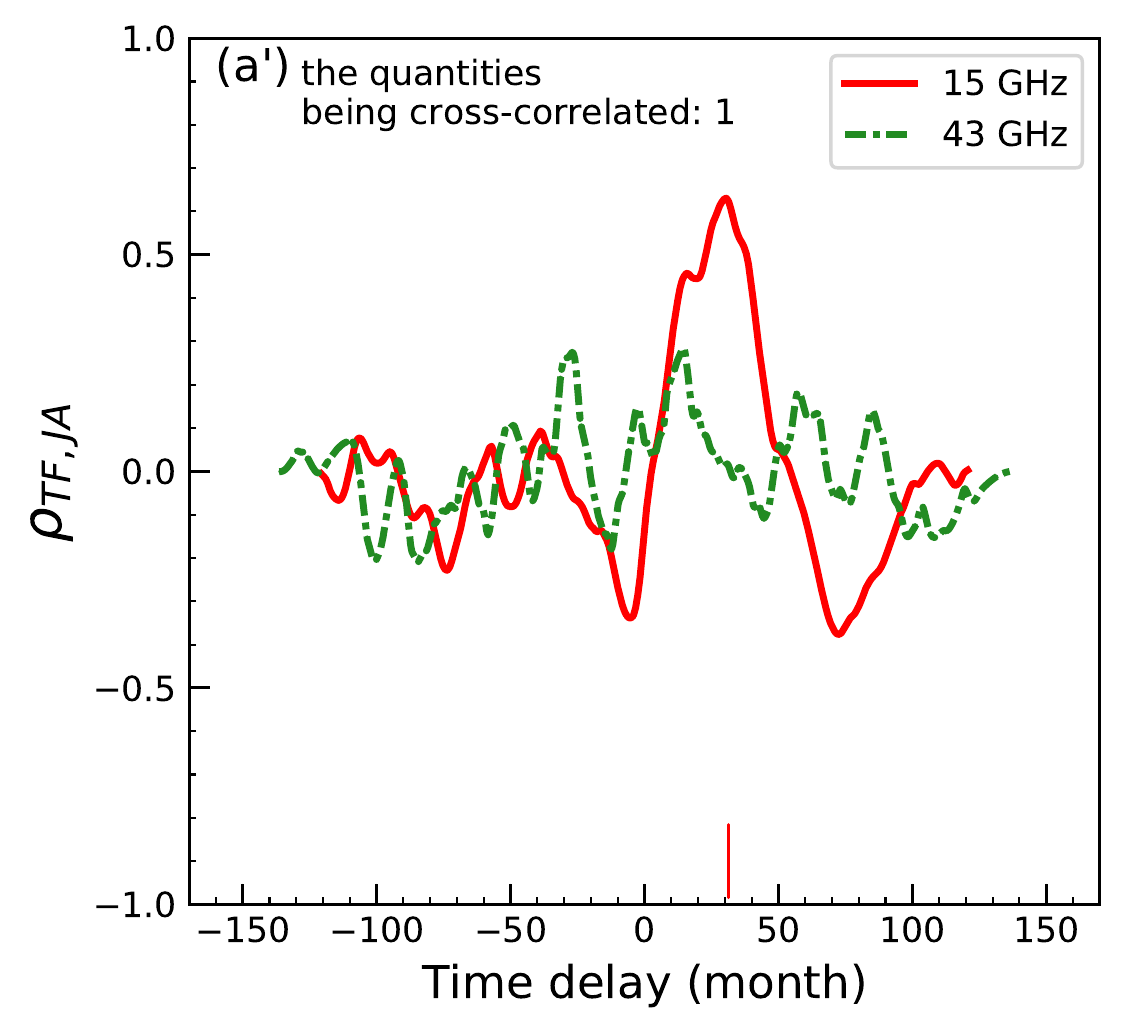}
\caption{Cross-correlation analysis of the total flux density and the jet position angle.
The short line marks the location of the cross-correlation peak of the core flux density and the jet position angle.
}
\label{fig:totalflux-jpa}
\end{figure*}

As mentioned in the previous subsection, we can infer that a correlation between the total flux density and the jet position angle for all bands if the source has a precession mechanism. 
To verify this inference, we calculated the correlation between the total flux density and the jet position angle in four bands.
In this calculation, for individual epoch observations, the total flux density is estimated as the sum of the flux densities of all fitted components.
Looking at Fig.~\ref{fig:totalflux-jpa}, it is apparent that the total flux density and the jet position angle show a significant positive correlation at 2.3~GHz and a clear negative correlation at 8.6~GHz and 15~GHz. 
The correlation coefficient values of the total flux density and the jet position angle at the cross-correlation peak of the core flux density and the jet position angle are 0.709, $-0.685$ and $-0.546$ (segment \Rmnum{1})/0.624 (segment \Rmnum{2})/ at the 2.3~GHz, 8.6~GHz,and 15~GHz bands, respectively.
These correlation results (Fig.~\ref{fig:totalflux-jpa}) indicate that 2.3~GHz, 8.6~GHz, 15~GHz bands completely match the above-mentioned physical inference.

\begin{figure*}
\centering
\includegraphics[width=4.2cm]{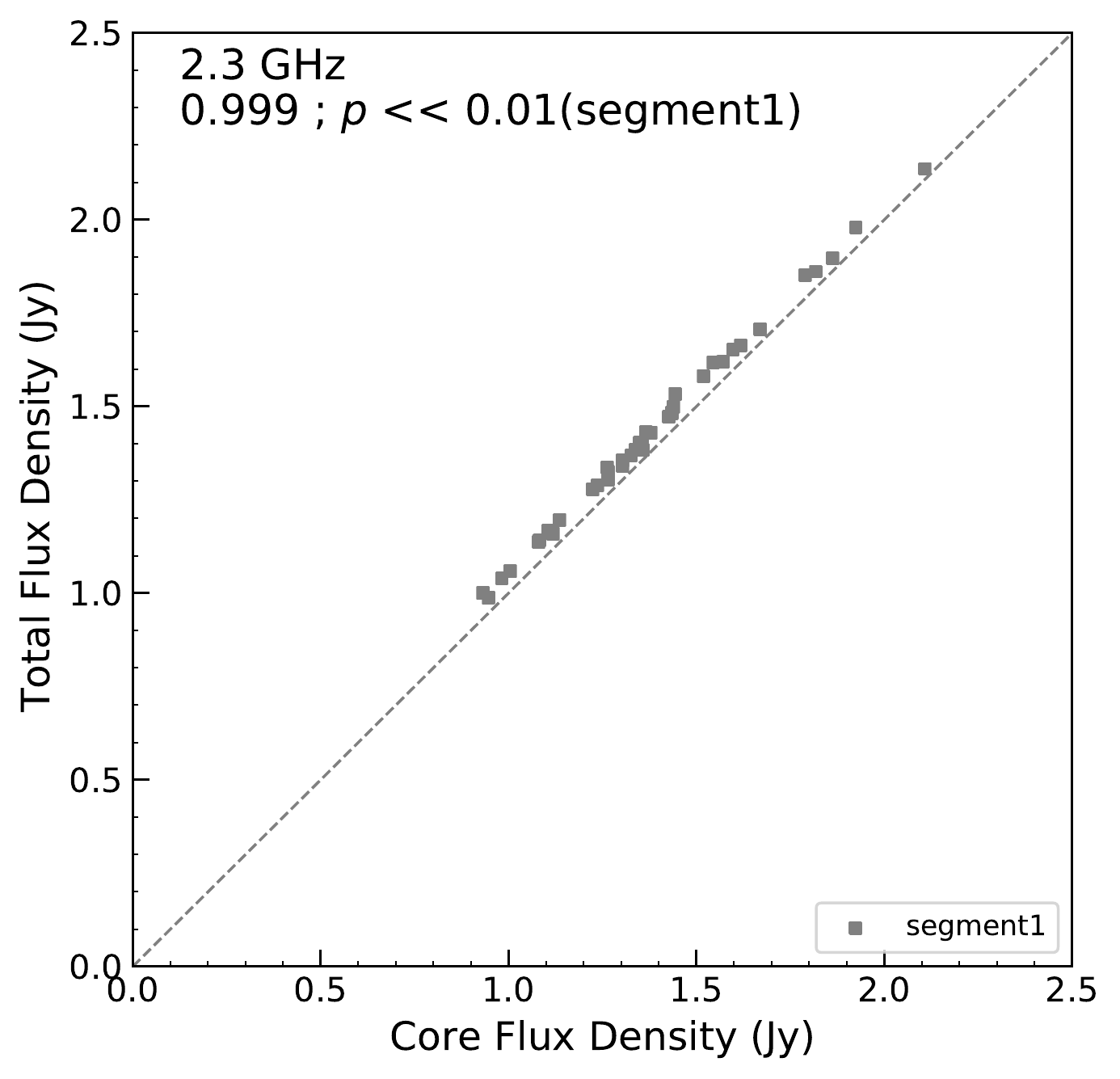}
\includegraphics[width=4.2cm]{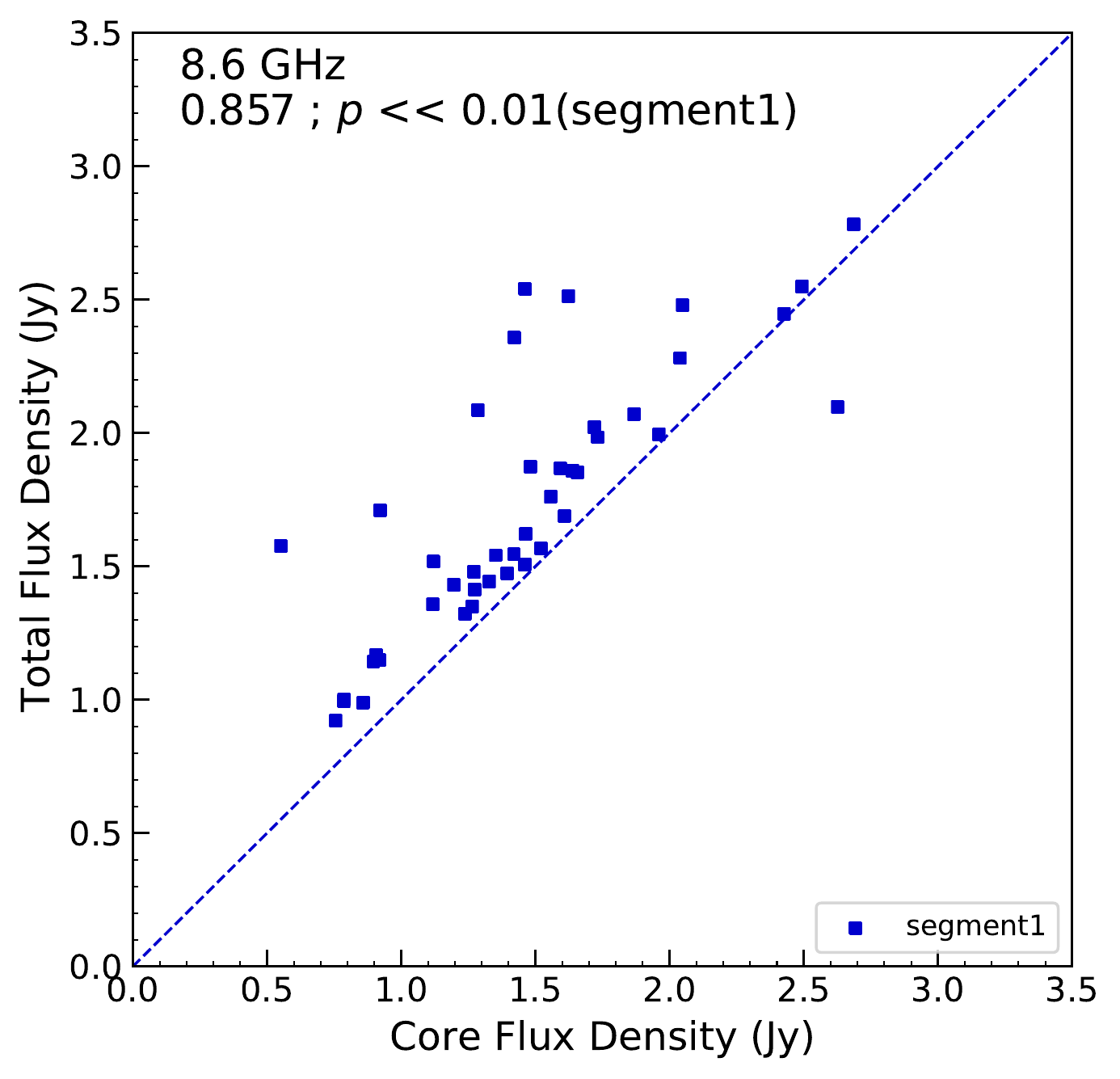}
\includegraphics[width=4.2cm]{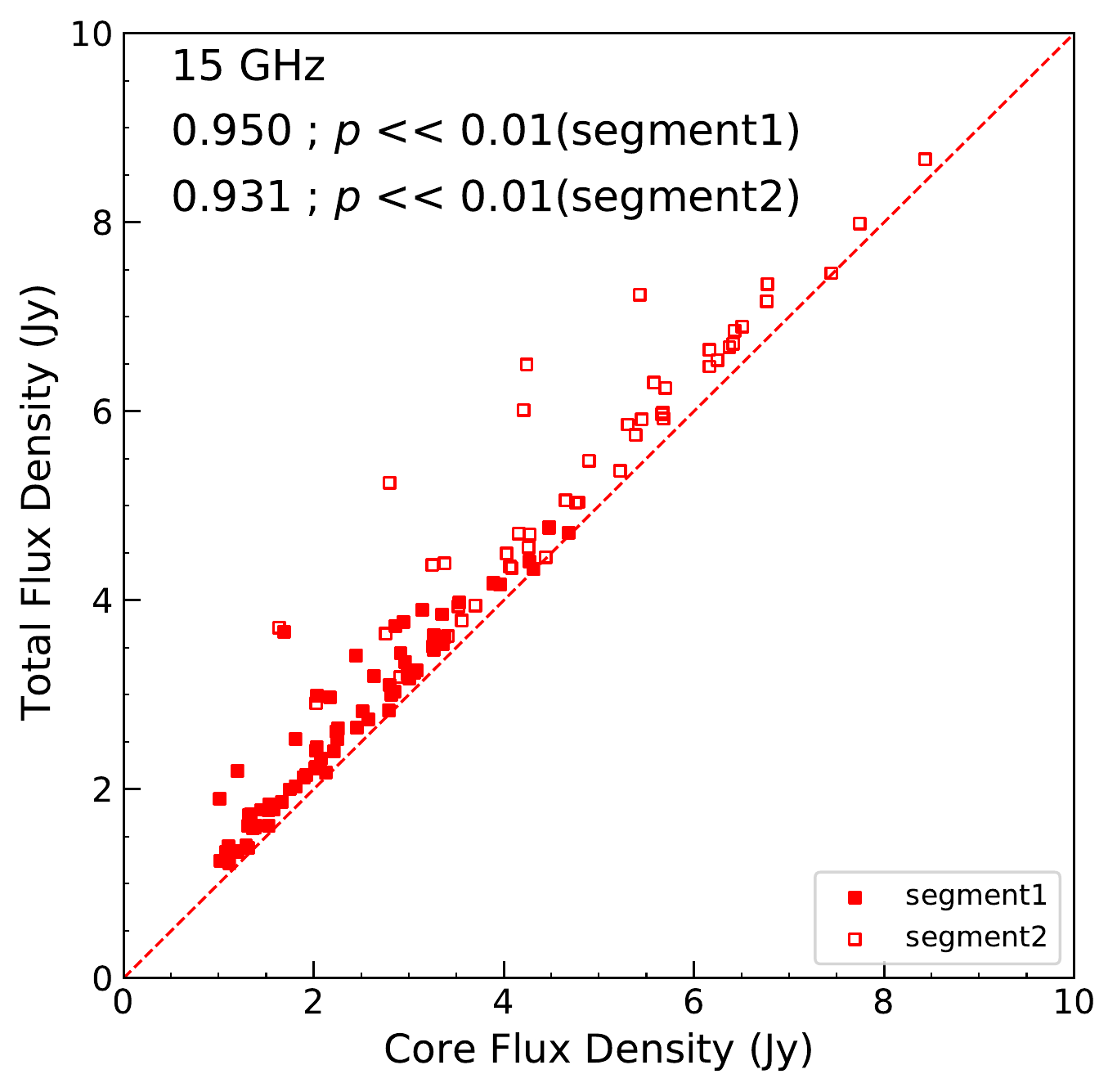}
\includegraphics[width=4.2cm]{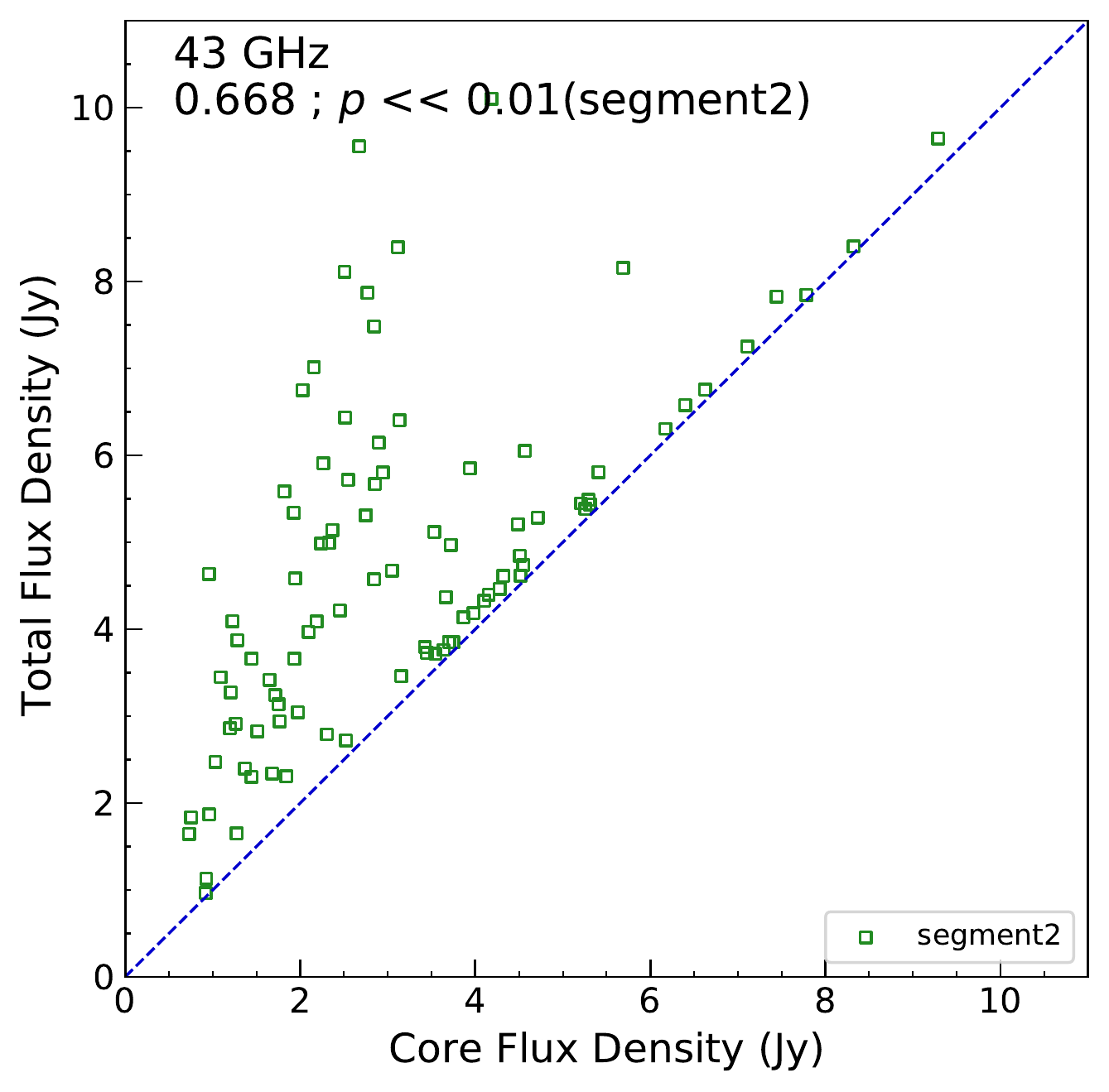}\\
\caption{Core flux density versus total flux density at 2.3 GHz, 8.6 GHz, 15 GHz, 43 GHz.
The correlation coefficient and p-value are marked in the figure.
An equality line is added to each subplot.
}
\label{fig:pearson}
\end{figure*}

\begin{figure}
    \centering
    \includegraphics[width=8cm]{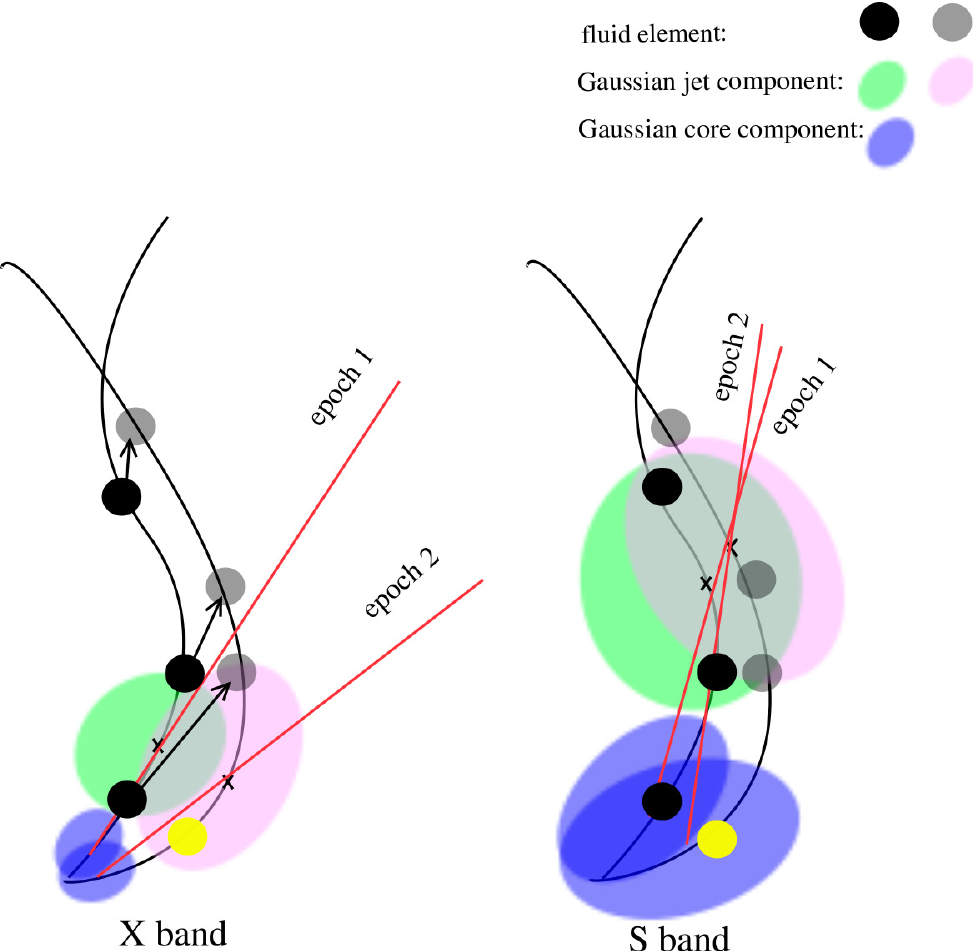}
    \caption{Sketch of a projection of the sky plane of a helical jet caused by precession.
    The 8.6~GHz (X-band) and 2.3~GHz (S-band) jet configurations are located in the left and right panels, respectively.
The black and gray circles represent the individual fluid elements in epoch 1 and epoch 2, respectively.
The yellow circle represents the emerging fluid element, we only draw an emerging fluid element for clear visualization.
Each black arrow connects each black circle with the corresponding grey circle, indicating the projected direction of the ballistic motion of each fluid element in space.
The blue oval indicates the fitted core component, and the green oval and pink oval indicate the fitted jet component at epochs 1 and 2.
The red line roughly represents the direction of the jet, i.e., jet position angle.}
    \label{fig:helical-jet}
\end{figure}

At 2.3~GHz, 8.6~GHz, and 15~GHz, we notice that the direction of the correlation between the core flux density and the jet position angle is consistent with that between the total flux density and the jet position angle.
At 2.3~GHz, the amplitude of the correlation between the core flux density and the jet position angle is basically the same as that between the total flux density and the jet position angle.
However, at 8.6~GHz and 15~GHz, the amplitude of the correlation between the core flux density and the jet position angle differs from that between the total flux density and the jet position angle.

We interpret these results as the emission of core dominance increases towards lower frequencies.
The main reason for this core dominance property is the limited jet length detected at 43~GHz, 15~GHz, 8.6~GHz, and 2.3~GHz and progressively decreasing angular resolution that results in blending in the core region swallowing more and more jet emissions.
In other words, the contribution of jet components flux density to the total flux density is minimal at 2.3~GHz and becomes greater at higher frequencies.
Two pieces of evidence support this contribution characteristic, i.e., the core dominance gradually increases with decreasing frequency.
The first is a near-perfect linear relationship between the core flux density and the total flux density appearing at 2.3 GHz, not at relatively higher frequencies (see Fig.~\ref{fig:pearson}).
The second is the source compactness factor, defined as the ratio of core flux density to total flux density, which decreases with increasing frequency.
The mean values of the compactness factor are 0.964, 0.875, 0.905, 0.755 for 2.3~GHz, 8.6~GHz, 15~GHz and 43~GHz, respectively.

For the core dominance gradually decreases as the frequency increases, the correlation between the total flux density and the jet position angle at 8.6~GHz and 15~GHz is not transferred to the correlation between the core flux density and the jet position angle.
Mathematically, suppose the correlation coefficients among three random variables A, B, C are
$\rho_{AB}$, $\rho_{AC}$, $\rho_{BC}$, respectively. If two correlation coefficients are known (e.g., $\rho_{AB}$ and $\rho_{AC}$),  
the value range of the third correlation coefficient ($\rho_{BC}$) is:
$\rho_{B C} \geqslant \rho_{A B} \rho_{A C}-\sqrt{1-\rho_{A B}^2} \sqrt{1-\rho_{A C}^2}$,
$\rho_{B C} \leqslant \rho_{A B} \rho_{A C}+\sqrt{1-\rho_{A B}^2} \sqrt{1-\rho_{A C}^2}$.
So through the correlation coefficient sequence of the total flux density and the jet position angle and the correlation coefficient of the core flux density and the total flux density at the time delay equal to 0, 
the value range of the correlation coefficient sequence of the core flux density and the jet position angle can be limited. 
We verify that our correlation coefficient sequence of the core flux density and the jet position angle is entirely within the limited range.

Interestingly, the correlation between the core flux density and the jet position angle shows the reverse correlation direction of 2.3~GHz and 8.6~GHz. 
We propose a plausible jet configuration based on the precession mechanism to explain this reverse correlation~(see Fig.~\ref{fig:helical-jet}).
Because the resolution of each observing frequency is different, the positions and scales of the fitted components for each band are also different, and the emerging components observed in some bands may not be resolved at lower resolutions.
Generally speaking, the precession model considered that the superluminal fluid elements are ejected from a jet 'nozzle' that precesses around a fixed axis.
\citet{stirling.03.mn} have found that some bright components evolve in straight paths, and their ejection angles are consistent with the 'nozzle' direction at the ejection time. 

Obviously, the curvature of the upstream jet is greater, while that of the downstream jet is less. It is conceivable that the curvature of the downstream jet is decreased by the previous components moving in the ejection direction.
As shown in Fig.~\ref{fig:helical-jet}, the geometric effect caused by the above physical reasons can lead to a configuration where the jet position angle changes in the opposite direction when observing the 2.3~GHz and 8.6~GHz.
However, the angle variation between the whole jet and our line of sight is consistent in each band.
Therefore, we can capture that the correlation direction between the core flux density and the jet position angle is exactly opposite at 2.3~GHz and 8.6~GHz, i.e., one positive correlation and one negative correlation.

\begin{figure*}
    \centering
    \includegraphics[width=17cm]{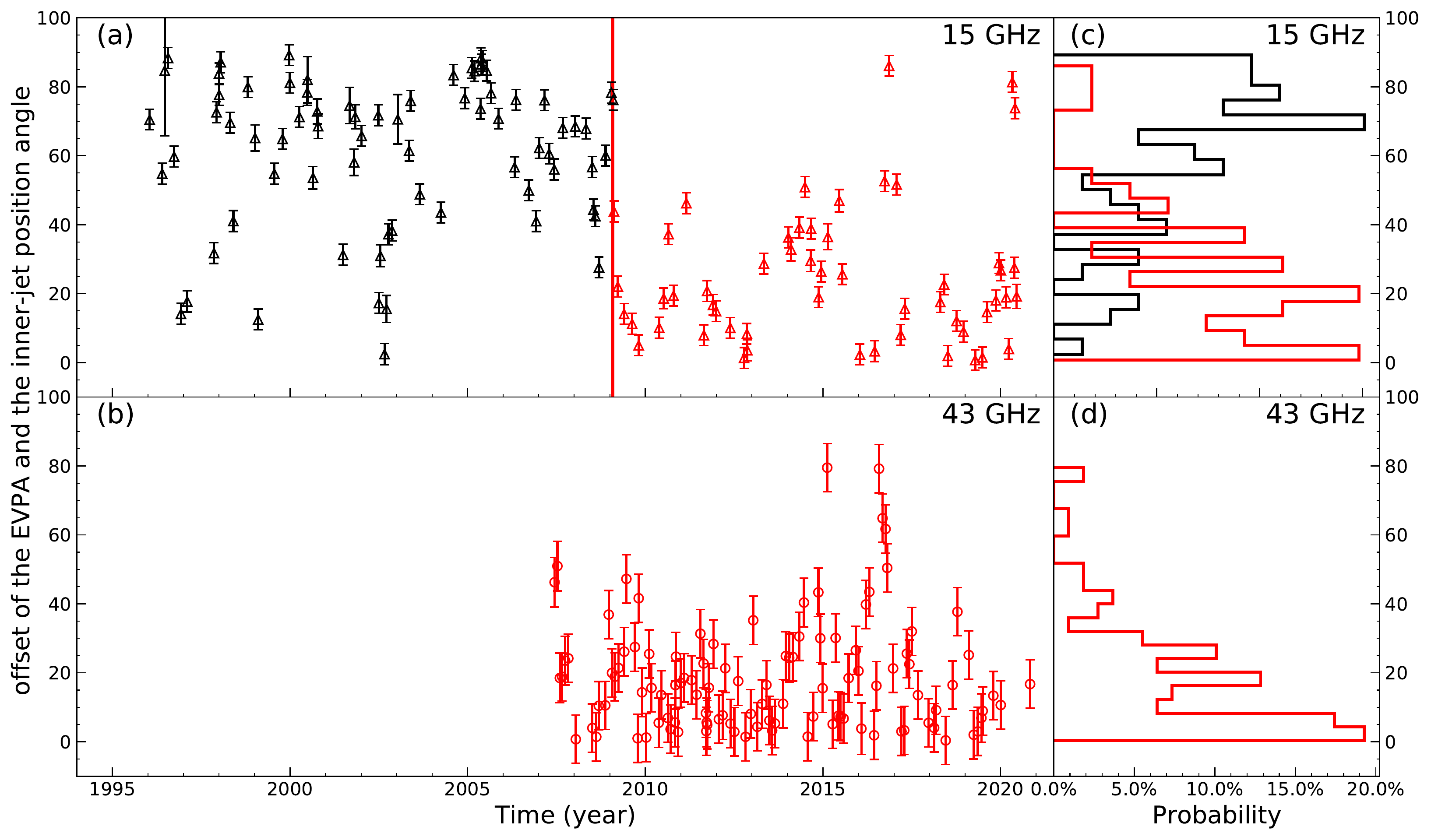}
    \caption{
    Time evolution of the offset of the core EVPA and the inner-jet position angle for OJ~287 at 15 GHz and 43 GHz.
    At 15 GHz, before January 2009, the offset values are marked in black triangles, and the offset values after January 2009 are marked in red triangles.
    Distributions of the offset of the EVPA and the inner-jet position angle at 15 GHz (Panel (c)) and 43 GHz (Panel (d)).
    At 15 GHz, the black border histogram shows the offset distribution of all data between January 1996 and January 2009 and the red border histogram shows the offset distribution of all data between February 2009 and June 2020.
    }
    \label{fig:offset}
\end{figure*}

\subsubsection{The inner-jet position angle and the jet position angle}

Distinctive features of the precessing jet include a gradually-curved helical jet. 
The definition of the inner-jet and main jet direction depends on the resolution and thus on the frequency of observation. 
The fitted components for different epochs on different observing frequencies are located at different positions of the S-like shaped apparent jet with complex curvature.
Hence the evolution of the jet position angle is also complicated; this situation does apply to the inside of the fitted core.
The ad hoc configuration of the unresolved core components may lead to an assembled kinematics of the inner-jet, which is either in-phase or out-of-phase with those of the main jet. 
For this reason, if the jet has a direction change near the base, the correlations with changing direction of the jet may be positive or negative depending on the viewing angle and the effective resolution, as new features emerge from the core region.

\subsection{Correlation analysis between the core EVPA and the inner-jet position angle}

The observed polarization could be attributed to the nature of the locally ordered magnetic field. 
If the magnetic field is tangled and isotropic, the linear polarization integrated and averaged in some parts of the jet would be absent, and the direction of the polarization is also disorganized.
In theory, the EVPA is orthogonal to the synchrotron magnetic field in optical thin regions. 
Virtually, all polarized emissions we observe in radio jet images arise in the optically thin regions, including polarization from the core region~\citep{gabuzda.21.galaxies}.

Usually, the AGN core EVPA is moderately alignment with the local jet direction, especially the BL Lac objects~\citep{gabuzda.00.mn,hodge.18.apj}.
However, it should be mentioned that some research findings on whether the EVPA and jet orientations in the core region tend to align are inconsistent, OJ~287 is no exception.
For OJ~287, some studies investigated the offset between core EVPA and local jet position angle for individual epoch~\citep{darcangleo.09.apj}, while others explored the average results of the evolution of their offset over time~\citep{sasada.18.apj,hodge.18.apj}. 
\citet{sasada.18.apj} reported that the differences between the core EVPA and the local jet position angle are scattered between 0$^\circ$ to 90$^\circ$ from June 2007 to March 2016 at 43 GHz.
\citet{hodge.18.apj} reported that the median absolute value and variance of the differences between EVPA and jet position angle in the core region is 70$^\circ$ and 30$^\circ$, respectively. 
These two statistics were calculated from a sample of 97 observation epochs over 11 years at 15 GHz.
The results above consistently conclude that the angle between the core EVPA and the local jet position angle of OJ~287 is not a relatively stable value for OJ~287.

We re-investigated the long-term variation of offset between the core EVPA and the inner-jet position angle\footnote{The local jet direction at the core refers to the inner-jet direction as defined previously in this paper.}, which is shown as a function of time (left panels of Fig.~\ref{fig:offset}). 
These results show that the offset is not always a stable value in either band, which is consistent with the previous studies.
Moreover, it can be noticed that the sudden jump in the offset occurred in early 2009 at 15 GHz. 
The inner-jet position angle caused this sudden jump in the offset because it underwent rapid rotation in January 2009, the core EVPA only shows erratic rather than abrupt variation over time, especially during 2009.
The offset distributions of the two frequencies are visualized by histograms (right panels of Fig.~\ref{fig:offset}), where the 15~GHz offset dataset is divided into two groups.
At 43 GHz, the offset's probability distribution exhibits J-shaped distribution;
and at 15 GHz, it consists of two J-shaped distributions that are mirror images of each other.

We have not found a strong evidence correlation between these two observables because the variations of the core EVPA and the inner-jet position angle cannot keep constant during long-term monitoring.
 This sign shows that EVPAs in the 15~GHz and 43~GHz unresolved core region are less ordered.

\subsection{Significance of correlation analysis}

The correlation analysis is crucial for further understanding the radio radiation variability causes and the jet magnetic field configurations.

This fact that the changing flux density and jet position angle are correlated may be of geometric origin.
Namely, the jet direction change can explain the variability of the radio flux density via viewing angle changes and the Doppler beaming.
The precession is a special and important mechanism that causes a change in the jet direction that brings out this geometric connection.
A common method for researchers to determine the presence of jet precession is to extract the jet position angles from the VLBI radio observations and use the obtained data to estimate precession model parameters~\citep{britzen.18.mn,dominik.21.mn}.
The better and more convenient way to search for AGNs with geometrically connected characteristics is to search for correlations between multiple observables, which can greatly improve the efficiency of searching for candidate sources with the jet precession.

The previous study has shown that the jet position angles of MOJAVE AGNs typically do not vary on the level of the core EVPAs, OJ~287 is an exceptional source that shows jet position angle change at the core scale in the plane of the sky~\citep{lister.13.aj,hodge.18.apj}.
In our study, the inner-jet position in OJ~287 shows time-varying characteristics at both 15 GHz and 43 GHz.
In the long-term multi-band observations, rotation in the inner-jet and core polarization directions provides unique opportunities to gain insight into the polarization and magnetic properties of the innermost jet.

\section{Summary}
\label{conclu}
This paper carries out a multivariate correlation analysis of observables of the OJ~287 which is a candidate with precessing jet.
We analyzed the correlation between the core flux density, the inner-jet position angle, and the jet position angle for four frequency bands and that between the core EVPA and the inner-jet position angle for 15 and 43 frequency bands.
The primary conclusions are summarized as follows.

In the four bands, there is a clear correlation between the core flux density, the inner-jet position angle and the jet position angle. 
The correlations that have been found can be explained by changes in the jet direction possibly due to precession.
We emphasize that the correlations between observables found in OJ~287 are not an accidental property of this particular dataset, and the above correlations can be found if a source has precession characteristics.
Admittedly, interference with correlation from relatively poor data quality is inevitable.

The offset of the core EVPA and the inner-jet position angle varies with time;
at 15 GHz, the offset jumps rapidly at this moment because of the sudden rotation of the inner-jet position angle in 2009.
In the 15~GHz and 43~GHz bands, there are no significant correlation between the core EVPA and the inner-jet position angle.
These results may not support the assertion that the magnetic field becomes more ordered down the jet, even though this property is clearly demonstrated by~\citet{pushkarev.17.galaxies}, who reported an increase in fractional polarization with core separation.
Nevertheless, we emphasize that longer-term VLBI monitoring is needed to confirm or refute this assertion.

\section*{Acknowledgments}

This work is supported by the National Key R\&D Intergovernmental Cooperation Program of China (2022YFE0133700), the Regional Collaborative Innovation Project of Xinjiang Uyghur Autonomous Region (2022E01013), the National Natural Science Foundation of China (12173078, 11773062), and the Light of West China Program of the Chinese Academy of Sciences (2017-XBQNXZ-A-008). 
This research has made use of data from the MOJAVE database that is maintained by the MOJAVE team~\citep{lister.18.apjs}.
This study makes use of VLBA data from the VLBA-BU BLAZAR Monitoring Program (BEAM-ME and VLBA-BU-BLAZAR;
http://www.bu.edu/blazars/BEAM-ME.html), funded by NASA through the Fermi Guest Investigator Program. The VLBA is an instrument of the National Radio Astronomy Observatory.
The National Radio Astronomy Observatory is a facility of the National Science Foundation operated by Associated Universities, Inc.

\software{
\textsc{AIPS}~\citep{greisen.03.software}, Python, \textsc{SAND}~\citep{zhangm.16.sand}.
}

\bibliography{main}{}
\bibliographystyle{aasjournal}

\end{document}